\documentclass[12pt]{article}
\pdfoutput=1
\usepackage{graphics}
\usepackage[dvips]{graphicx}
\usepackage{mathrsfs}
\usepackage{amssymb}
\usepackage{amsbsy}
\usepackage{amsfonts}
\usepackage{amsmath}
\usepackage{verbatim}
\usepackage{float}
\usepackage{cancel}
\usepackage{multicol}
\usepackage{multirow}
\usepackage{wrapfig}
\usepackage[scaled=.88]{helvet}
\usepackage{cite}
\usepackage{mcite}

\newcommand{\be}{\begin{equation}}
\newcommand{\ee}{\end{equation}}
\newcommand{\bea}{\begin{eqnarray}}
\newcommand{\eea}{\end{eqnarray}}

\def\4vol{{\int d^4x \sqrt{-g}}}

\def\beq{\begin{equation}}
\def\eeq{\end{equation}}
\def\bitem{\begin{itemize}}
\def\eitem{\end{itemize}}

\newcommand{\nc}{\newcommand}

\nc{\nt}{\tilde{N}}
\nc{\ra}{\rightarrow}
\nc{\lsim}{\begin{array}{c}\,\sim\vspace{-21pt}\\< \end{array}}
\nc{\gsim}{\begin{array}{c}\sim\vspace{-21pt}\\> \end{array}}
\nc{\tnt}{\tilde{N}}
\nc{\tst}{\tilde{t}}

\nc{\LL}{L}
\nc{\vv}{\tilde{v}}
\nc{\ve}{{\textrm{\fontfamily{phv}\selectfont v}}_{\textrm{\fontfamily{phv}\selectfont e}}}

\title{
\vspace*{-1cm} 
\Large\textbf{``$L=R$" -- $U(1)_R$ Lepton Number at the LHC}
\vspace*{1cm} 
\author{
\textbf{Claudia Frugiuele$^{a,b}$, Thomas Gr\'egoire$^a$,}\\[0.2cm]  
\textbf{Piyush Kumar$^{c,d}$, Eduardo Pont\'on$^c$}\\ \\[0.5cm]
$^a$\normalsize\emph{Ottawa-Carleton Institute for Physics, Department of Physics, Carleton University} \\
\normalsize\emph{1125 Colonel By Drive, Ottawa, K1S 5B6 Canada
}\\ [0.3em]
$^b$\normalsize\emph{Theoretical Physics Department, Fermilab} \\
\normalsize\emph{P.O. Box 500, Batavia, IL 60510 USA} \\ [0.3em]
$^c$\normalsize\emph{Department of Physics \& ISCAP, Columbia University} \\ 
\normalsize\emph{538 W. 120th St, New York, NY 10027, USA} \\ [0.3em]
$^{d}$\normalsize\emph{Department of Physics,Yale University, New Haven, CT 06520 USA} 
}
}
\begin{document}
\setcounter{page}{0}
\maketitle
\begin{abstract}

We perform a detailed study of a variety of LHC signals in
supersymmetric models where lepton number is promoted to an
(approximate) $U(1)_R$ symmetry.  Such a symmetry has interesting
implications for naturalness, as well as flavor- and CP-violation,
among others.  Interestingly, it makes large sneutrino vacuum
expectation values phenomenologically viable, so that a slepton
doublet can play the role of the down-type Higgs.  As a result, (some
of) the leptons and neutrinos are incorporated into the chargino and
neutralino sectors.  This leads to characteristic decay patterns that
can be experimentally tested at the LHC. The corresponding collider
phenomenology is largely determined by the new approximately conserved
quantum number, which is itself closely tied to the presence of
``leptonic R-parity violation".  We find rather loose bounds on the
first and second generation squarks, arising from a combination of
suppressed production rates together with relatively small signal
efficiencies of the current searches.  Naturalness would indicate that
such a framework should be discovered in the near future, perhaps
through spectacular signals exhibiting the lepto-quark nature of the
third generation squarks.  The presence of fully visible decays, in
addition to decay chains involving large missing energy (in the form
of neutrinos) could give handles to access the details of the spectrum
of new particles, if excesses over SM background were to be observed.
The scale of neutrino masses is intimately tied to the source of
$U(1)_R$ breaking, thus opening a window into the $R$-breaking sector
through neutrino physics.  Further theoretical aspects of the model
have been presented in the companion paper \protect\cite{FGKP}.

\end{abstract}

\thispagestyle{empty}
\newpage
\setcounter{page}{1}

\tableofcontents

\section{Introduction}
\label{Intro}

The recent discovery at the LHC of a Higgs-like signal at $\sim
125~{\rm GeV}$ has put the general issue of electroweak symmetry
breaking under a renewed perspective.  In addition, the absence of
other new physics signals is rapidly constraining a number of
theoretically well-motivated scenarios.  One of these concerns
supersymmetry, which in its minimal version is being tested already
above the TeV scale.  In view of this, it is pertinent to consider
alternate realizations that could allow our prejudices regarding
e.g.~naturalness to be consistent with the current experimental
landscape, within a supersymmetric framework.  At the same time, such
scenarios might motivate studies for non-standard new physics signals.

One such non-standard realization of supersymmetry involves the
possible existence of an approximately conserved $R$-symmetry at the
electroweak scale~\cite{Hall:1990hq, Nelson:2002ca, Fox:2002bu,
Chacko:2004mi, Kribs:2007ac, Benakli:2008pg,
Choi:2008pi,Kribs:2010md,Abel:2011dc,Davies:2011mp,Kumar:2011np,Bertuzzo:2012su}.
It is known that one of the characteristics of such scenarios, namely
the Dirac character of the gauginos (in particular, gluinos), can
significantly soften the current exclusion
bounds~\cite{Heikinheimo:2011fk,Kribs:2012gx}.  At the same time, an
approximate $R$-symmetry which extends to the matter sector, could end
up playing a role akin to the GIM mechanism in the SM, thereby
allowing to understand the observed flavor properties of the light
(SM) particles.  As advocated in Ref.~\cite{Frugiuele:2011mh}, a
particularly interesting possibility is that the $R$-symmetry be an
extension of lepton number (see also \cite{Gherghetta:2003he}).  In a companion paper \cite{FGKP}, we
classify the phenomenologically viable R-symmetric models, and present
a number of theoretical and phenomenological aspects of the case in
which R-symmetry is tied to the lepton number.  Such a realization
involves the ``R-parity violating (RPV) superpotential operators", $W
\supset \lambda LLE^c + \lambda' LQD^c$ where, unlike in standard RPV
scenarios, there is a well-motivated structure for the new $\lambda$
and $\lambda'$ couplings, some of them being related to (essentially)
known Yukawa couplings.  Although, at first glance, one might think
that such a setup, possibly with a preponderance of leptonic signals
should be rather constrained, we shall establish here that this is not
the case.  In fact, the scenario is easily consistent with most of the
superparticles lying below the TeV scale.  Only the Dirac gauginos are
expected to be somewhat above the TeV scale, which may be completely
consistent with naturalness considerations.  As we will see, the light
spectrum is particularly simple: there is no LR mixing in the scalar
sector, and there is only one light (Higgsino-like)
neutralino/chargino pair.  At the same time, it turns out that
the~\textit{ (electron) sneutrino vev} can be sizable, since it is not
constrained by neutrino masses (in contrast to that in standard RPV
models).  This is because the Lagrangian (approximately) respects
lepton number, which is here an $R$ symmetry, and the sneutrinos do
not carry lepton number.  Such a sizable vev leads to a mixing of the
neutralinos/charginos above with the neutrino and charged lepton
sectors ($\nu_e$ and $e^-$ to be precise), which results in novel
signatures and a rather rich phenomenology.  Although the flavor
physics can in principle also be very rich, we will not consider this
angle here.

We give a self-contained summary of all the important physics aspects
that are relevant to the collider phenomenology in
Section~\ref{review}.  This will also serve to motivate the specific
spectrum that will be used as a basis for our study.  
In Section~\ref{sec:decaymodes}, we put together
all the relevant decay widths, as a preliminary step for exploring the
collider phenomenology.  In Section~\ref{12Pheno} we discuss the
current constraints pertaining to the first and second generation
squarks, concluding that they can be as light as $500-700~{\rm GeV}$.
We turn our attention to the third generation phenomenology in
Section~\ref{3Pheno}, where we show that naturalness considerations
would indicate that interesting signals could be imminent, if this
scenario is relevant to the weak scale.  In Section~\ref{predictions},
we summarize the most important points, and discuss a number of
experimental handles that could allow to establish the presence of a
leptonic $R$ symmetry at the TeV scale.

\section{$U(1)_R$ Lepton Number: General Properties}
\label{review}

Our basic assumption is that the \textit{Lagrangian} at the TeV scale
is approximately $U(1)_R$ symmetric, with the scale of $U(1)_R$
symmetry breaking being negligible for the purpose of the
phenomenology at colliders.  Therefore, we will concentrate on the
exact $R$-symmetric limit, which means that the patterns of production
and decays are controlled by a new (approximately) conserved quantum
number.  We will focus on the novel case in which the $R$-symmetry is
an extension of the SM lepton number.  Note that this means that the
extension of lepton number to the new (supersymmetric) sector is
non-standard.

\subsection{The Fermionic Sector}
\label{sec:inos}

As in the MSSM, the new fermionic sector is naturally divided into
strongly interacting fermions (gluinos), weakly interacting but
electrically charged fermions (charginos) and weakly interacting
neutral fermions (neutralinos).  However, in our framework there are
important new ingredients, and it is worth summarizing the physical
field content.  This will also give us the opportunity to introduce
useful notation.

\subsubsection{Gluinos}
\label{sec:gluinos}

One of the important characteristics of the setup under study is the
Dirac nature of gauginos.  In the case of the gluon superpartners,
this means that there exists a fermionic colored octet (arising from a
chiral superfield) that marries the fermionic components of the
$SU(3)_C$ vector superfield through a Dirac mass term: $M^D_3
\tilde{g}_\alpha^a \tilde{o}^{a \alpha} + {\rm h.c.}$, where $a$ is a
color index in the adjoint representation of $SU(3)_C$ and $\alpha$ is
a Lorentz index (in 2-component notation).  Whenever necessary, we
will refer to $\tilde{o}$ as the \textit{octetino components}, and to
$\tilde{g}$ as the \textit{gluino components}.  For the most part, we
will focus directly on the 4-component fermions $\tilde{G}^a =
\{\tilde{g}^a_\alpha, \bar{\tilde{o}}^{a \dot{\alpha}}\}$ and we will
refer to them as \textit{(Dirac) gluinos}, since they play a role
analogous to Majorana gluinos in the context of the MSSM. However,
here the Majorana masses are negligible (we effectively set them to
zero) and, as a result, the Dirac gluinos carry an approximately
conserved ($R$) charge.  In particular, $R(\tilde{g}) = - R(\tilde{o})
= 1$, so that $R(\tilde{G}) = 1$.  $R$-charge (approximate)
conservation plays an important role in the collider phenomenology.

The Dirac gluino pair-production cross-section is about twice as large
as the Majorana gluino one, due to the larger number of degrees of
freedom.  Assuming heavy squarks, and within a variety of simplified
model scenarios, both ATLAS~\cite{Okawa:2011xg,Aad:2012hm,Aad:2012hn}
and CMS~\cite{:2012jx,:2012mf,CMS-PAS-SUS-11-016,ICHEP-CMS} have set
limits on Majorana gluinos in the $0.9-1~{\rm TeV}$ range.  As
computed with Prospino2~\cite{Beenakker:1996ed} in this limit of
decoupled squarks, the NLO Majorana gluino pair-production
cross-section is $\sigma^{\tilde{g}\tilde{g}}_{\rm
Majorana}(M_{\tilde{g}} = 1~{\rm TeV}) \approx 8~{\rm fb}$ at the
$7~{\rm TeV}$ LHC run.  Although, for the same mass, the Dirac gluino
production cross-section is significantly larger, it also falls very
fast with the gluino mass so that the above limits, when interpreted
in the Dirac gluino context, do not change qualitatively.  Indeed,
assuming a similar K-factor in the Dirac gluino case, we find a NLO
pair-production cross section of $\sigma^{\tilde{g}\tilde{g}}_{\rm
Dirac}(M^D_3 = 1.08~{\rm TeV}) \approx 8~{\rm fb}$.  Nevertheless,
from a theoretical point of view the restrictions on Dirac gluinos
coming from naturalness considerations are different from those on
Majorana gluinos, and allow them to be significantly heavier.  We will
take $M_{\tilde{g}} \equiv M^D_3 = 2~{\rm TeV}$ to emphasize this
aspect.  This is sufficiently heavy that direct gluino pair-production
will play a negligible role in this study.\footnote{However, at 14
TeV, with $\sigma^{\tilde{g}\tilde{g}}_{\rm Dirac}(M^D_3 = 2~{\rm
TeV}) \approx 3~{\rm fb}$, direct gluino pair-production may become
interesting.  The K-factor ($\approx 2.6$) is taken from the Majorana
case, as given by Prospino2.  This production cross-section is
dominated by gluon fusion, and is therefore relatively insensitive to
the precise squark masses.} At the same time, such gluinos can still
affect the pair-production of squarks through gluino t-channel
diagrams, as discussed later (for the gluinos to be effectively
decoupled, as assumed in e.g.~\cite{Kribs:2012gx}, they must be
heavier than about $5~{\rm TeV}$).

\subsubsection{Charginos}
\label{sec:charginos}

We move next to the chargino sector.  This includes the charged
fermionic $SU(2)_L$ superpartners (winos) $\tilde{w}^\pm$ and the
charged \textit{tripletino} components, $\tilde{T}^+_u$ and
$\tilde{T}^-_d$, of a fermionic adjoint of $SU(2)_L$ (arising from a
triplet chiral superfield).  It also includes the charged components
of the Higgsinos, $\tilde{h}^+_u$ and $\tilde{r}^-_d$.  The use of the
notation $\tilde{r}^-_d$ instead of $\tilde{h}^-_d$ indicates that,
unlike in the MSSM, the neutral ``Higgs" component $R_d^0$ does not
acquire a vev.  Rather, in our setup, the role of the down-type Higgs
is played by the \textit{electron sneutrino} $\tilde{\nu}_e$ (we will
denote its vev by $\ve$).  As a result, the LH electron $e^-_L$ mixes
with the above charged fermions, and becomes part of the chargino
sector (as does the RH electron field $e^c_R$).  Besides the gauge
interactions, an important role is played by the superpotential
operator $W \supset \lambda_u^T H_u T R_d$, where $T$ is the $SU(2)_L$
triplet superfield~\cite{FGKP}.

The pattern of mixings among these fermions is dictated by the
conservation of the electric as well as the $R$-charges:
$R(\tilde{w}^\pm) = R(e^c_R) = R(\tilde{r}^-_d) = +1$ and
$R(\tilde{T}^+_u) = R(\tilde{T}^-_d) = R(e^-_L) = R(\tilde{h}^+_u) =
-1$.  In 2-component notation, we then have that the physical
charginos have the composition
\bea
\tilde{\chi}_i^{++} &=& V^+_{i \tilde{w}} \, \tilde{w}^+ + V^+_{i e} \, e^c_R~,
\nonumber \\ [0.4em]
\tilde{\chi}_i^{--} &=& U^+_{i \tilde{t}} \, \tilde{T}_{d}^- + U^+_{i e} \, e^-_L~,
\label{eq:ch1comp}
\nonumber \\ [0.4em]
\tilde{\chi}_i^{+-} &=&  V^-_{i \tilde{t}} \, \tilde{T}_{u}^+ + V^-_{i u} \, \tilde{h}_u^+~,
\nonumber \\ [0.4em]
\tilde{\chi}_i^{-+} &=& U^-_{i \tilde{w}} \, \tilde{w}^- + U^-_{i d} \,  \tilde{r}_d^-~,
\nonumber
\label{eq:ch2comp}
\eea
where $i=1,2$.  The notation here emphasizes the conserved electric
and $R$-charges, by indicating them as superindices,
e.g.~$\tilde{\chi}_i^{+-} $ denoting the two charginos with $Q = +1$
and $R = -1$.  The $U^\pm$, $V^\pm$ are $2\times 2$ unitary matrices
that diagonalize the corresponding chargino mass matrices.  The
superindex denotes the product $R \times Q$, while the subindices in
the matrix elements should have an obvious interpretation.  We refer
the reader to Ref.~\cite{FGKP} for further details.  In this work we
will not consider the possibility of CP violation, and therefore all
the matrix elements will be taken to be real.  The above states are
naturally arranged into \textit{four} 4-component Dirac fields,
$\tilde{X}^{++}_i = (\tilde{\chi}^{++}_i,
\overline{\tilde{\chi}^{--}_i})$ and $\tilde{X}^{+-}_i =
(\tilde{\chi}^{+-}_i, \overline{\tilde{\chi}^{-+}_i})$, for $i=1,2$,
whose charge conjugates will be denoted by $\tilde{X}^{--}_i$ and
$\tilde{X}^{-+}_i$.  In this notation, $e = \tilde{X}^{--}_1$
corresponds to the physical electron (Dirac) field.

As explained in the companion paper~\cite{FGKP}, precision
measurements of the electron properties place bounds on the allowed
admixtures $V^+_{1 \tilde{w}}$ and $U^+_{1 \tilde{t}}$, that result in
a lower bound on the Dirac masses, written as $M^D_2 (\tilde{w}^+
\tilde{T}^{-}_d + \tilde{w}^- \tilde{T}^{+}_u) + {\rm h.c.}$.  This
lower bound can be as low as $300~{\rm GeV}$ for an appropriate choice
of the sneutrino vev.  However, a sizably interesting range for the
sneutrino vev requires that $M^D_2$ be above about $1~{\rm TeV}$.  For
definiteness, we take in this work $M^D_2 = 1.5~{\rm TeV}$, which
implies that $10~{\rm GeV} \lesssim \ve \lesssim 60~{\rm GeV}$.  Thus,
the heaviest charginos are the $\tilde{X}^{++}_2 \approx (\tilde{w}^+,
\overline{\tilde{T}_{d}^-})$ and $\tilde{X}^{+-}_2 \approx
(\tilde{T}_{u}^+, \overline{\tilde{w}^-})$ Dirac fields, which we will
simply call ``winos".  The lightest chargino is the electron, $e
\approx (e^-_L, \overline{e^c_R})$, with non-SM admixtures below the
$10^{-3}$ level.  The remaining state is expected to be almost pure
$\tilde{h}_u$-$\tilde{r}_d$, with a mass set by the
$\mu$-term.\footnote{\label{muterm}In the companion paper~\cite{FGKP}
we have denoted this $\mu$-term as $\mu_u$ to emphasize that it is
different from the ``standard" $\mu$-term: the former is the
coefficient of the $H_u R_d$ superpotential operator, where $R_d$ does
not get a vev and, therefore, does not contribute to fermion masses,
while the role of the latter in the present scenario is played by
$\mu' H_u L_e$, with $L_e$ being the electron doublet whose sneutrino
component gets a non-vanishing vev.  While the first one is allowed by
the $U(1)_R$ symmetry, the second one is suppressed.  However, for
notational simplicity, in this paper we will denote the $U(1)_R$
preserving term simply by $\mu$, since the ``standard", $U(1)_R$
violating one, will not enter in our discussion.} Naturalness
considerations suggest that this parameter should be around the EW
scale, and we will take $\mu = 200-300~{\rm GeV}$.  However, it is
important to note that the gaugino component of this Higgsino-like
state, $U^-_{1 \tilde{w}}$, although small, should not be neglected.
This is the case when considering the $\tilde{X}_1^{+-}$ couplings to
the first two generations, which couple to the Higgsino content only
through suppressed Yukawa interactions.  In the left panel of
Fig.~\ref{fig:NeutralinoComposition}, we exhibit the mixing angles of
the two lightest chargino states as a function of the sneutrino vev,
$\ve$, for $M^D_{2} = 1.5$~TeV, $\mu=200$~GeV, $\lambda^S_u = 0$ and
$\lambda^T_u = 1$.  The $V$-type matrix elements are shown as solid
lines, while the $U$-type matrix elements are shown as dashed lines
(sometimes they overlap).  In the right panel we show the chargino
composition as a function of $\lambda^T_u$ for $\ve = 10~{\rm GeV}$.
This illustrates that there can be accidental cancellations, as seen
for the $\tilde{w}^-$ component of $\tilde{X}^{-+}_1$ at small values
of $\lambda^T_u$.  For the most part, we will choose parameters that
avoid such special points, in order to focus on the ``typical" cases.
\begin{figure}[t]
\begin{center}
\begin{tabular}{cc}
\includegraphics[scale=0.61]{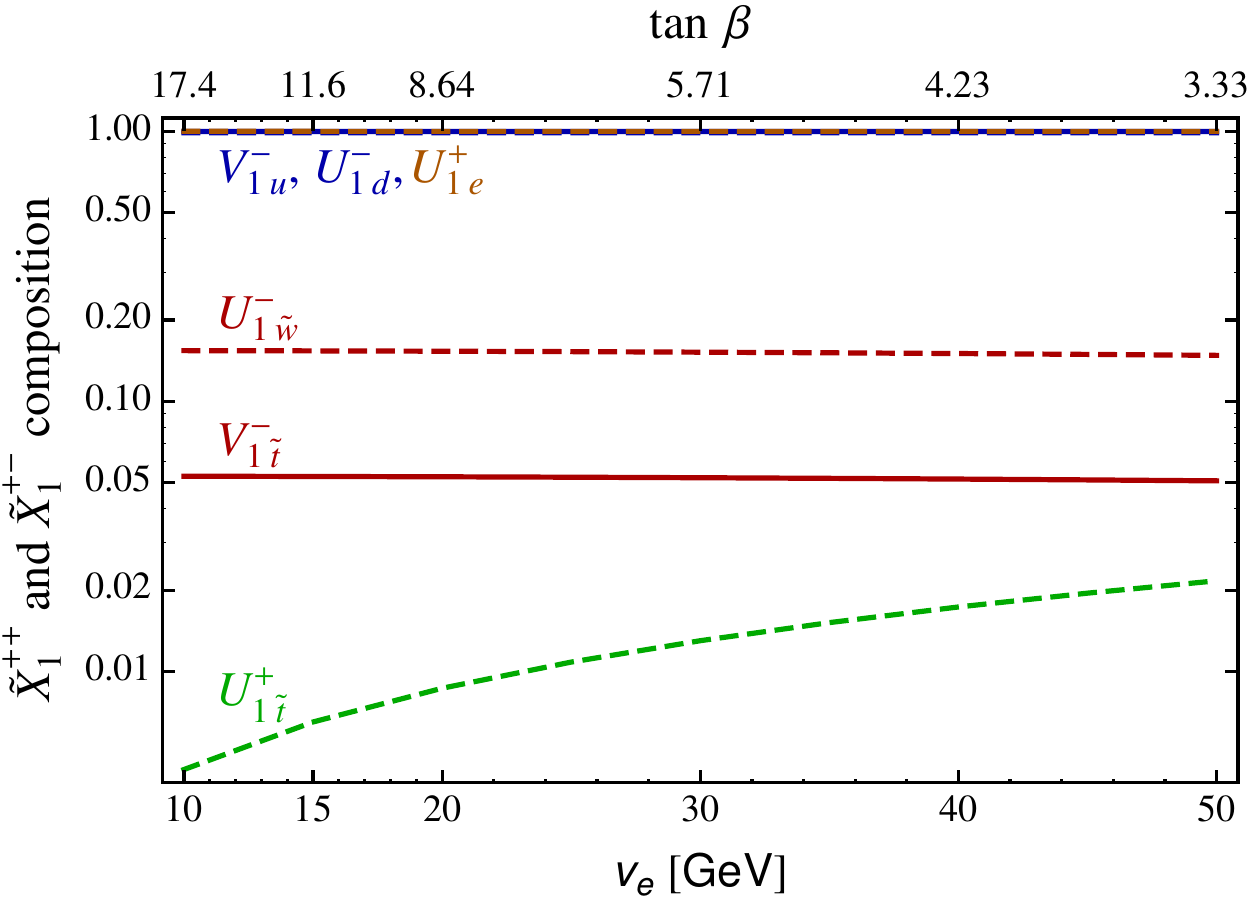}  & 
\includegraphics[scale=0.61]{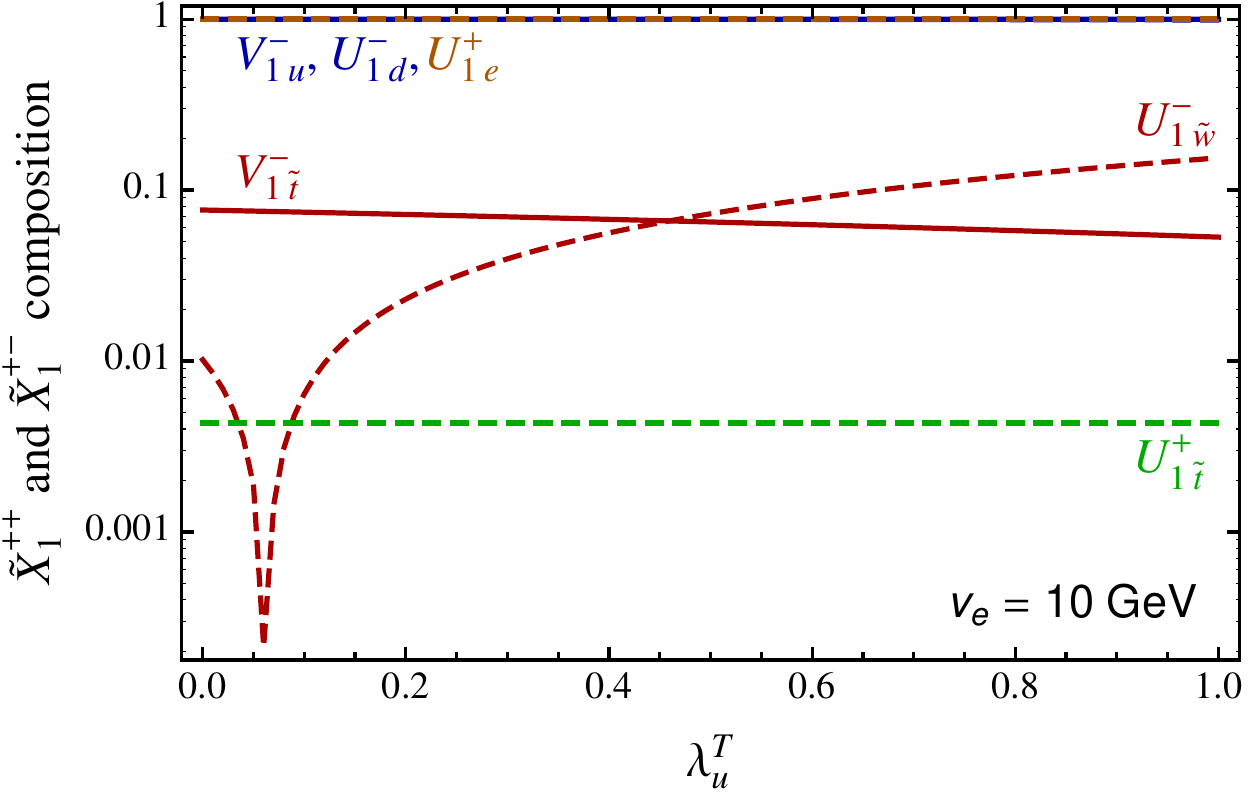} 
\end{tabular}
\end{center}
\caption{\footnotesize{Composition of the two lightest chargino states
as a function of the sneutrino vev (left panel) and as a function of
$\lambda^T_u$ (right panel).  We fix $M^D_{2} = 1.5$~TeV,
$\mu=200$~GeV and $\lambda^S_u = 0$.  In the left panel we take
$\lambda^T_u = 1$ and in the right panel we take $\ve = 10~{\rm GeV}$.
We plot the absolute magnitude of the rotation matrix elements
$V^\pm_{ik}$ (solid lines) and $U^\pm_{i k}$ (dashed lines).  Not
plotted are $V^+_{1\tilde{w}} = 0$ and $V^+_{1e} = 1$.
$\tilde{X}^{++}_1$ is the physical (charge conjugated) electron, and
$\tilde{X}^{+-}_1$ is the lightest BSM chargino state (which is
Higgsino-like).  For reference, we also show in the upper horizontal
scale the values of $\tan\beta = v_u/\ve$.}}
\label{fig:CharginoComposition}
\end{figure}  
It is also important to note that the quantum numbers of these two
lightest chargino states (the lightest of which is the physical
electron) are different.  This has important consequences for the
collider phenomenology, as we will see.

\subsubsection{Neutralinos}
\label{sec:neutralinos}

The description of the neutralino sector bears some similarities to
the chargino case discussed above.  In particular, and unlike in the
MSSM, it is natural to work in a Dirac basis.  The gauge eigenstates
are the hypercharge superpartner (bino) $\tilde{b}$, the neutral wino
$\tilde{w}$, a SM singlet, $\tilde{s}$, the neutral tripletino
$\tilde{T}^0$, the neutral Higgsinos, $\tilde{h}_u^0$ and
$\tilde{r}_d^0$ and, finally, the electron-neutrino $\nu_e$ (which
mixes with the remaining neutralinos when the electron sneutrino gets
a vev).  If there were a right-handed neutrino it would also be
naturally incorporated into the neutralino sector.  In principle, due
to the neutrino mixing angles (from the PMNS mixing matrix) the other
neutrinos also enter in a non-trivial way.  However, for the LHC
phenomenology these mixings can be neglected, which we shall do for
simplicity in the following.  Besides the gauge interactions and the
$\lambda^T_u$ superpotential coupling introduced in the previous
subsection, there is a second superpotential interaction, $W \supset
\lambda_u^S S H_u R_d$, where $S$ is the SM singlet superfield, that
can sometimes be relevant~\cite{FGKP}.

In two-component notation, we have neutralino states of
definite $U(1)_R$ charge
\bea
\label{eq:n1comp}
\tilde{\chi}_i^{0+} &=& V^N_{i \tilde{b}} \, \tilde{b} + V^N_{i \tilde{w}} \, \tilde{w} + V^N_{id} \, \tilde{h}_d^0~,\\
\tilde{\chi}_i^{0-} &=& U^N_{i \tilde{s}} \, \tilde{s} + U^N_{i \tilde{t}} \, \tilde{T}^0 + U^N_{i u} \, \tilde{h}_u^0 + U^N_{i \nu} \, \nu_e~,
\eea
where $V^N_{ik}$ and $U^N_{i k}$ are the unitary matrices that
diagonalize the neutralino mass matrix (full details are given in
Ref.~\cite{FGKP}).  These states form Dirac fermions $\tilde{X}^{0+}_i
= (\tilde{\chi}^{0+}_i, \overline{\tilde{\chi}^{0-}_i})$, for $i =
1,2,3$, where, as explained in the previous subsection, the
superindices indicate the electric and $R$-charges.  In addition,
there remains a massless Weyl neutralino:
\bea
\tilde{\chi}_4^{0-} &=& U^N_{4 \tilde{s}} \, \tilde{s} + U^N_{4 \tilde{t}} \, 
\tilde{T}^0 + U^N_{4u} \, \tilde{h}_u^0+ U^N_{4 \nu} \, \nu_e~, 
\label{eq:neucomp}
\eea
which corresponds to the \textit{physical} electron-neutrino.  With
some abuse of notation we will refer to $\tilde{\chi}_4^{0-}$ as
``$\nu_e$" in subsequent sections, where it will always denote the
above mass eigenstate and should cause no confusion with the original
gauge eigenstate.  Similarly, we will refer to $\tilde{X}^{0+}_1$ as
the ``lightest neutralino", with the understanding that strictly
speaking it is the second lightest.  Nevertheless, we find it more
intuitive to reserve the nomenclature ``neutralino" for the states not
yet discovered.  The heavier neutralinos are labeled accordingly.

\begin{figure}[t]
\begin{center}
\begin{tabular}{cc}
\includegraphics[scale=0.61]{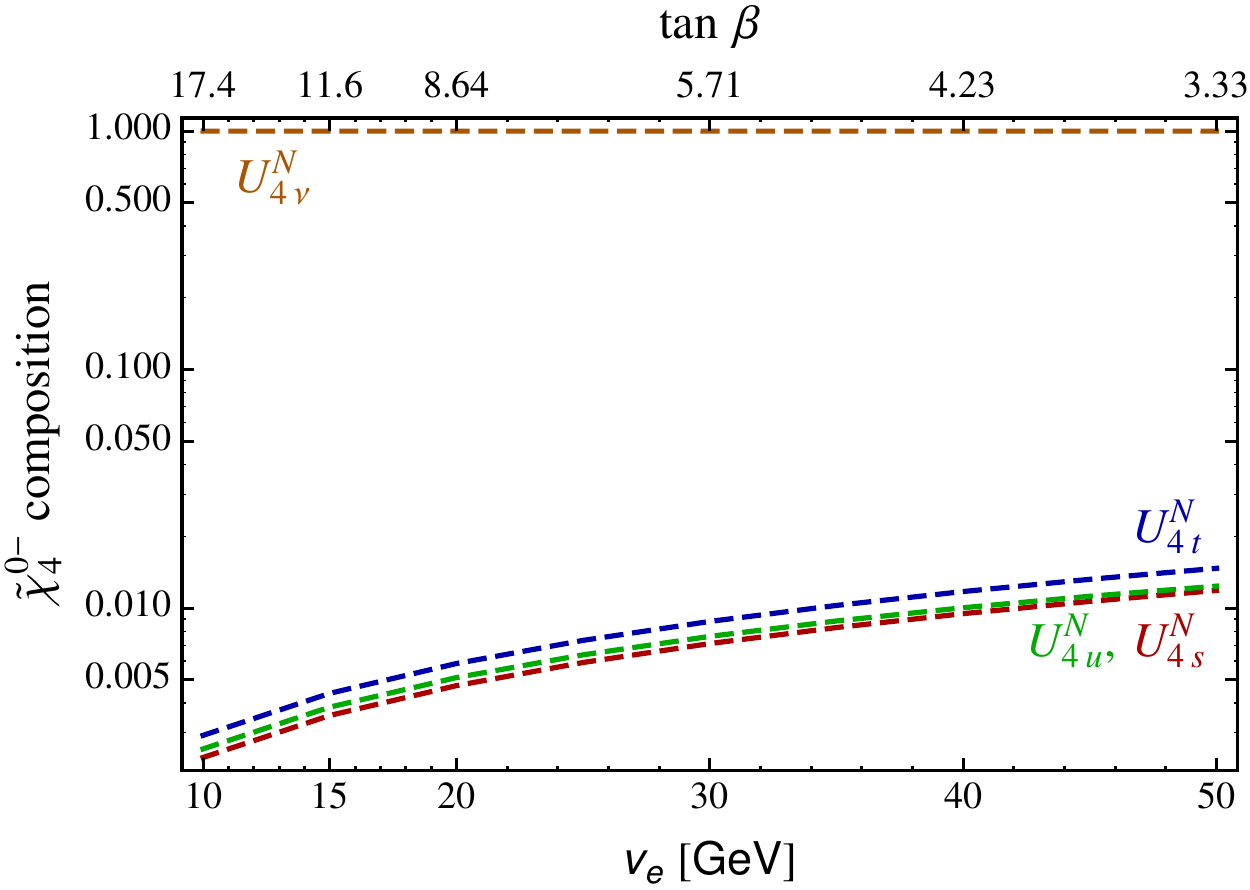}  & 
\includegraphics[scale=0.6]{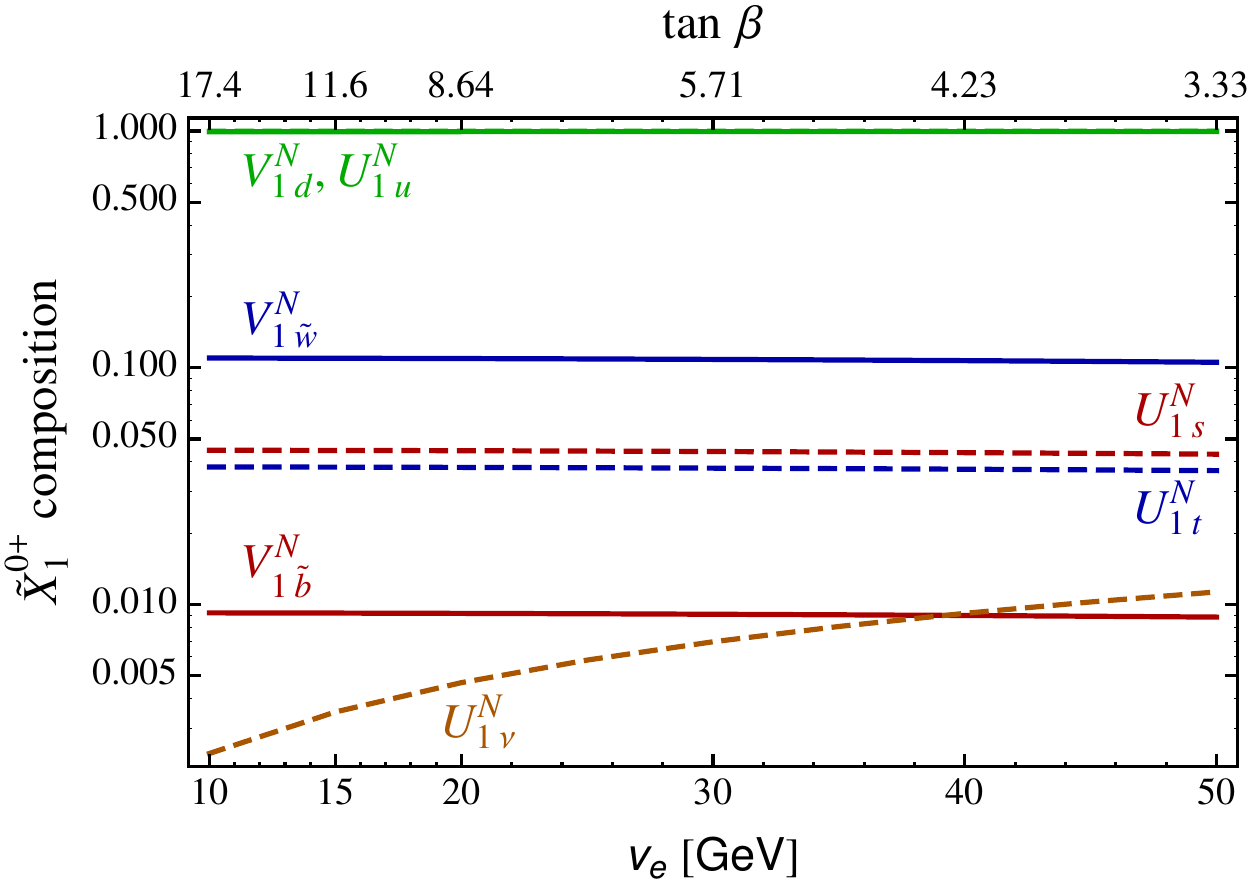} 
\end{tabular}
\end{center}
\caption{\footnotesize{Left panel: $\tilde{\chi}_4^{0-}$ (neutrino)
and right panel: $\tilde{X}_1^{0+}$ (Higgsino-like) composition for
$M^D_{1} = 1$~TeV, $M^D_{2} = 1.5$~TeV, $\mu=200$~GeV, $\lambda^S_u =
0$ and $\lambda^T_u = 1$, as a function of the sneutrino vev.  We plot
the absolute magnitude of the rotation matrix elements $V^N_{ik}$ and
$U^N_{i k}$.}}
\label{fig:NeutralinoComposition}
\end{figure}  
Given that both the gluino and wino states are taken to be above a
TeV, we shall also take the Dirac bino mass somewhat large,
specifically $M^D_1 = 1~{\rm TeV}$.  This is mostly a simplifying
assumption, for instance closing squark decay channels into the
``second" lightest neutralino (which is bino-like).  Thus, the
lightest (non SM-like) neutralino is Higgsino-like, and is fairly
degenerate with the lightest (non SM-like) chargino,
$\tilde{X}^{+-}_1$.

In Fig.~\ref{fig:NeutralinoComposition}, we show the composition of
the physical neutrino ($\tilde{\chi}^{0-}_4$) and of the Higgsino-like
neutralino state ($\tilde{X}^{0+}_1$).  Note that, as a result of
$R$-charge conservation, the neutrino state has no wino/bino
components.  In addition, its (up-type) Higgsino component is rather
suppressed.  As a result, the usual gauge or Yukawa induced
interactions are very small.  Instead, the dominant couplings of
$\tilde{\chi}^{0-}_4$ to other states will be those inherited from the
neutrino content itself.  The associated missing energy signals will
then have a character that differs from the one present in mSUGRA-like
scenarios.  However, it shares similarities with gauge mediation,
where the gravitino can play a role similar to the neutrino in our
case.\footnote{There is also a light gravitino in our scenario, but
its couplings are suppressed, and plays no role in the LHC
phenomenology.} By contrast, the ``lightest" neutralino,
$\tilde{X}^{0+}_1$, typically has non-negligible wino/bino components
that induce couplings similar to a more standard (massive) neutralino
LSP. Nevertheless, here this state decays promptly, and is more
profitably thought as a neutralino LSP in the RPV-MSSM (but with
2-body instead of 3-body decays).

\subsection{The Scalar Sector}
\label{sec:scalars}

In this section, we discuss the squark, slepton and Higgs sectors,
emphasizing the distinctive features compared to other supersymmetric
scenarios.

\subsubsection{Squarks}
\label{sec:squarks}

Squarks have interesting non-MSSM properties in the present setup.
They are charged under the $R$-symmetry ($R=+1$ for the LH squarks and
$R = -1$ for the RH ones), and as a result they also carry lepton
number.  Thus, they are scalar \textit{lepto-quarks} (strongly
interacting particles carrying both baryon and lepton number).  This
character is given by the superpotential RPV operator $\lambda'_{ijk}
L_i Q_j D^c_k$, which induces decays such as $\tilde{t}_L \to b_R
e^+_L$.  In addition, and unlike in more familiar RPV scenarios, some
of these couplings are not free but directly related to Yukawa
couplings: $\lambda'_{111} = y_d$, $\lambda'_{122} = y_s$ and
$\lambda'_{133} = y_b$.  The full set of constraints on the $\lambda'$
couplings subject to these relations was analyzed in Ref.~\cite{FGKP}.
The $\lambda'_{333}$ coupling is the least unconstrained, being
subject to
\bea
\lambda'_{333} &\lesssim& (2.1 \times 10^{-2})/y_b ~\approx~ 1.4 \cos\beta~,
\label{lambdap333}
\eea
where $y_b = m_b/ \ve$, $\tan\beta = v_u/\ve$ and we took $m_b(\mu
\approx 500~{\rm GeV}) \approx 2.56~{\rm GeV}$~\cite{Xing:2007fb}.  In
this work, we will assume that the only non-vanishing $\lambda'$
couplings are those related to the Yukawa couplings, together with
$\lambda'_{333}$.  We will often focus on the case that the upper
limit in Eq.~(\ref{lambdap333}) is saturated, but should keep in mind
that $\lambda'_{333}$ could turn out to be smaller, and will comment
on the relevant dependence when appropriate.

It is also important to keep in mind that the $R$-symmetry forbids any
LR mixing.  As a result, the squark eigenstates coincide with the
gauge eigenstates, at least if we neglect intergenerational
mixing.\footnote{This assumptions is not necessary, given the mild
flavor properties of $U(1)_R$-symmetric
models~\cite{Kribs:2007ac,Fok:2010vk,Kribs:2010md}.  This opens
up the exciting prospect of observing a non-trivial flavor structure
at the LHC, that we leave for future work.} We will assume in this
work that the first two generation squarks are relatively degenerate.
As we will see, the current bound on their masses is about
$500-700~{\rm GeV}$.  We will also see that the
third generation squarks can be lighter, possibly consistent with
estimates based on naturalness from the Higgs sector.

\subsubsection{Sleptons}
\label{sec:sleptons}

The sleptons are expected to be among the lightest sparticles in the
new physics spectrum.  This is due to the intimate connection of the
slepton sector with EWSB, together with the fact that a good degree of
degeneracy between the three generation sleptons is expected.  The
possible exception is the LH third generation slepton doublet, if the
RPV coupling $\lambda'_{333}$ turns out to be sizable.  As a result,
due to RG running, the LH stau can be several tens of GeV lighter than
the selectron and smuon, while the latter should have masses within a
few GeV of each other.  Note that the sleptons are $R$-neutral, hence
do not carry lepton number.  This is an important distinction compared
to the standard extension of lepton number to the new physics sector.

Since the electron sneutrino plays the role of the down-type Higgs,
naturalness requires its soft mass to be very close to the electroweak
scale.  To be definite, we take $m^2_{\tilde L} \sim m^2_{\tilde E}
\sim ($200-300$~{\rm GeV})^2$.  Depending on how this compares to the
$\mu$-term, the sleptons can be heavier or lighter than the lightest
neutralino, $\tilde{X}^{+-}_1$.  When $\tilde{X}^{+-}_1$ is lighter
than the sleptons we will say that we have a ``neutralino LSP
scenario".  The other case we will consider is one where the LH third
generation slepton doublet is lighter than $\tilde{X}^{+-}_1$, while
the other sleptons are heavier.  Given the possible mass gap of
several tens of GeV between the $(\tilde{\nu}_\tau, \tilde{\tau}_L)$
pair and the other sleptons, this is a rather plausible situation.  We
will call it the ``stau LSP scenario", although the $\tau$-sneutrino
is expected to be up to ten GeV lighter than the stau.\footnote{Again,
we remind the reader that we are using standard terminology in a
non-standard setting.  In particular, a rigorous separation of the SM
and supersymmetric sectors is not possible, due to the mixings in the
neutralino and chargino sectors.  Also, the supersymmetric particles
end up decaying into SM ones, similar to RPV-MSSM scenarios.
Furthermore, the light gravitino could also be called the LSP, as in
gauge-mediation.  However, unlike in gauge mediation, here the
gravitino is \textit{very} rarely produced in superparticle decays,
hence not phenomenologically relevant at the LHC. Thus, we will refer
to either the $(\tilde{\nu}_\tau, \tilde{\tau}_L)$ pair or
$\tilde{X}^{+-}_1$ as the ``LSP", depending on which one is lighter.
Our usage emphasizes the allowed decay modes.} The possibility that
several or all the sleptons could be lighter than $\tilde{X}^{+-}_1$
may also deserve further study, but we will not consider such a case
in this work.

We also note that some of the couplings in the RPV operator
$\lambda_{ijk} L_i L_j E^c_k$ are related to lepton Yukawa couplings:
$\lambda_{122} = y_\mu$ and $\lambda_{133} = y_\tau$.  The bounds on
the remaining $\lambda_{ijk}$'s under these restriction have been
analyzed in~\cite{FGKP}, and have been found to be stringent.  We note
that, in principle, it could be possible to produce sleptons singly at
the LHC through the $\lambda'_{ijk} L_i Q_j D^c_k$ operator, with
subsequent decays into leptons via the $\lambda_{ijk} L_i L_j E^c_k$
induced interactions.  We have studied this possibility in
Ref~\cite{FGKP}, and found that there may be interesting signals in
the $\mu^+\mu^-$ and $e^\mp\mu^\pm$ channels.  However, in this work
we do not consider such processes any further, and set all $\lambda$
couplings to zero, with the exception of the Yukawa ones.  The tau
Yukawa, in particular, can play an important role.
 
\subsubsection{The Higgs Sector}
\label{sec:higgs}

The ``Higgs" sector is rather rich in our scenario.  The EW symmetry
is broken by the vev's of the neutral component of the up-type Higgs
doublet, $H^0_u$, and the electron sneutrino, $\tilde{\nu}_e$, which
plays a role akin to the neutral down-type Higgs in the MSSM. We have
also a scalar SM singlet and a scalar $SU(2)_L$ triplet, the
superpartners of the singlino and tripletino discussed in the previous
section.  These scalars also get non-vanishing expectation values.
However, it is well known that constraints on the Peskin-Takeuchi
$T$-parameter require the triplet vev to be small, $v_T \lesssim
2~{\rm GeV}$.  We will also assume that the singlet vev is in the few
GeV range.  This means that these two scalars are relatively heavy,
and not \textit{directly} relevant to the phenomenology discussed in
this paper.  Note that all of these states are $R$-neutral.

There is another doublet, $R_d$, the only state with non-trivial
$R$-charge ($=+2$).  It does not acquire a vev, so that the
$R$-symmetry is not spontaneously broken, and therefore this state
does not mix with the previous scalars.  Its (complex) neutral and
charged components are relatively degenerate, with a mass splitting of
order 10~GeV, arising from EWSB as well as the singlet vev.  For
simplicity, we will assume its mass to be sufficiently heavy (few
hundred GeV) that it does not play a role in our discussion.
Nevertheless, it would be interesting to observe such a state, due to
its special $R$-charge.

The upshot is that the light states in the above sector are rather
similar to those in the MSSM: a light CP-even Higgs, a heavier CP-even
Higgs, a CP-odd Higgs and a charged Higgs pair.  The CP-even and
CP-odd states are superpositions of the real and imaginary components
of $h^0_u$ and the electron sneutrino (with a small admixture of the
singlet and neutral triplet states).  Given our choice for the slepton
soft masses, the heavy CP-even, CP-odd and charged Higgses are
expected to be relatively degenerate, with a mass in the $200-300~{\rm
GeV}$ range (the charged Higgs being slightly heavier than the neutral
states).  The charged Higgs is an admixture of $H^+_u$ and the LH
selectron $\tilde{e}_L$ (and very suppressed charged tripletino
components).  The RH selectron, as well as the remaining neutral and
charged sleptons do not mix with the Higgs sector, and can be cleanly
mapped into the standard slepton/sneutrino terminology.

The light CP-even Higgs, $h$, is special, given the observation of a
Higgs-like signal by both the ATLAS~\cite{:2012gk} and
CMS~\cite{:2012gu} collaborations at about $125~{\rm GeV}$.  This
state can also play an important role in the decay patterns of the
various super-particles.  Within our scenario, a mass of $m_h \approx
125~{\rm GeV}$ can be obtained from radiative corrections due to the
triplet and singlet scalars, even if \textit{both} stops are
relatively light (recall the suppression of LR mixing due to the
$R$-symmetry).  This is an interesting distinction from the MSSM. A
more detailed study of these issue will be dealt with in a separate
paper~\cite{Kumar:2012yy}.  Here we point out that these arguments
suggest that $\lambda^S_u$ should be somewhat small, while
$\lambda^T_u$ should be of order one.  This motivates our specific
benchmark choice: $\lambda^S_u = 0$ and $\lambda^T_u = 1$ (although
occasionally we will allow $\lambda^S_u$ to be non-vanishing).  These
couplings affect the neutralino/chargino composition and are therefore
relevant for the collider phenomenology.

\subsection{Summary}
\label{sec:particlesummary}

Let us summarize the properties of the superpartner spectrum in our
scenario, following from the considerations in the previous
sections.  All the gauginos (``gluino", ``wino" and ``bino") are
relatively heavy, in particular heavier than all the sfermions.  The
first two generation squarks can be below $1~{\rm TeV}$, while the
third generation squarks can be in the few hundred GeV range.
These bounds will be discussed more fully in the remaining of the
paper.  The sleptons, being intimately connected to the Higgs sector,
are in the couple hundred GeV range.  So are the ``lightest"
neutralino and chargino states, which are Higgsino-like.  Mixing due
to the electron sneutrino vev, induces interesting couplings of the
new physics states to the electron-neutrino and the electron, while
new interactions related to the lepton and down-quark Yukawa couplings
give rise to non-MSSM signals.  The collider phenomenology is largely
governed by a new (approximately) conserved $R$-charge, and will be
seen to be extremely rich, even though the spectrum of light states
does not seem, at first sight, very complicated or unconventional.
Finally, we mention that there is also an $SU(3)_C$ octet scalar
(partner of the octetinos that are part of the physical gluino states)
that will not be studied here (for studies of the octet scalar
phenomenology, see~\cite{Plehn:2008ae, Choi:2009ue}).

\section{Sparticle Decay Modes}
\label{sec:decaymodes}

In this section we discuss the decay modes of the superparticles
relevant for the LHC collider phenomenology.  We have checked that
three-body decays are always negligible and therefore we focus on the
two-body decays.

\subsection{Neutralino Decays}
\label{sec:decayneutralinos}

From our discussion in the previous section, the lightest (non
SM-like) neutralino is a Higgsino-like state (that we call
$\tilde{X}_1^{0+}$), while the truly stable neutralino state is none
other than the electron-neutrino.  It was also emphasized that
$\tilde{X}_1^{0+}$ has small, but not always negligible, gaugino
components.  The other two (Dirac) neutralino states are heavy.  We
therefore focus here on the decay modes of $\tilde{X}_1^{0+}$.

As explained in Subsection~\ref{sec:sleptons}, we consider two
scenarios: a ``neutralino LSP scenario", where $\tilde{X}_1^{0+}$ is
lighter than the LH third generation slepton doublet, and a ``stau LSP
scenario" with the opposite hierarchy.  The decay modes of the
lightest neutralino depend on this choice and we will consider them
separately.

\medskip
\noindent
\underline{\textit{Neutralino LSP Scenario}}:
\medskip

\noindent 
If $\tilde{X}_1^{0+} $ is lighter than the $(\tilde{\nu}_\tau, \tilde
\tau^-_L)$ pair, the possible decay modes for $\tilde{X}_1^{0+}$ have
partial decay widths [in the notation of
Eqs.~(\ref{eq:n1comp})-(\ref{eq:neucomp})]:
\bea
\Gamma(\tilde{X}^{0+}_1  \rightarrow  W^- e^+_L) &=&  \frac{g^{2} m_{\tilde{X}^0_1}}{128  \pi} \, (U^+_{1 e} U^N_{1 \nu} + \sqrt{2} \, U^+_{1 \tilde{t}} U^N_{1 \tilde{t}} )^2  \left(1- \frac{M_W^2}{ m_{\tilde{X}^0_1}^2}\right )^2  \left(2+ \frac{ m_{\tilde{X}^0_1}^2} {M_W^2}\right)~,
\label{XToWe} 
\\
\Gamma(\tilde{X}^{0+}_1  \rightarrow  Z \bar{\nu}_e) &=&  \frac{g^{2} m_{\tilde{X}^0_1}}{512 \pi c_W^2}    (U^N_{1 \nu} U^N_{4 \nu} -  U^N_{1 u} U^N_{4 u} )^2  \left(1- \frac{M_Z^2}{ m_{\tilde{X}^0_1}^2}\right )^2  \left(2+ \frac{ m_{\tilde{X}^0_1}^2} {M_Z^2}\right)~, 
\label{XToZnu} 
\eea
%
%
\bea
\Gamma(\tilde{X}^{0+}_1  \rightarrow  h \bar{\nu}_e) &=& 
\frac{m_{\tilde{X}^0_1} }{256  \pi } \left(1- \frac{m_h^2}{ m_{\chi^0_1}^2} \right)^2 \times
\label{XTohnu} \\ 
& &  \left[  \left(-g V^N_{1 \tilde{w}} U^N_{4 u} + g' V^N_{1 \tilde{b}}  U^N_{4 u} \right) R_{1u} 
+ \left( g V^N_{1 \tilde{w}} U^N_{4 \nu}- g' V^N_{1 \tilde{b}} U^N_{4 \nu} \right) R_{1 \tilde{\nu}}
\right.
\nonumber \\ 
& &  \left. \mbox{} + 
\sqrt{2} \left(\lambda^S_u U^N_{4\tilde{s}} + \lambda^T_u U^N_{4 \tilde{t}} \right) V^N_{1 d} R_{1 u} 
+
\sqrt{2} V^N_{1 d} U^N_{4 u} \left(\lambda^S_u R_{1 s} + \lambda^T_u R_{1 t} \right) \right]^2~,
\nonumber
\eea
where we denote the $\tilde{X}_1^{0+}$ mass by $m_{\tilde{X}^0_1}$,
and $R_{1 i} $ are the mixing angles characterizing the composition of
the lightest Higgs, $h$.  In our scenario all the other Higgs bosons
are heavier than the lightest neutralino.  We note that the above
expressions contain an explicit factor of $1/\sqrt{2}$ for each
occurrence of a neutralino mixing angle, compared to the standard
ones~\cite{Baltz:1997gd,Djouadi:2000bx,Djouadi:2001fa}.  This is
because the mixing matrix elements, $U^N_{ij}$ and $V^N_{ij}$ are
defined in a Dirac basis, whereas in the usual approach the
neutralinos are intrinsically Majorana particles.  Recall also that,
for simplicity, we are assuming here that all quantities are real.
The generalization of these and subsequent formulas to the complex
case should be straightforward.
\begin{figure}[t]
\begin{center}
\begin{tabular}{cc}
\includegraphics[scale=0.6]{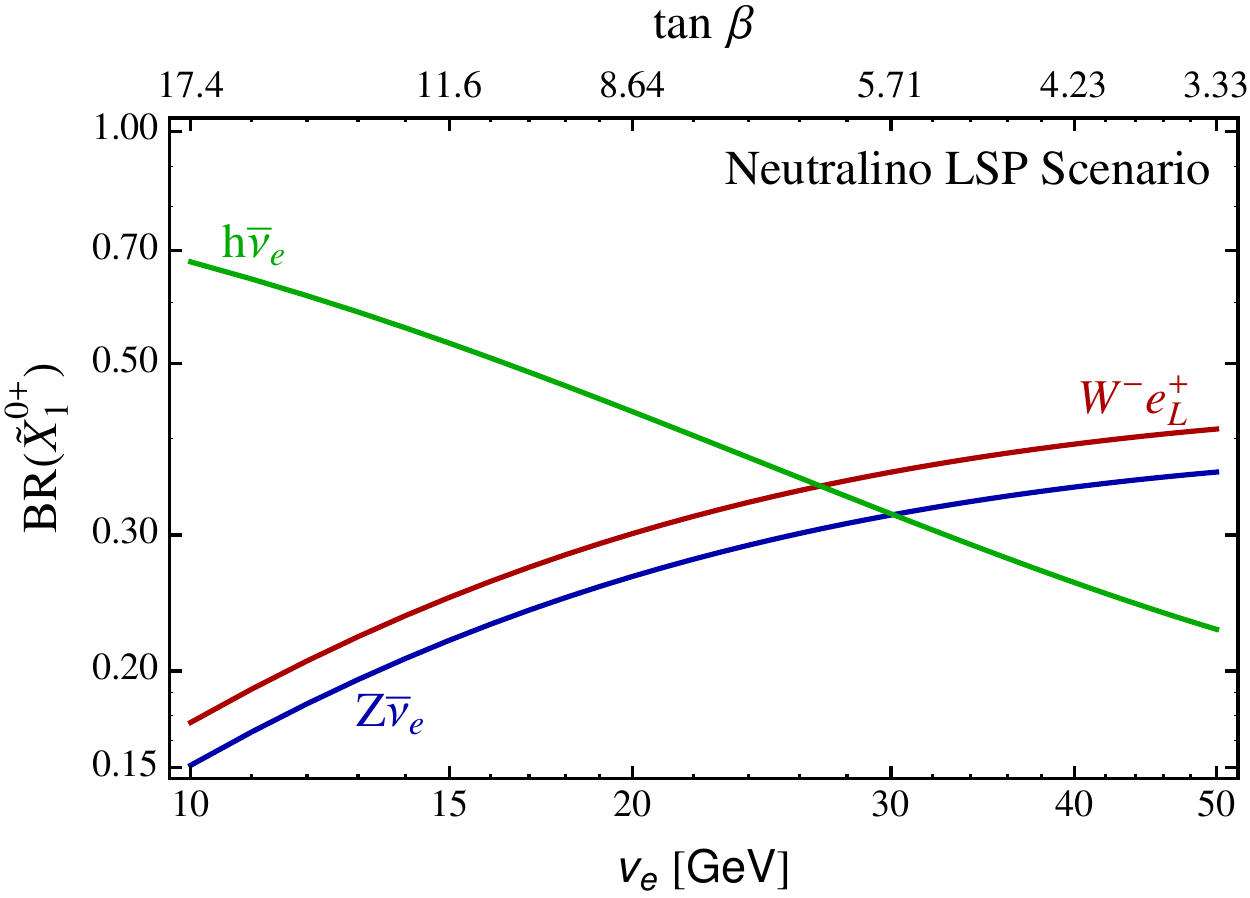}  & 
\includegraphics[scale=0.6]{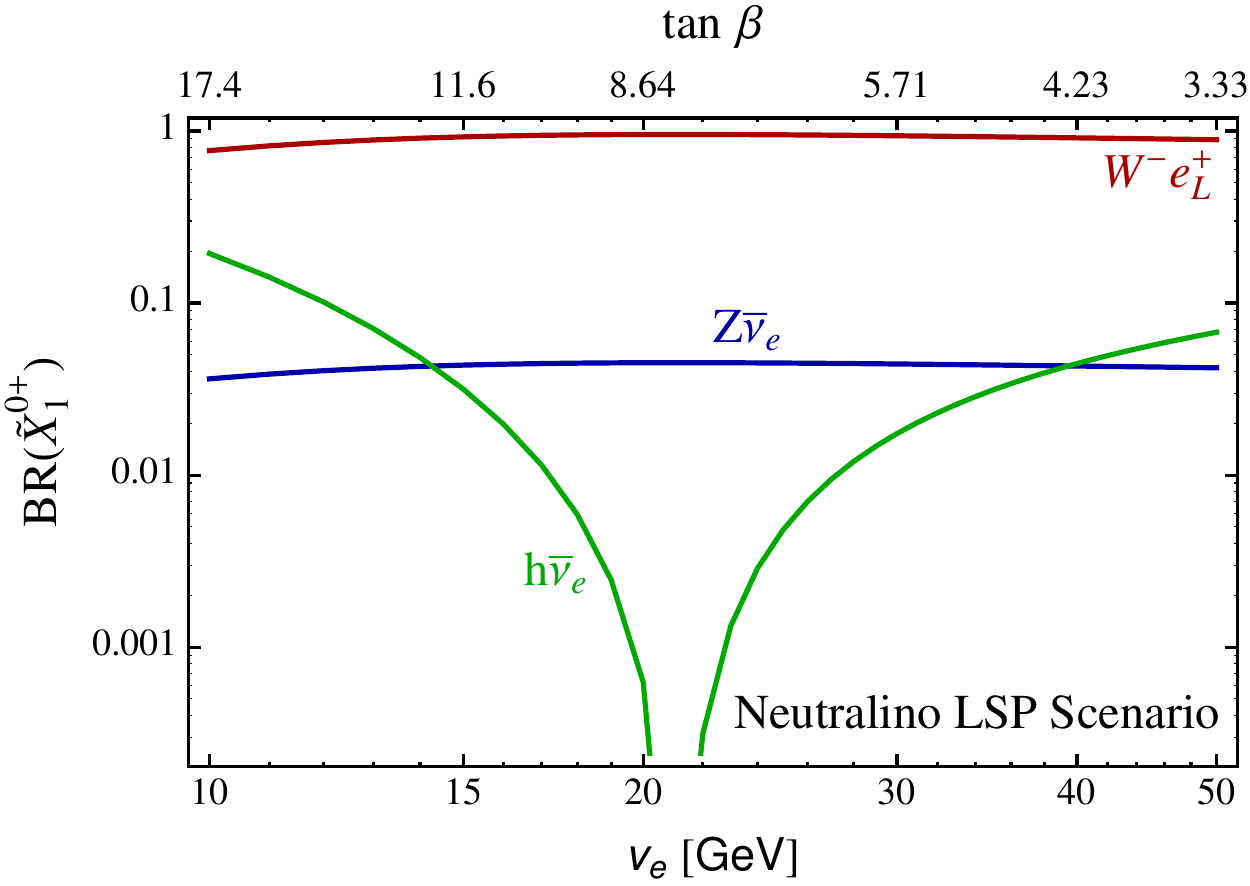}  
\end{tabular}
\end{center}
\caption{\footnotesize{$\tilde{X}^{0+}_1$ branching fractions in the
``neutralino LSP scenario" for $M^D_{1} = 1$~TeV $M^D_{2} = 1.5$~TeV, 
and $\mu=200$~GeV. In the left panel we take $\lambda^S_u = 0$ and
$\lambda^T_u = 1$, and in the right panel we take $\lambda^S_u =
\lambda^T_u = 0.4$.  The former case might be favored by the
observation of a Higgs-like state at $m_h \approx 125~{\rm GeV}$.  We
also take the Higgs mixing angles as $R_{1u} \approx 0.98$,
$R_{1\tilde{\nu}} \approx 0.2$ and $R_{1s}, R_{1t} \ll 1$.}}
\label{fig:N1decayNeutrLSP}
\end{figure}  

The above decay modes can easily be dominated by the
neutrino-neutralino mixing angles, since the contributions due to the
higgsino ($U^N_{4 u}$) and tripletino components are highly suppressed.
This mixing angles, in turn, are controlled by the sneutrino vev.
Note that in the RPV-MSSM such decay modes are typically characterized
by displaced vertices due to the extremely stringent bounds on the
sneutrino vev arising from neutrino physics
\cite{Banks:1995by}.  By contrast, in our scenario
the sneutrino vev is allowed to be sizable (tens of GeV), and is in
fact bounded from below from perturbativity/EWPT arguments, so that
these decays are prompt.

The left panel of Fig.~\ref{fig:N1decayNeutrLSP} shows that the decay
width into $h \bar{\nu}_e$ is the dominant one in the small sneutrino
vev limit, while in the large sneutrino vev limit the channels
involving a gauge boson can be sizable.  We also note that it is
possible for the $W^- e^+_L$ decay channel to be the dominant one, as
shown in the right panel of Fig.~\ref{fig:N1decayNeutrLSP}.  In this
case we have chosen $\lambda^S_u=\lambda^T_u=0.4$, which leads to a
cancellation between the mixing angles such that $Z \bar{\nu}_e $ is
suppressed compared to $W^- e^+_L$.  For such small couplings, the
radiative contributions to the lightest CP-even Higgs are not large
enough to account for the observed $m_h \approx 125~{\rm GeV}$, while
stops (due to the absence of LR mixing) are also not very effective
for this purpose.  Therefore, without additional physics such a
situation may be disfavored.  We mention it, since it is tied to a
striking signal, which one should nevertheless keep in mind.

\medskip
\noindent
\underline{\textit{Stau LSP Scenario}}:
\medskip

\noindent
If instead the $(\tilde{\nu}_\tau, \tilde \tau^-_L)$ pair is lighter
than $\tilde{X}_1^{0+} $, the $\tilde{\tau}^-_L \tau^+_L$ and
$\tilde{\nu}_{\tau} \bar{\nu}_{\tau}$ channels open up with partial
decay widths given by
\bea
\label{tau}
\Gamma(\tilde{X}^{0+}_1  \rightarrow  \tilde{\tau}^-_L \tau^+_L) &\approx&  \frac{g^2}{64  \pi}   
\left(V^N_{1 \tilde{w}} + \tan{\theta_W} V_{1 \tilde{b}}^N \right)^2 m_{\tilde{X}^0_1}  \left(1- \frac{m_{\tilde{\tau}_L}^2}{ m_{\tilde{X}^0_1}^2} \right)^2~, \\
\Gamma(\tilde{X}^{0+}_1  \rightarrow  \tilde{\nu}_{\tau} \nu_{\tau}) &=&  \frac{ g^{2}}{64  \pi}   
\left(V^N_{1 \tilde{w}} - \tan{\theta_W} V_{1 \tilde{b}}^N \right)^2  m_{\tilde{X}^0_1}  \left(1- \frac{m_{\tilde{\nu}_{\tau} }^2}{ m_{\tilde{X}^0_1}^2} \right)^2~.
\label{sneu}
\eea
\begin{wrapfigure}{r}{0.53\textwidth}
\vspace{-15pt}
\centering
\begin{tabular}{c}
\includegraphics[scale=0.65]{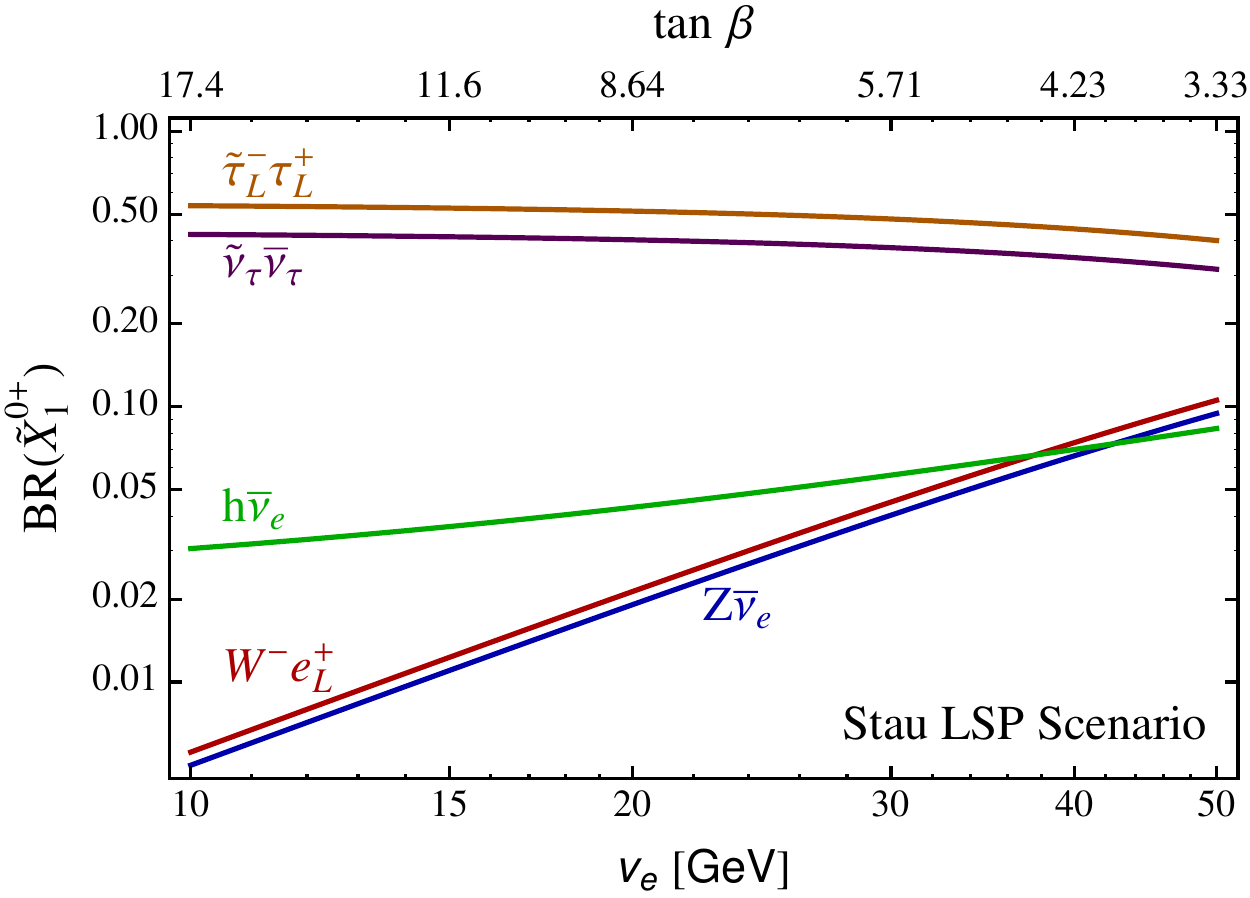}
\end{tabular}
\caption{\footnotesize{$\tilde{X}^{0+}_1$ branching fractions in the
``stau LSP scenario" for $M^D_{1} = 1$~TeV, $M^D_{2} = 1.5$~TeV, 
$\mu=250$~GeV, $\lambda^S_u = 0$ and $\lambda^T_u = 1$.  We also take
$m_{\tilde{\tau}_L} \approx ~m_{\tilde{\nu}_\tau} = 200~{\rm GeV}$.
The Higgs mixing angles are as in Fig.~\ref{fig:N1decayNeutrLSP}.}}
\label{fig:N1decayStauLSP}
\end{wrapfigure}

In Eq.~(\ref{tau}) we have suppressed additional terms proportional to
the $\tau$ Yukawa coupling, that give negligible contributions
compared to the ones displayed.  Although we have included the full
expressions in the numerical analysis, we choose to not display such
terms to make the physics more transparent.  The only cases where
contributions proportional to the Yukawa couplings are not negligible
occur when the top Yukawa is involved.\footnote{Even the contribution
from the bottom Yukawa coupling (with possible large $\tan\beta$
enhancements) is negligible, given the typical mixing angles in the
scenario.} We then see that Eqs.~(\ref{tau}) and (\ref{sneu}) are
controlled by the gaugino components, even for the suppressed $V^N_{1
\tilde{w}}$ and $V^N_{1 \tilde{b}}$ shown in
Fig.~\ref{fig:NeutralinoComposition}.  Thus, these decay channels
dominate over the ones driven by the neutrino-neutralino mixing, as
shown in Fig.~\ref{fig:N1decayStauLSP}.  Here the $\tilde{\nu}_{\tau}
\bar{\nu}_{\tau}$ channel is slightly suppressed compared to the one
into the charged lepton and slepton due to a cancellation between the
mixing angles in Eq.~(\ref{tau}).  In other regions of parameter space
such a cancellation may be more or less severe.

\subsection{Chargino  Decays}
\label{sec:decaycharginos}

The lightest of the charginos (other than the electron) is
$\tilde{X}^{+-}_1$.  It is Higgsino-like, which follows from its
$R=-Q$ nature, and the fact that the winos are heavy.  Note that, in
contrast, the electron and the other charged leptons have $R=Q$.
\textit{Therefore, the two-body decays of $\tilde{X}^{+-}_1 $ can
involve a charged lepton only when accompanied with an electrically
neutral, $|R| = 2$ particle, the only example of which is the $R^0_d$
scalar.} However, this state does not couple directly to the
leptons.\footnote{Recall that the $R_d$ $SU(2)$ doublet does not play
any role in EWSB.} We take it to be heavier than $\tilde{X}^{+-}_1$,
which has important consequences for the allowed chargino decay modes.
For instance, in the region where $\tilde{\tau}_L $ is heavier than
$\tilde{X}^{+-}_1$ the potentially allowed decay modes of
$\tilde{X}^{+-}_1$ are into $W^+ \nu_e$ and $W^+ \tilde{X}^{0-}_1$,
where $\tilde{X}^{0-}_1$ denotes the antiparticle of
$\tilde{X}^{0+}_1$.  However, the second channel is closed in most of
the parameter space since $\tilde{X}^{0+}_1 $ and $\tilde{X}^{+-}_1$
are relatively degenerate (with a mass splitting of order ten ${\rm
GeV}$).  The dominant decay mode in this ``neutralino LSP scenario"
has a partial decay width given by:
\bea
\Gamma(\tilde{X}^{+-}_1  \rightarrow  W^+ \nu_e) &=&  \frac{g^{2}}{128  \pi}   
(V^-_{1 u} U^N_{4 u} - \sqrt{2} V^-_{1 \tilde{t}} U^N_{4 \tilde{t}} )^2  
m_{\tilde{X}^\pm_1} \left (1- \frac{M_{W}^2}{ m_{\tilde{X}^\pm_1}^2} \right)^2   
\left(2+ \frac{m_{\tilde{X}^\pm_1}^2}{ M_{W}^2} \right)~,
\label{fig:C1W}
\eea
where we denote the mass of $\tilde{X}^{+-}_1$ by $m_{\tilde{X}^\pm_1}$.
Therefore, for sufficiently heavy sleptons the chargino always decays
into $W^+ \nu_e.$

If instead $\tilde \tau_L $ is lighter than $\tilde{X}^{+-}_1$ one can
also have $\tilde{X}^{+-}_1 \rightarrow \tilde{\tau}^+_L \nu_{\tau}$
with
\bea
\Gamma(\tilde{X}^{+-}_1  \rightarrow \ \tilde{\tau}^+_L  \nu_\tau) &=&  
\frac{g^{2}}{32  \pi}   (U^{-}_{1 \tilde{w}})^2 m_{\tilde{X}^\pm_1}  
\left(1- \frac{m_{\tilde{\tau}_L}^2}{ m_{\tilde{X}^\pm_1}^2} \right)^2~.
\eea
Typically, this decay channel dominates, but the $W^+ \nu_e$ can still
have an order one branching fraction.

\subsection{Slepton Decays}
\label{sec:sleptonsdecays}

%
\begin{figure}[t]
\begin{center}
\begin{tabular}{cc}
\includegraphics[scale=0.6]{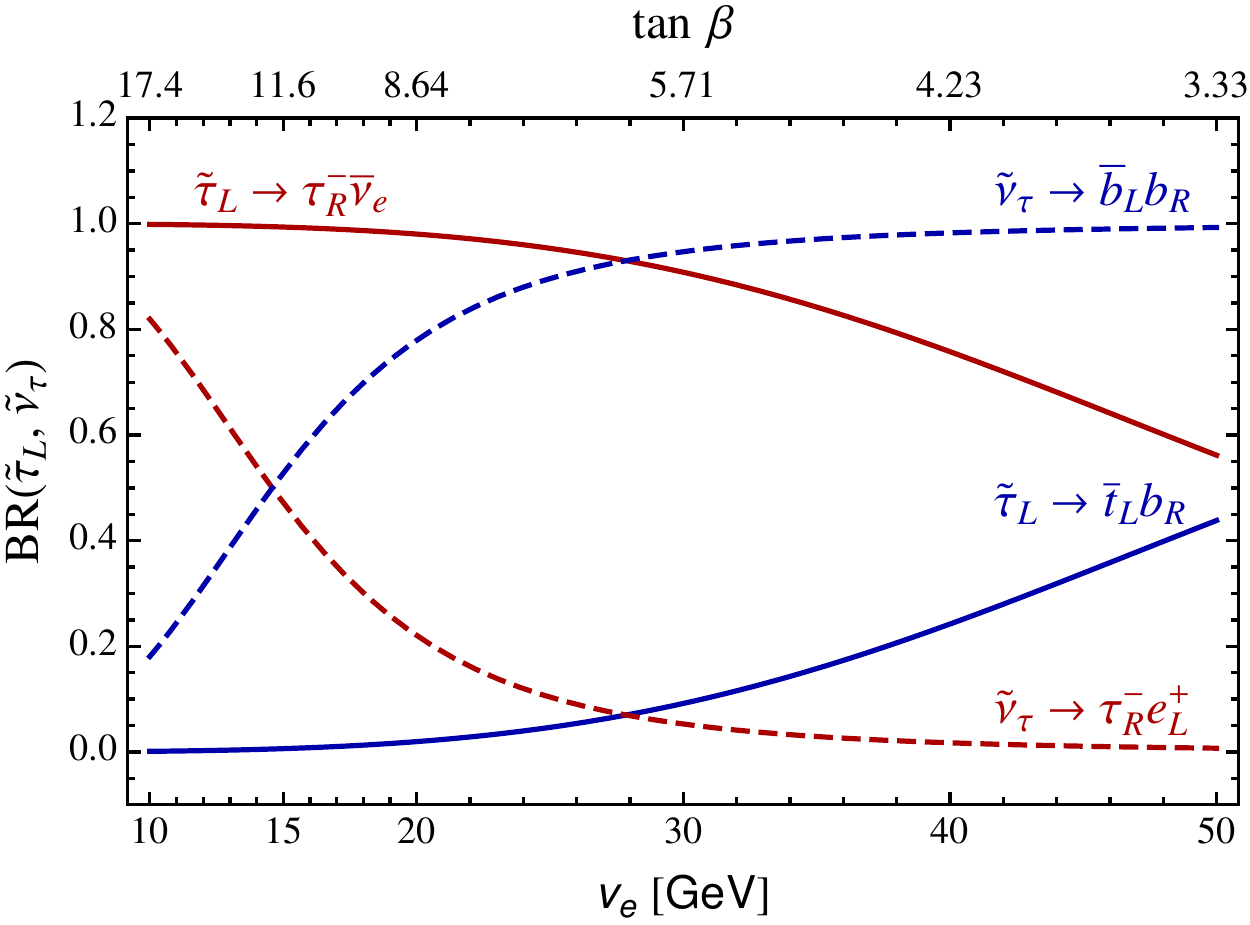}  & 
\includegraphics[scale=0.6]{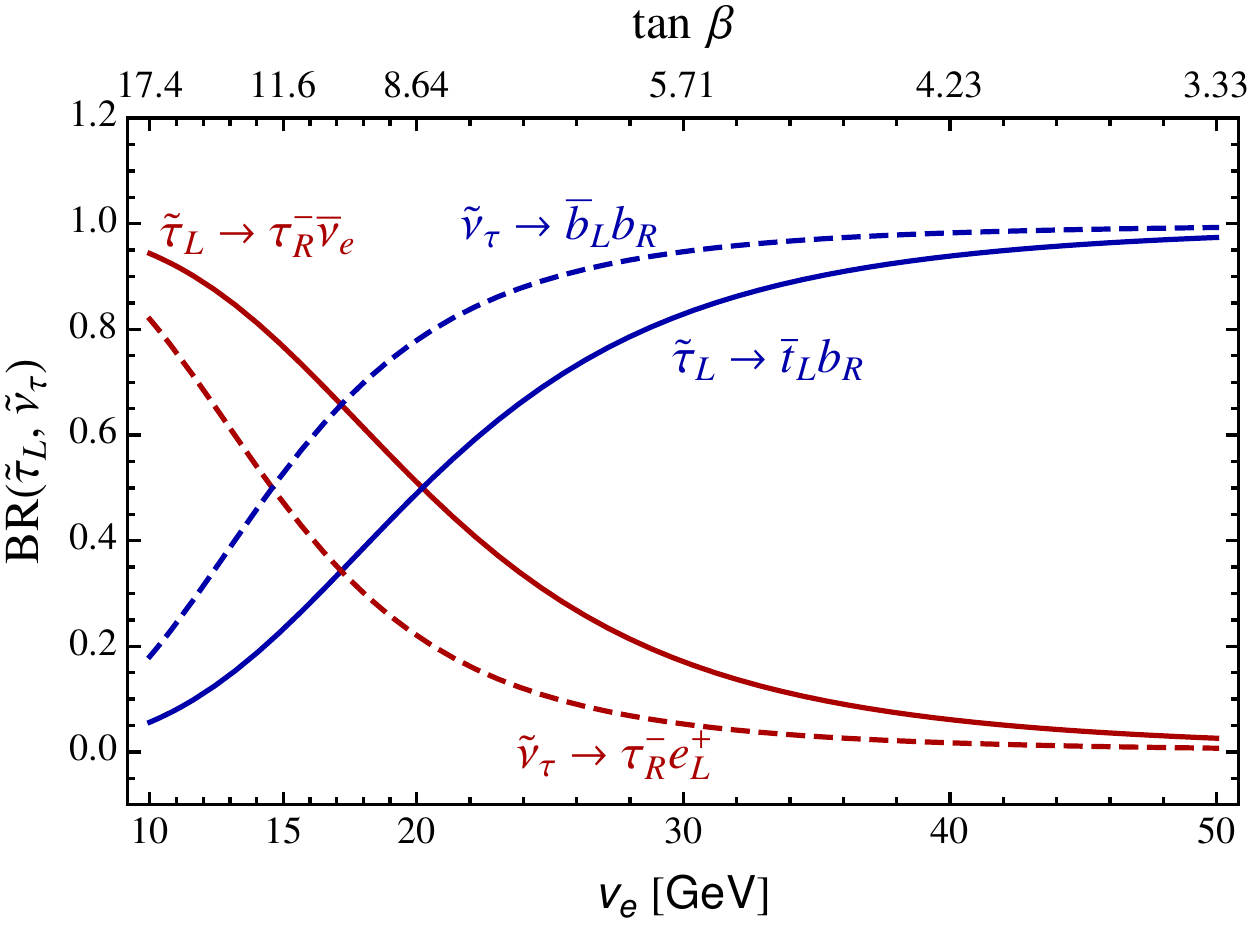}
\end{tabular}
\end{center}
\caption{\footnotesize{$\tilde \tau_L$ (solid lines) and
$\tilde{\nu}_\tau$ (dashed lines) decay modes for two masses:
$m_{\tilde{\tau}_L} \approx m_{\tilde{\nu}_\tau} = 180~{\rm GeV}$
(left panel) and $m_{\tilde{\tau}_L} \approx m_{\tilde{\nu}_\tau} =
250~{\rm GeV}$ (right panel).  It is assumed that $\tilde{X}^{0+}_1$
is heavier than the $(\tilde{\nu}_\tau, \tilde{\tau}_L)$ pair, and
that $\lambda'_{333}$ saturates Eq.~(\ref{lambdap333}).}}
\label{fig:stau}
\end{figure}  
We focus on the decays of the $(\tilde{\nu}_\tau, \tilde{\tau}_L)$
pair since it may very well be the ``LSP", i.e. the last step in a
cascade decade to SM particles.  In this case the charged slepton
decay modes are $\tilde{\tau}^-_L \to \tau^-_R \bar{\nu}_{e}$ and
$\tilde{\tau}^-_L \to \bar{t}_L b_R$, with partial decay widths given
by:
\bea
\Gamma( \tilde \tau^-_L \rightarrow \tau^-_R \bar \nu_{e} ) &=& \frac{ m_{\tilde \tau_L}}{16 \pi} \, y_\tau^2~, 
\label{stautaunu} \\
\Gamma( \tilde \tau^-_L \rightarrow  \bar{t}_L b_R ) &=& \frac{ m_{\tilde \tau_L}}{16 \pi} \,  (\lambda'_{333})^2 \left(1-\frac{m_t^2}{m_{\tilde \tau_L^2}}\right)^2~.
\label{stautb}
\eea
The decay widths for the $SU(2)_L$ related processes,
$\tilde{\nu}_{\tau} \to \tau^-_R e^+_L$ and $\tilde{\nu}_\tau \to
\bar{b}_L b_R$, are obtained from Eqs.~(\ref{stautaunu}) and
(\ref{stautb}) with the replacements $m_{\tilde \tau_L} \to m_{\tilde
\nu_\tau}$ and $m_t \to m_b$.  In Fig.~\ref{fig:stau}, we show the
branching fractions as a function of the sneutrino vev, assuming that
$\lambda'_{333}$ saturates Eq.~(\ref{lambdap333}), and taking $m_\tau
= 1.7~{\rm GeV}$.  We see that the $\bar{t}_L b_R$ channel can be
sizable in the large sneutrino vev/small $\tan\beta$ limit, in spite
of the phase space suppression when $m_{\tilde{\tau}_L} \sim m_t +
m_b$ (left panel).  Away from threshold, it can easily dominate (right
panel).

If, on the other hand, $\tilde{X}^{0+}_1$ and $\tilde{X}^{+-}_1$ are
lighter than the LH third generation sleptons, their dominant decay
modes would be $\tilde{\tau}^-_L \to \tilde{X}^{0+}_1 \tau^-_L$ or
$\tilde{\tau}^+_L \to \tilde{X}^{+-}_1 \bar{\nu}_{\tau}$, for the
charged lepton, and $\tilde{\nu}_\tau \to \tilde{X}^{0+}_1 \nu_\tau$
or $\tilde{\nu}_\tau \to \tilde{X}^{+-}_1 \tau^+_L$ for the sneutrino.

\subsection{Squark Decays}
\label{sec:squarkdecays}

As already explained, we focus on the case where the gluinos are
heavier than the squarks and, therefore, the squark decay mode into a
gluino plus jet is kinematically closed.  The lightest neutralinos and
charginos are instead expected to be lighter than the squarks since
naturalness requires the $\mu$-term to be at the electroweak scale,
while we will see that the first and second generation squarks have to
be heavier than about $600~{\rm GeV}$.  Thus, the squark decays into a
quark plus the lightest neutralino or into a quark plus the lightest
chargino should be kinematically open.  However, the decay mode of the
left handed up-type squarks, which have $Q = 2/3$ and $R = 1$, into
the lightest chargino $\tilde{X}_1^{+-}$ plus a ($R$-neutral) jet is
forbidden by the combined conservation of the electric and R-charges:
$\tilde{u}_L~\slash{\!\!\!\!  \to}~\tilde{X}_1^{+-} j$.  The decay
mode into the second lightest neutralino, which can be of the $(++)$
type, could be allowed by the quantum numbers, but our choice $M^D_{1}
> m_{\tilde q}$ ensures that it is kinematically closed.  Note also
that since $u_R$ has $Q = 2/3$ and $R = -1$, one can have $\tilde{u}_R
\to \tilde{X}_1^{+-} j$.

\subsubsection{First and Second Generation Squarks}
\label{sec:squark12decays}

%
\begin{itemize}
\item The left-handed up-type squarks, $\tilde{u}_L $ and
$\tilde{c}_L$, decay into $\tilde{X}^{0+}_1 j $ and $e^+_L j$ with:
\bea
\Gamma( \tilde{u}_L \rightarrow \tilde{X}^{0+}_1 j) &\approx& \frac{ m_{\tilde q}}{32 \pi} 
\left[ \frac{1}{18} \left( g' V^N_{1 \tilde{b}} + 3 g V^N_{1 \tilde{w}} \right)^2 \right]   \left( 1- \frac{m_{\tilde{X}^0_1}^2}{m_{\tilde q}^2}\right)^2~,  
\label{SU2Diff}
\\
\Gamma( \tilde u_L \rightarrow  e^+_L j)  &=& \frac{ m_{\tilde q}}{16 \pi} \, y_d^2 \, (U^+_{1e})^2~,
\eea
and analogous expressions for $\tilde{c}_L$ (in Eq.~(\ref{SU2Diff}),
we do not display subleading terms proportional to the Yukawa
couplings).  The second decay is an example of a lepto-quark decay
mode.  However, taking into account the smallness of the Yukawa
couplings for the first two generations, together with the
$\tilde{X}_1^{0+}$ composition shown in
Fig.~\ref{fig:NeutralinoComposition}, one finds that the dominant
decay mode is the one into neutralino and a jet.  Therefore, in the
region of parameter space we are interest in, $\tilde{u}_L $ and
$\tilde{c}_L$ decay into $\tilde{X}_1^{0+} j$ with almost $100\%$
probability.

\item The down-type left-handed squarks, $\tilde{d}_L $ and
$\tilde{s}_L$, have the following decay channels:
\bea
\Gamma( \tilde d_L \rightarrow \tilde{X}_1^{0+} j) &\approx& \frac{ m_{\tilde q}}{32 \pi} \left[  \frac{1}{18} \left( g' V^N_{1 \tilde b}-3 g V^N_{1 \tilde{w}} \right)^2 \left( U^N_{1\nu} \right)^2 \right]  \left ( 1- \frac{m_{\tilde{X}^0_1}^2}{m_{\tilde q}^2}\right)^2~, 
\\
\Gamma( \tilde d_L \rightarrow \tilde{X}^{-+}_1 j) &\approx& \frac{ m_{\tilde q}}{16 \pi}  \left[  g^2 (U^-_{1 \tilde{w}})^2 \right]  \left( 1- \frac{m_{\tilde{X}^\pm_1}^2}{m_{\tilde q}^2} \right)^2~,  
\\
\Gamma( \tilde d_L \rightarrow  \bar{\nu}_e j)  &=& \frac{ m_{\tilde q}}{32 \pi} \, y_d^2 (U^N_{ 4 \nu})^2~,
\eea
with analogous expressions for $\tilde{s}_L$.  The relative minus sign
in the gaugino contributions to the neutralino decay channel is due to
the $SU(2)$ charge of the down-type squarks, and should be compared to
the up-type case, Eq.~(\ref{SU2Diff}).  This leads to a certain degree
of cancellation between the contributions from the bino and wino
components, which together with the factor of $1/18$ results in a
significant suppression of the neutralino channel.  Since the Yukawa
couplings are very small, it follows that the chargino channel is the
dominant decay mode of the down-type squarks of the first two
generations.

\item The right-handed up-type squarks, $ \tilde u_R $ and $\tilde
c_R$, decay according to
\bea
\Gamma( \tilde u^*_R \rightarrow \tilde{X}_1^{0+} j)  &\approx& \frac{ m_{\tilde q}}{32 \pi} \, \left[ \frac{8}{9} \, (g' V^N_{1 \tilde b})^2    \right] \left( 1- \frac{m_{\tilde{X}^0_1}^2}{m_{\tilde q}^2} \right)^2~,  
\label{suRN} \\
\Gamma( \tilde u_R \rightarrow \tilde{X}^{+-}_1 j)  &=&  \frac{ m_{\tilde q}}{16 \pi} \, (y_u V^-_{1 u})^2 \left( 1- \frac{m_{\tilde{X}^\pm_1}^2}{m_{\tilde q}^2} \right)^2~,
\label{suRC}
\eea
with analogous expressions for $\tilde c_R$.  The chargino decay mode
fo $\tilde{u}_R$ is suppressed since the up-type Yukawa coupling is
very small.  Therefore, the right-handed up-type squark decays into
$\tilde{X}^{0+}_1 j$ with almost $100\%$ probability.  However, the
charm Yukawa coupling is such that the various terms in
Eqs.~(\ref{suRN}) and (\ref{suRC}) are comparable when the mixing
angles are as in Figs.~\ref{fig:CharginoComposition} and
\ref{fig:NeutralinoComposition}.  For this benchmark scenario, both
decay channels happen to be comparable, as illustrated in the left
panel of Fig.~\ref{fig:cRsR}.  Here we used $y_c = m_c/\sqrt{v^2 -
v^2_e}$ with $m_c(\mu \approx 600~{\rm GeV}) \approx 550~{\rm
MeV}$~\cite{Xing:2007fb}.

\begin{figure}[t]
\begin{center}
\begin{tabular}{cc}
\includegraphics[scale=0.6]{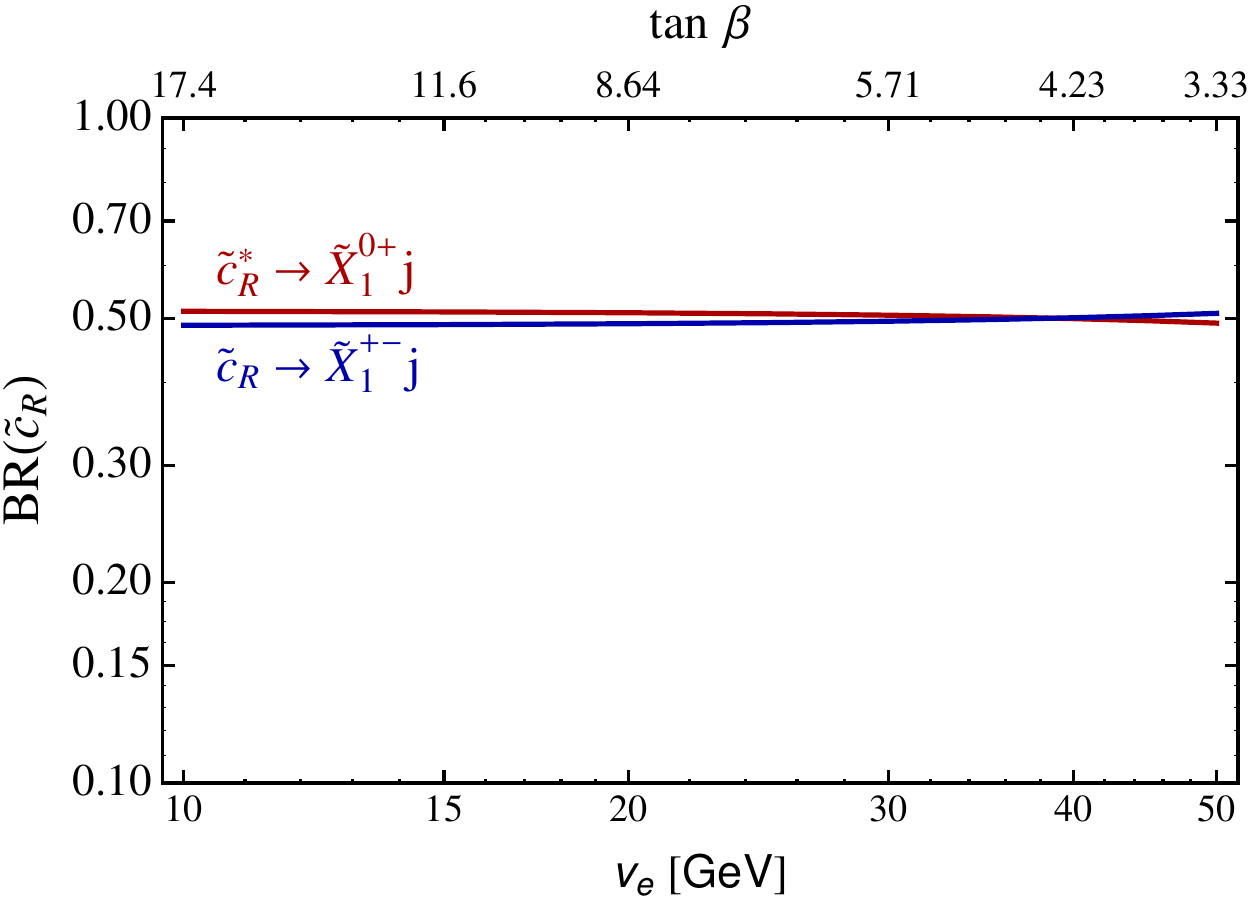} & 
\includegraphics[scale=.6]{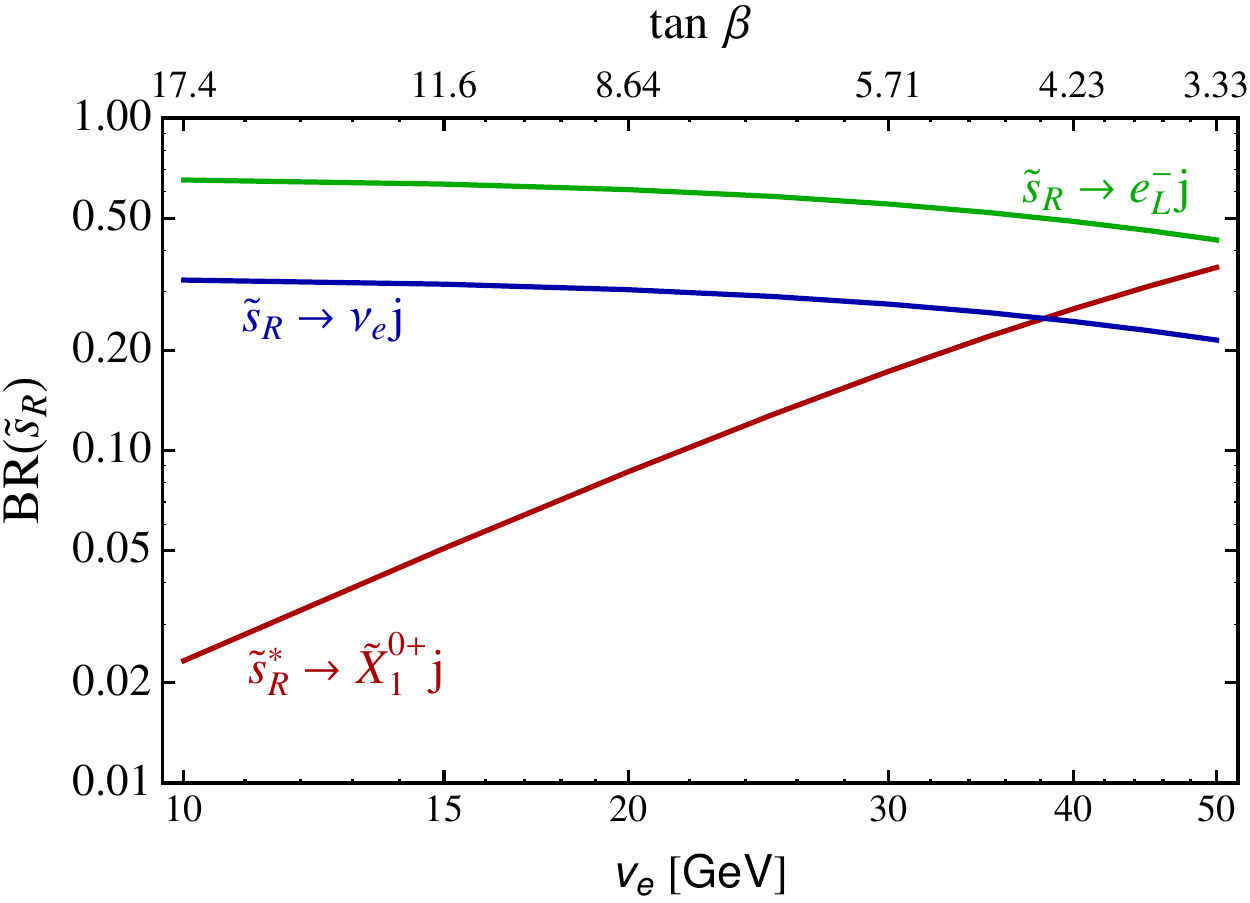} 
\end{tabular}
\end{center}
\caption{\footnotesize{Branching fractions for $\tilde c_R$ (left
panel) and $\tilde s_R$ (right panel) taking $M^D_{1} = 1$~TeV, $M^D_{2} =
1.5$~TeV, $\mu=200$~GeV, $\lambda^S_u = 0$ and
$\lambda^T_u = 1$.}}
\label{fig:cRsR}
\vspace{-10pt}
\end{figure}  

\item The right-handed down-type squarks, $\tilde d_R$ and $\tilde
s_R$, decay according to
\bea
\Gamma( \tilde d^*_R \rightarrow \tilde{X}_1^{0+} j) &\approx& \frac{ m_{\tilde q}}{32 \pi}  \left[ \frac{2}{9} \, (g' \, V^N_{1 \tilde b})^2 \right]  \left( 1- \frac{m_{\tilde{X}^0_1}^2}{m_{\tilde q}^2}\right)^2~,  
\\
\Gamma( \tilde d_R \rightarrow  e^-_L  j) &=&  \frac{ m_{\tilde q}}{16 \pi} \, y_d^2 \, (U^+_{1e})^2~,
\\ [0.4em]
\Gamma( \tilde d_R \rightarrow  \nu_e j)    &=&  \frac{ m_{\tilde q}}{32 \pi} \, y_d^2 \, (U^N_{4 \nu})^2~,
\eea
with analogous expressions for $\tilde s_R$.  Again, for the down
squark the Yukawa couplings are negligible so that it decays
dominantly into neutralino plus jet.  For the strange squark, however,
the various channels can be competitive as illustrated in the right
panel of Fig.~\ref{fig:cRsR}.  Here we used $y_s = m_s/\ve$ with
$m_s(\mu \approx 600~{\rm GeV}) \approx 49~{\rm
MeV}$~\cite{Xing:2007fb}.
\end{itemize}
%

\subsubsection{Third Generation Squarks}
\label{sec:squark3decays}

For the third generation we expect the lepto-quark signals to be
visible in all of our parameter space, although they may be of
different types.  The point is that the bottom Yukawa coupling can be
sizable in the small sneutrino vev/large $\tan\beta$ limit (as in the
MSSM), thus leading to a signal involving first generation leptons
through the $\lambda'_{133} \equiv y_b \approx 1.15 \times 10^{-2}
\sec\beta$ coupling.  In the large sneutrino vev/small $\tan\beta$
limit, on the other hand, the RPV coupling $\lambda'_{333} \lesssim
1.4 \cos\beta$ can be of order of $g'$, and may lead to third
generation leptons in the final state.

\begin{figure}[t]
\begin{center}
\begin{tabular}{cc}
\hspace{-5mm}
\includegraphics[scale=0.65]{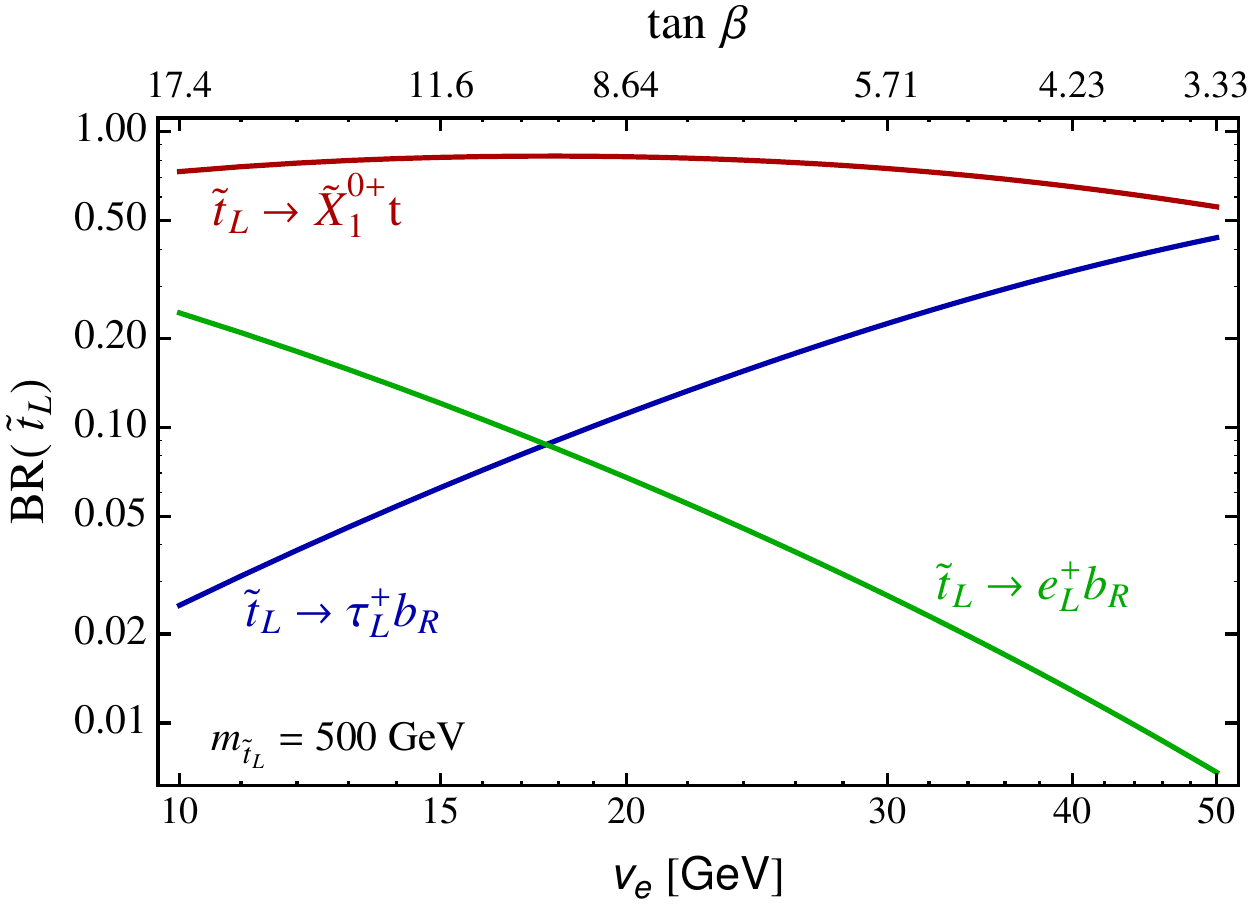} &
\includegraphics[scale=0.65]{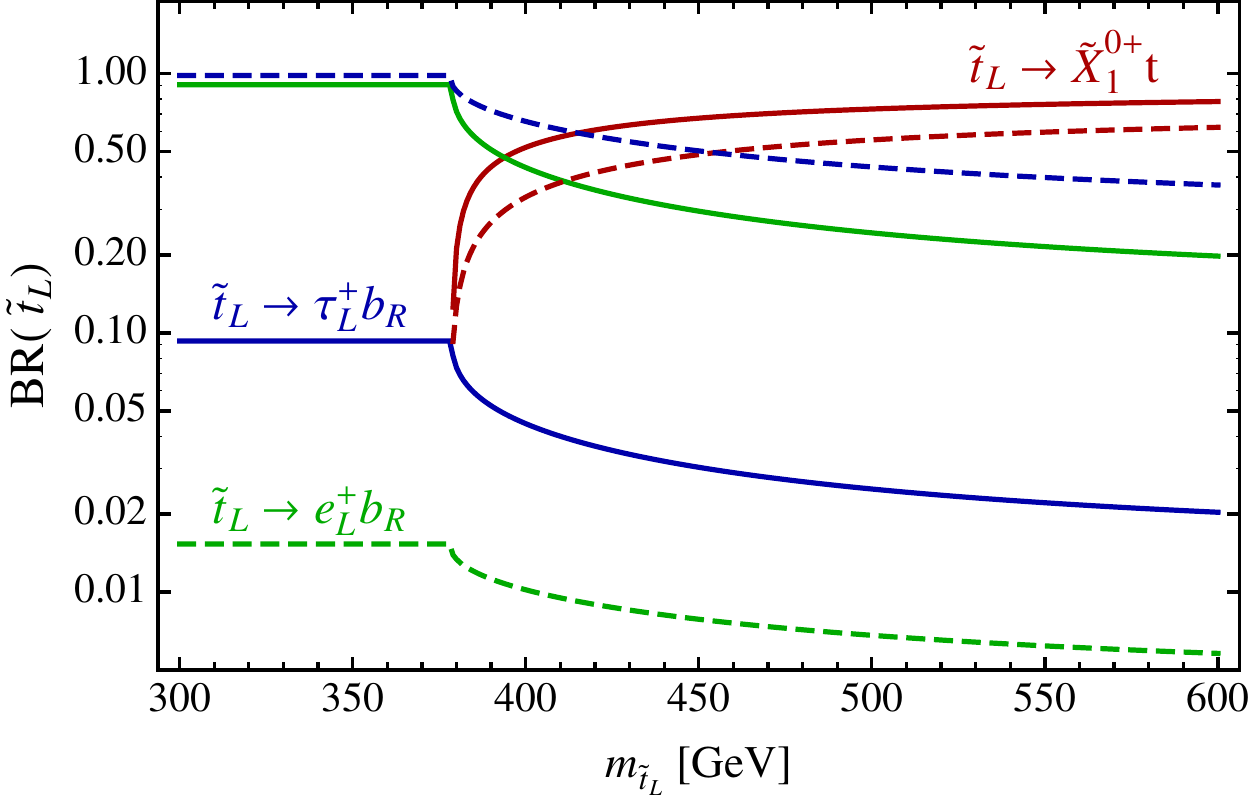}
\end{tabular}
\end{center}
\caption{\footnotesize{Branching fractions for the $\tilde t_L$ decay
modes computed for $M^D_{1} = 1$~TeV, $M^D_{2} = 1.5$~TeV, 
$\mu=200$~GeV, $\lambda^S_u = 0$ and $\lambda^T_u = 1$.  We also
assume $\lambda_{333}' = (2.1 \times 10^{-2}) / y_b$.  In the left
panel we take $m_{\tilde{t}_L} = 500~{\rm GeV}$, and show the
dependence on the sneutrino vev.  In the left panel we show the
dependence on $m_{\tilde{t}_L}$ for $\ve = 10~{\rm GeV}$ (solid lines)
and $\ve = 50~{\rm GeV}$ (dashed lines).}}
\vspace{-10pt}
\label{fig:tL}
\end{figure}  
\begin{itemize}
\item The left-handed stop, $\tilde{t}_L$, has the following decay
modes:
\bea
\Gamma( \tilde{t}_L \rightarrow \tilde{X}_1^{0+} t) &=&  \frac{m_{\tilde{t}_L}}{32 \pi}   
\left\{ \left[ \frac{1}{18} \left( g' V^N_{1 \tilde{b}} + 3 g V^N_{1 \tilde{w}} \right)^2 + y_t^2 (U^N_{1 u})^2 \right] 
\left( 1 - \frac{m^2_{\tilde X_1^0}}{m^2_{\tilde{t}_L}} - \frac{m^2_{t}}{m^2_{\tilde{t}_L}} \right) \right.
\nonumber \\ 
&& \hspace{3mm}
\left. \mbox{} - \frac{2}{3} \sqrt{2} \, y_t U^N_{1 u} \left( g' V^N_{1 \tilde{b}} + 3 g V^N_{1 \tilde{w}} \right) \frac{m_t m_{\tilde X_1^0}}{m^2_{\tilde{t}_L}} \rule{0mm}{7mm} \right\}
\lambda( m_{\tilde{t}_L}, m_{\tilde X_1^0},m_t)~,
\\ [0.3em]
\Gamma( \tilde{t}_L \rightarrow  e^+_L b_R)  &=& \frac{m_{\tilde{t}_L}}{16 \pi} \, y_b^2 \, (U^+_{1e})^2~, 
\\[0.4em]
\Gamma( \tilde{t}_L \rightarrow  \tau^+_L b_R)  &=& \frac{m_{\tilde{t}_L}}{16 \pi} \, (\lambda_{333}')^{2}~,
\eea
where
\bea \lambda( m_1, m_2,m_3
)= \sqrt{1+ \frac{m_2^4}{m_{1}^4} + \frac{m_3^4}{m_{1}^4} - 2 \left (
\frac{m_2^2}{m_{1}^2}+ \frac{m_3^2}{m_{1}^2} + \frac{m_2^2
m_3^2}{m_{1}^4} \right)}~.  
\eea 
When kinematically allowed, the decay mode into neutralino plus top is
the dominant one since it is driven by the top Yukawa coupling, as
shown in Fig.~\ref{fig:tL}.  However, this figure also shows that the
two lepto-quark decay modes can have sizable branching
fractions.\footnote{Here we used $y_t = m_t/\sqrt{v^2 - v^2_e}$ and
$y_b = m_b/\ve$ with $m_t(\mu \approx 500~{\rm GeV}) \approx 157~{\rm
GeV}$ and $m_b(\mu \approx 500~{\rm GeV}) \approx 2.56~{\rm
GeV}$~\cite{Xing:2007fb}.} In particular, at small sneutrino vev the
electron-bottom channel is the dominant lepto-quark decay mode (since
it is proportional to the bottom Yukawa), while in the large vev limit
the third generation lepto-quark channel dominates [we have assumed
that $\lambda_{333}'$ saturates the upper bound in
Eq.~(\ref{lambdap333})].  The existence of lepto-quark channels with a
sizable (but somewhat smaller than one) branching fraction is a
distinctive feature of our model, as will be discussed in more detail
in the following section.  We also note that in the case that
$\lambda'_{333}$ is negligible and does not saturate the bound in
Eq.~(\ref{lambdap333}), the $\tilde{t}_L \rightarrow \tau^+_L b_R$
channel is no longer present, so that the ${\rm BR}(\tilde{t}_L
\rightarrow e^+_L b_R)$ and ${\rm BR}(\tilde{t}_L \rightarrow
\tilde{X}_1^{0+} t)$ increase in the large sneutrino vev limit (but
are qualitatively the same as the left panel of
Fig.~\ref{fig:tL}).

\item The left-handed sbottom, $\tilde b_L $, has several decay modes
as follows:
\bea
 \Gamma( \tilde b_L \rightarrow \tilde{X}_1^{0+} b) &\approx& \frac{ m_{\tilde{b}_L}}{32 \pi} \left[ \frac{1}{18} \left( g' V^N_{1 \tilde{b}} - 3 g V^N_{1 \tilde{w}} \right)^2 \right]  \left( 1- \frac{m_{\tilde{X}^0_1}^2}{m_{\tilde{b}_L}^2} \right)^2~, 
 \\ 
\Gamma( \tilde b_L \rightarrow \tilde{X}^{-+}_1 t) &=& \frac{ m_{\tilde{b}_L}}{16 \pi}  \left\{ \left[  g^2 (U^-_{1 \tilde{w}})^2 + y_t^2 (V^-_{1 u})^2 \right] \left( 1 - \frac{m^2_{\tilde X_1^\pm}}{m^2_{\tilde{b}_L}} - 
\frac{m^2_{t}}{m^2_{\tilde{b}_L}} \right) \right.
\nonumber \\ 
&& \hspace{1cm}
\left. \mbox{} 
+ 4 g y_t U^-_{1 \tilde{w}} V^-_{1 u} \frac{m_{\tilde X_1^\pm} m_t}{m^2_{\tilde{b}_L}}
\right\}  \lambda( m_{\tilde{b}_L}, m_{\tilde X_1^{\pm}},m_t)~,  
\\[0.3em]
\Gamma( \tilde b_L \rightarrow  \bar{\nu}_e b_R)  &=& \frac{ m_{\tilde{b}_L}}{32 \pi}  \, y_b^2 \, (U^N_{4 \nu})^2~,  
\\[0.3em]
\Gamma( \tilde b_L \rightarrow  \bar{\nu}_{\tau} b_R)  &=& \frac{ m_{\tilde{b}_L}}{16 \pi} \, (\lambda_{333}')^{2}~. 
\eea
\begin{figure}[t]
\begin{center}
\begin{tabular}{ccc}
\hspace{-5mm}
\includegraphics[scale=0.65]{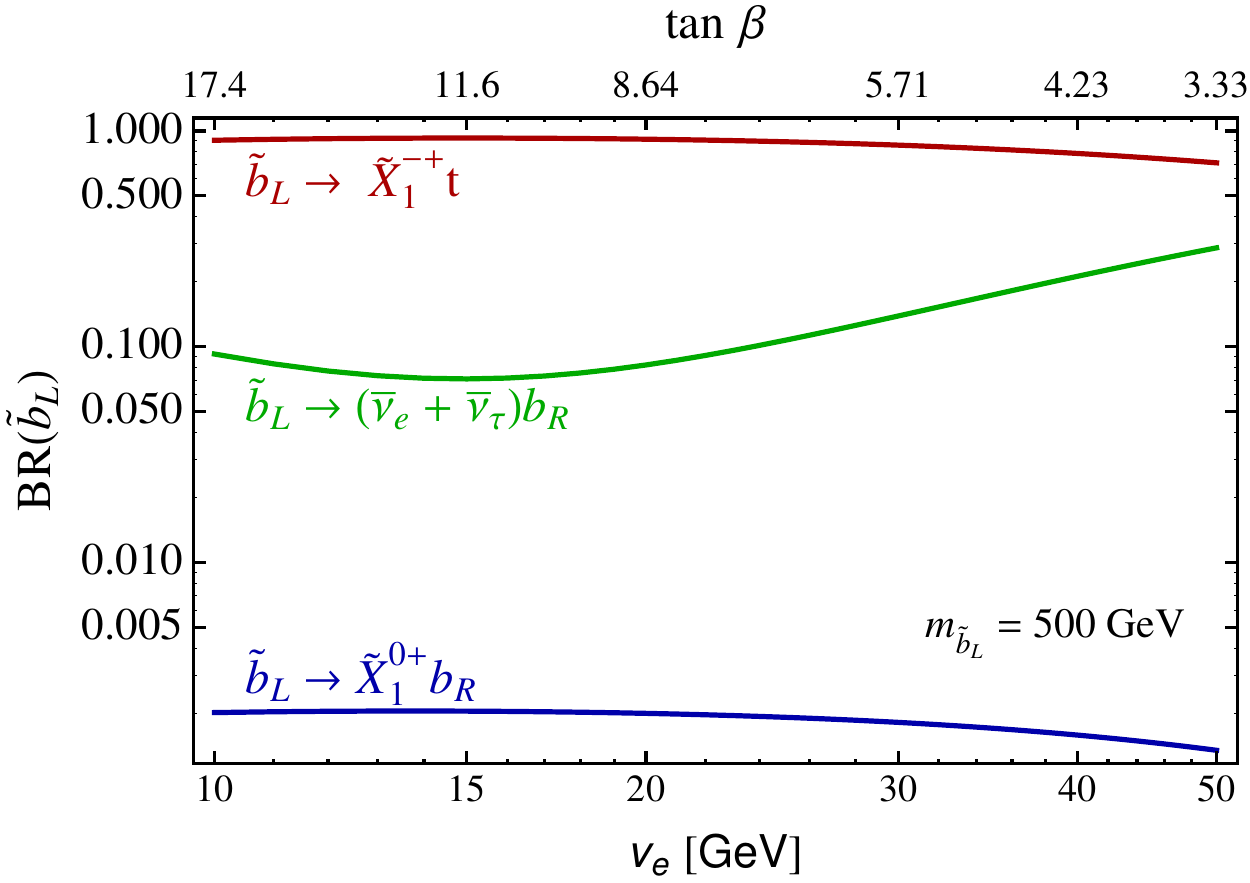}
\includegraphics[scale=0.65]{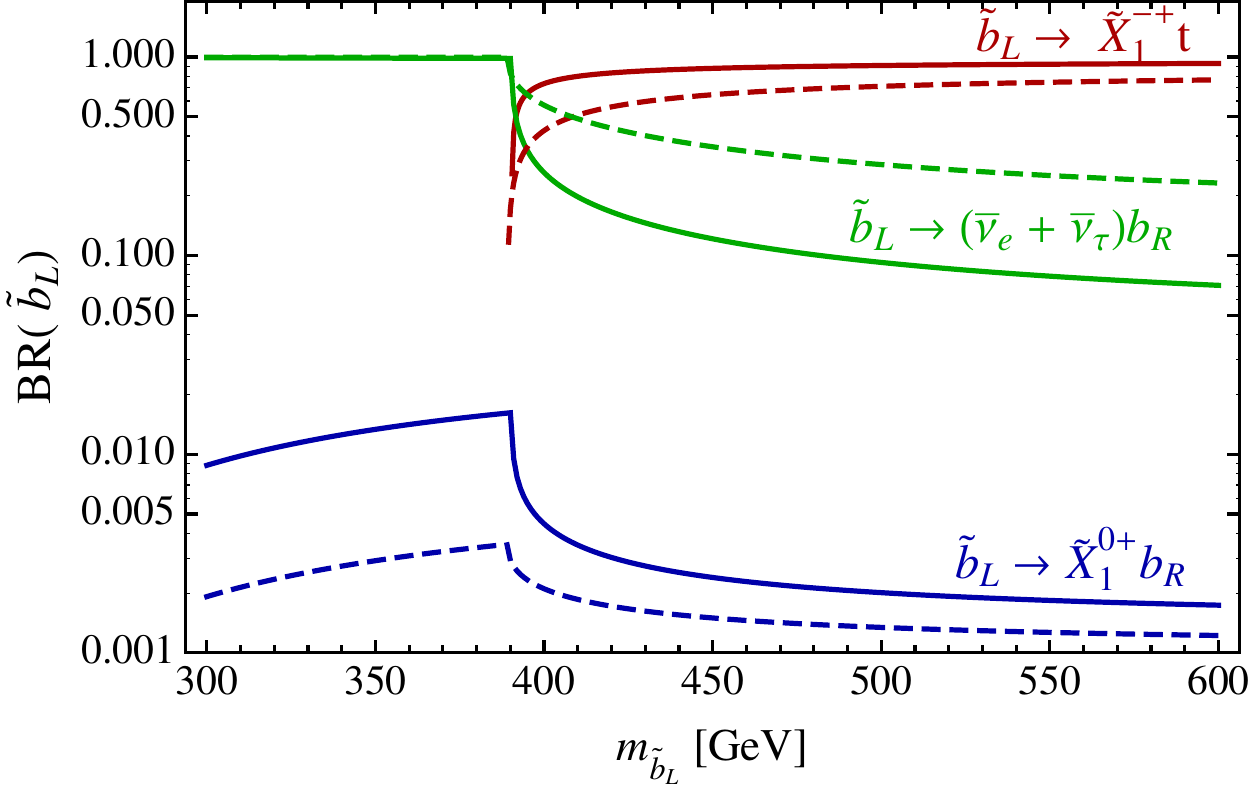}
\end{tabular}
\end{center}
\caption{\footnotesize{Branching fractions for $\tilde b_L$ computed
for $M^D_{1} = 1$~TeV, $M^D_{2} = 1.5$~TeV, $\mu=200$~GeV,
$\lambda^S_u = 0$ and $\lambda^T_u = 1$.  We also take $\lambda_{333}'
= (2.1 \times 10^{-2}) / y_b$, and add together the two neutrino
channels ($\bar{\nu}_e$ and $\bar{\nu}_\tau$).  In the left panel we
take $m_{\tilde{b}_L} = 500~{\rm GeV}$, and show the dependence on the
sneutrino vev.  In the left panel we show the dependence on
$m_{\tilde{b}_L}$ for $\ve = 10~{\rm GeV}$ (solid lines) and $\ve =
50~{\rm GeV}$ (dashed lines).}}
\label{fig:bL}
\end{figure}  
When kinematically open, the dominant decay mode is into a chargino
plus top since it is controlled by the top Yukawa coupling.  The
decays into neutrino plus bottom have always a sizable branching
fraction, as can be seen in Fig.~\ref{fig:bL}.  However, one should
note that when $\lambda'_{333}$ is negligible, so that the $\tilde b_L
\rightarrow \bar{\nu}_{\tau} b_R$ channel is unavailable, the decay
involving a neutrino ($\nu_e$ only) decreases as the sneutrino vev
increases (being of order 0.3\% at $\ve = 50~{\rm GeV}$).  The other
two channels adjust accordingly, but do not change qualitatively.

\begin{figure}[t]
\begin{center}
\begin{tabular}{ccc}
\hspace{-5mm}
\includegraphics[scale=0.65]{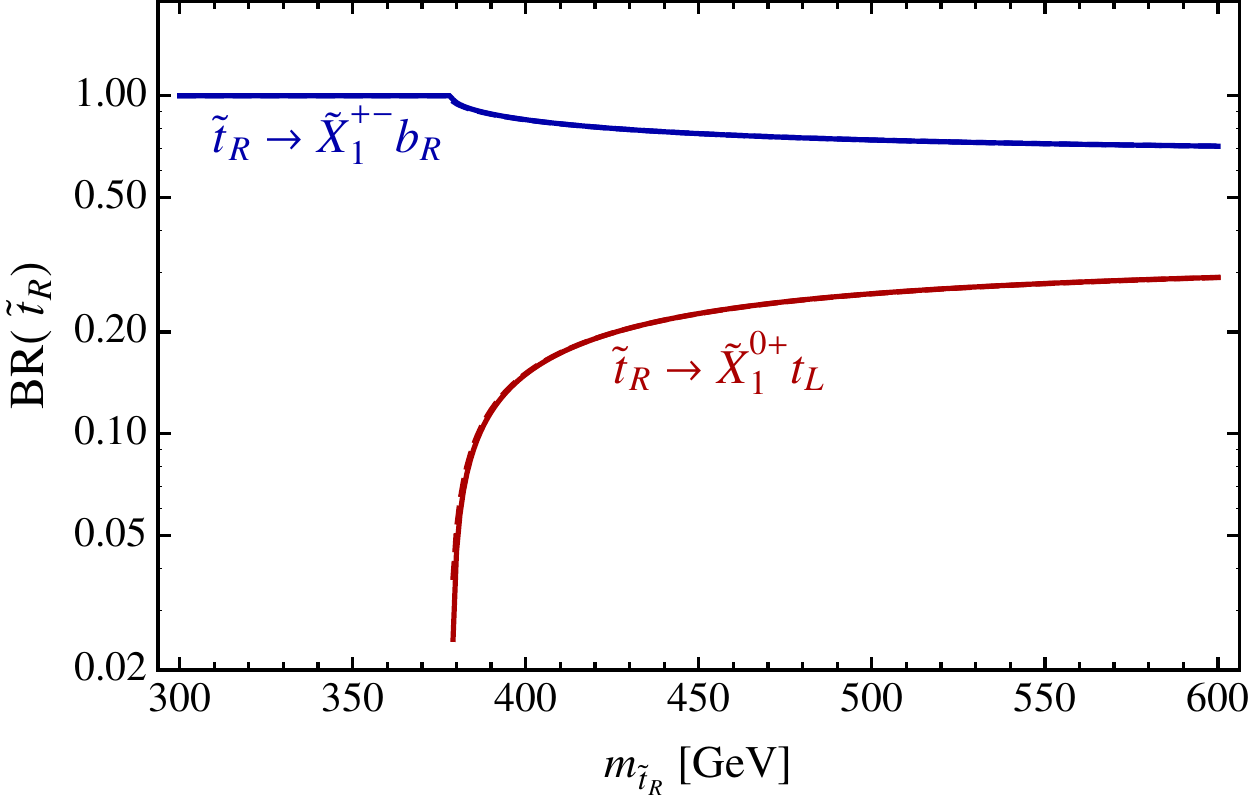}
\includegraphics[scale=0.65]{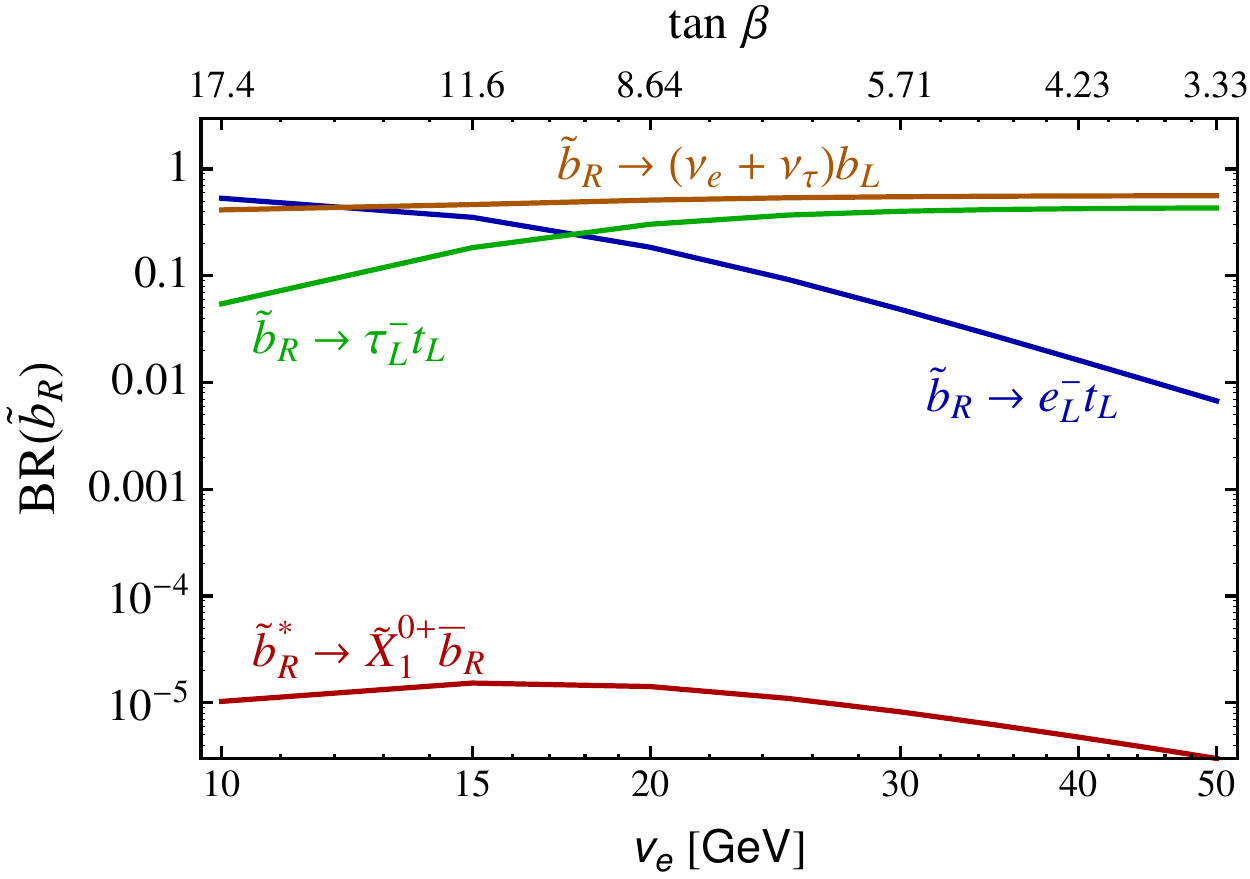}
\end{tabular}
\end{center}
\caption{\footnotesize{Branching fractions for $\tilde t_R$ as a
function of $m_{\tilde{t}_R}$ (left panel), and for $\tilde{b}_R$ as a
function of $\ve$ (right panel) computed for 
$M^D_{1} = 1$~TeV, $M^D_{2} = 1.5$~TeV, $\mu=200$~GeV, $\lambda^S_u = 0$ and $\lambda^T_u =
1$.  For $\tilde{b}_R$, we take $\lambda_{333}' = (2.1 \times 10^{-2})
/ y_b$, assume $m_{\tilde{b}_R} \gg m_{\tilde{X}^0_1}, m_t$, and add
together the two neutrino channels ($\bar{\nu}_e$ and
$\bar{\nu}_\tau$).}}
\label{fig:tRbR}
\end{figure}  

\item For the right-handed stop, $\tilde t_R$, the decay widths are:
\bea
\Gamma( \tilde t^*_R \rightarrow \tilde{X}_1^{0+} \bar{t}_L)  &=& \frac{m_{\tilde{t}_R}}{32 \pi} \left\{ \left[ \frac{8}{9} \left(g' V^N_{1 \tilde{b}} \right)^2 + y_t^2 (V^-_{1 u})^2 \right] \left( 1 - \frac{m^2_{\tilde X_1^0}}{m^2_{\tilde{t}_L}} - \frac{m^2_{t}}{m^2_{\tilde{t}_L}} \right) \right.
\label{stRDecay}
\nonumber \\ 
&& \hspace{1cm}
\left. \mbox{} + \frac{8}{3} \sqrt{2} \, y_t g' V^N_{1 \tilde{b}} U^N_{1 u} \frac{m_t m_{\tilde X_1^0}}{m^2_{\tilde{t}_L}} \rule{0mm}{7mm} \right\} \lambda( m_{\tilde{t}_R}, m_{\tilde X_1^0},m_t)~, 
\eea
%
%
\bea
\Gamma( \tilde t_R \rightarrow \tilde{X}^{+-}_1 b_R)  &=&  \frac{m_{\tilde{t}_R}}{16 \pi}  \left(y_t V^-_{1 u} \right)^2  \left( 1- \frac{m_{\tilde{X}_1^0}^2}{m_{\tilde{t}_R}^2}\right)^2~.
\eea
For the benchmark choice of $M^D_{2} = 1.5$~TeV, $M^D_{1} = 1$~TeV,
$\mu=200$~GeV, $\lambda^S_u = 0$ and $\lambda^T_u = 1$, we have
$\Gamma( \tilde t^*_R \rightarrow \tilde{X}_1^{0+} t_L) = 26\%~(15\%)$
and $\Gamma( \tilde t_R \rightarrow \tilde{X}^{+-}_1 b_R) =
74\%~(85\%)$ for $m_{\tilde{t}_R} = 500~(400)~{\rm GeV}$,
independently of the sneutrino vev.  For $m_{\tilde{t}_R} <
m_{\tilde{X}^0_1} + m_t$, the RH stop decays into $\tilde{X}^{+-}_1
b_R$ essentially 100\% of the time.  See left panel of
Fig.~\ref{fig:tRbR}.

\item The right-handed sbottom, $\tilde b_R$, has a variety of decay
modes:
\bea
\Gamma( \tilde b^*_R \rightarrow \tilde{X}_1^{0+} \bar{b}_R) &\approx& \frac{ m_{\tilde{b}_R}}{32 \pi} \left[ \frac{2}{9} \left(g' V^N_{1 \tilde{b}} \right)^2 \right] \left( 1- \frac{m_{\tilde{X}_1^0}^2}{m_{\tilde{b}_R}^2}\right)^2~,  
\\
\Gamma( \tilde b_R \rightarrow  e^-_L  t_L) &=& \frac{ m_{\tilde{b}_R}}{16 \pi} \, y_b^2 \, (U^+_{1 e})^2 \left( 1- \frac{m_{t}^2}{m_{\tilde{b}_R}^2}\right)^2~,
\\
\Gamma( \tilde b_R \rightarrow  \nu_e b_L) &=& \frac{ m_{\tilde{b}_R}}{32 \pi} \, y_b^2 \, (U^N_{4 \nu})^2~,  
\eea
%
%
\bea
\Gamma( \tilde b_R \rightarrow  \tau^-_L t_L)  &=& \frac{ m_{\tilde{b}_R}}{16 \pi} \, (\lambda_{333}')^{2}   \left( 1- \frac{m_{t}^2}{m_{\tilde{b}_R}^2}\right)^2~, 
\\
\Gamma( \tilde b_R \rightarrow  \nu_{\tau} b_L)  &=& \frac{ m_{\tilde{b}_R}}{16 \pi} \, (\lambda_{333}')^{2}~.
\eea
The lepto-quark signals are the dominant ones.  Adding the two
neutrino channels, the decay mode into $\nu b$ has a branching
fraction of about $50\%$ as shown in the right panel of
Fig.~\ref{fig:tRbR}.  The charged lepton signals can involve a LH
electron or a $\tau$ plus a top quark.  Note also that the decay mode
into $\tilde{X}_1^{0+} b$ is very suppressed.  We finally comment on
the modifications when $\lambda'_{333}$ is negligible.  Once the
$\tilde b_R \rightarrow \tau^-_L t_L$ and $\tilde b_R \rightarrow
\nu_{\tau} b_L$ channels become unavailable, one has that ${\rm
BR}(\tilde b_R \rightarrow e^-_L t_L) \approx 0.6$ and ${\rm
BR}(\tilde b_R \rightarrow \nu_e b_L) \approx 0.4$, independent of the
sneutrino vev.  The $\tilde b^*_R \rightarrow \tilde{X}_1^{0+}
\bar{b}_R$ channel remains negligible.

\end{itemize}

\section{$1^{\rm st}$ and $2^{\rm nd}$ Generation Squark Phenomenology}
\label{12Pheno}

In the present section we discuss the LHC phenomenology of the first
and second generation squarks, which are expected to be the most
copiously produced new physics particles.  Although these squarks are
not required by naturalness to be light, flavor considerations may
suggest that they should not be much heavier than the third generation
squarks.  Therefore, it is interesting to understand how light these
particles could be in our scenario.  As we will see, current bounds
allow them to be as light as $500-700~{\rm GeV}$, while in the MSSM
the LHC bounds have already exceeded the $1~{\rm TeV}$ threshold.  The
bounds can arise from generic jets + $\cancel{E}_T$ searches, as well
as from searches involving leptons in the final state.

\subsection{Squark Production}

We compute the cross section to produce a given final state $X$ in our
model as follows:
\bea
\sigma(pp \to X) &=& \sum_i \sigma(pp \to i) \times \textrm{BR}(i \to X)~,
\label{squarkproductionBR}
\eea
where $i = \tilde{q}_1 \tilde{q}_2, \tilde{g} \tilde{q}, \tilde{g}
\tilde{g}$, and the squark pair production can in principle come in
several flavor and chirality combinations.  We generate the production
cross section for each independent $i$-th state with
MadGraph5~\cite{Alwall:2011uj}.  Here we note that, due to the
assumption of gluinos in the multi-TeV range, and the fact that we
will be interested in squarks below 1 TeV, our cross section is
dominated by the production of squark pairs.  We have also computed
the corresponding $K$-factor with Prospino2~\cite{Beenakker:1996ed},
as a function of the squark mass for fixed (Majorana) gluino masses of
$2-5$~TeV. We find that for squark masses below about 1~TeV, the
K-factor is approximately constant with $K \approx 1.6$.  Since, to
our knowledge, a NLO computation in the Dirac case is not available,
we will use the previous K-factor to obtain a reasonable estimate of
the Dirac NLO squark pair-production cross-section.
\begin{figure}[t]
\begin{center}
\begin{tabular}{cc}
\includegraphics[scale=0.6]{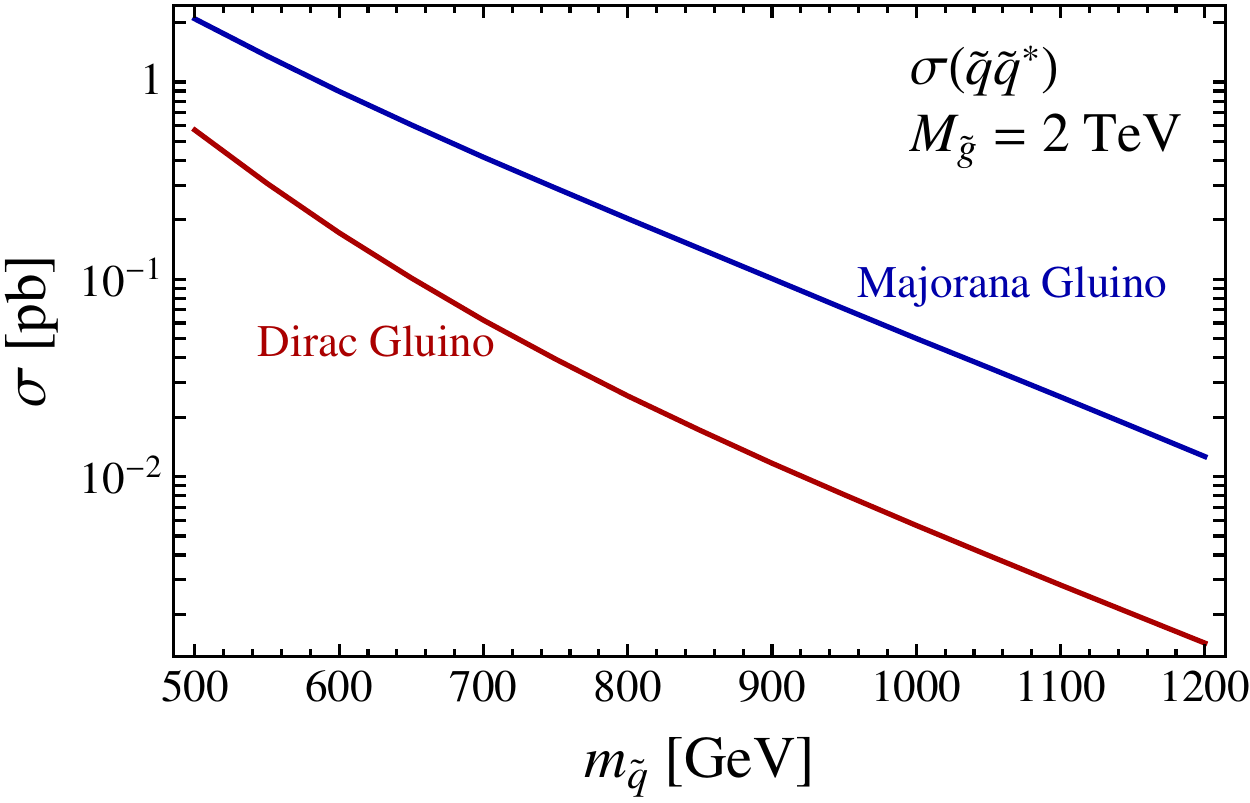} & 
\includegraphics[scale=.6]{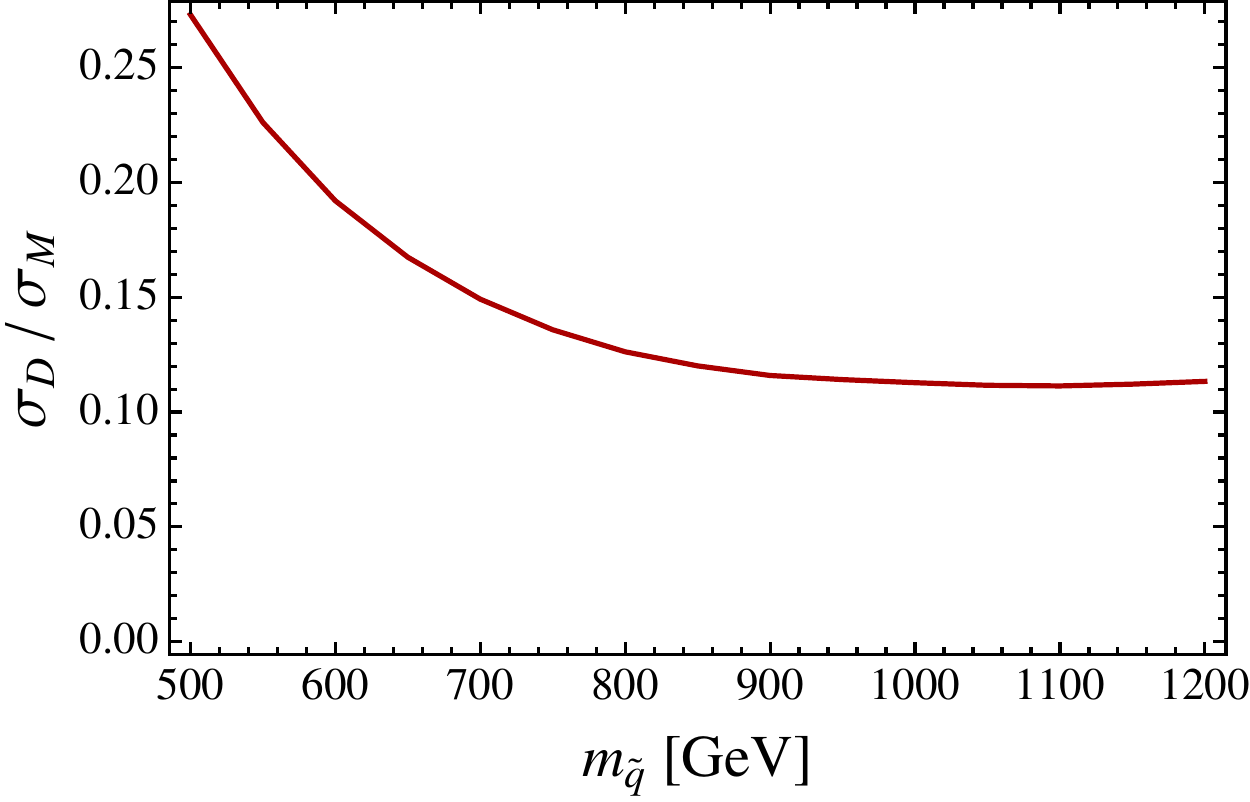} 
\end{tabular}
\end{center}
\caption{\footnotesize{Left Panel: $\tilde{q} \tilde{q}^*$ production
cross-section (all flavor combinations), for the 7 TeV LHC run,
computed for 2~TeV Dirac (red) and Majorana (blue) gluinos.  In the
right panel we plot the ratio between the two cross sections, showing
that the suppression in the Dirac case can be significant.}}
\label{cross}
\end{figure}  

One should note that the Dirac nature of the gluinos results in a
significant suppression of certain $t$-channel mediated gluino
diagrams compared to the Majorana (MSSM) case, as already emphasized
in~\cite{Heikinheimo:2011fk,Kribs:2012gx} (see also Fig.~\ref{cross}).
Nevertheless, at $M_{\tilde g}=2$~TeV such contributions are not
always negligible, and should be included.  For instance, we find that
for degenerate squark with $m_{\tilde{q}} = 800~{\rm GeV}$, the
production of $ \tilde{u}_L \tilde{u}_R$, $ \tilde{u}_L \tilde{d}_R$
and $ \tilde{u}_R \tilde{d}_L$ is comparable to the ``diagonal"
production of $\tilde{q}_L \tilde{q}^*_L$ and $\tilde{q}_R
\tilde{q}^*_R$ for all the squark flavors $\tilde{q} = \tilde{u},
\tilde{d}, \tilde{s}, \tilde{c}$ taken together.  As indicated in
Eq.~(\ref{squarkproductionBR}) we include separately the $\textrm{BR}$
for each $i$-th state to produce the final state $X$, since these can
depend on the squark flavor, chirality or generation.

\subsection{``Simplified Model" Philosophy}
\label{sec:Simplified}

We have seen that $\tilde{u}_L$, $\tilde{u}_R$, $\tilde{d}_R$ and
$\tilde{c}_L$ decay dominantly through the neutralino channel, the LH
down-type squarks, $\tilde{d}_L$ and $\tilde{s}_L$, decay dominantly
through the chargino channel, and $\tilde{c}_R$ and $\tilde{s}_R$ can
have more complicated decay patterns (see Fig.~\ref{fig:cRsR}).  The
striking lepto-quark decay mode, $\tilde{s}_R \to e^-_L j$, will be
treated separately. In this section we focus on the decays involving
neutralinos and charginos.  Since the signals depend on how the
neutralino/chargino decays, it is useful to present first an analysis
based on the simplified model (SMS) philosophy.  To be more precise,
we set bounds assuming that the neutralinos/charginos produced in
squark decays have a single decay mode with ${\rm BR}=1$.  We also
separate the ``neutralino LSP scenario", in which
$\tilde{X}^{0+}_1/\tilde{X}^{+-}_1$ decay into SM particles, from the
``stau LSP scenario", where they decay into $\tilde{\tau}^-_L
\tau^+_L$, $\tilde{\nu}_\tau \bar{\nu}_\tau$ or $\tilde{\tau}^+_L
\nu_\tau$.  We will give further details on these subsequent decays
below, where we treat the two cases separately.

\begin{wrapfigure}{r}{0.55\textwidth}
\centering
\begin{tabular}{cc}
\includegraphics[scale=0.65]{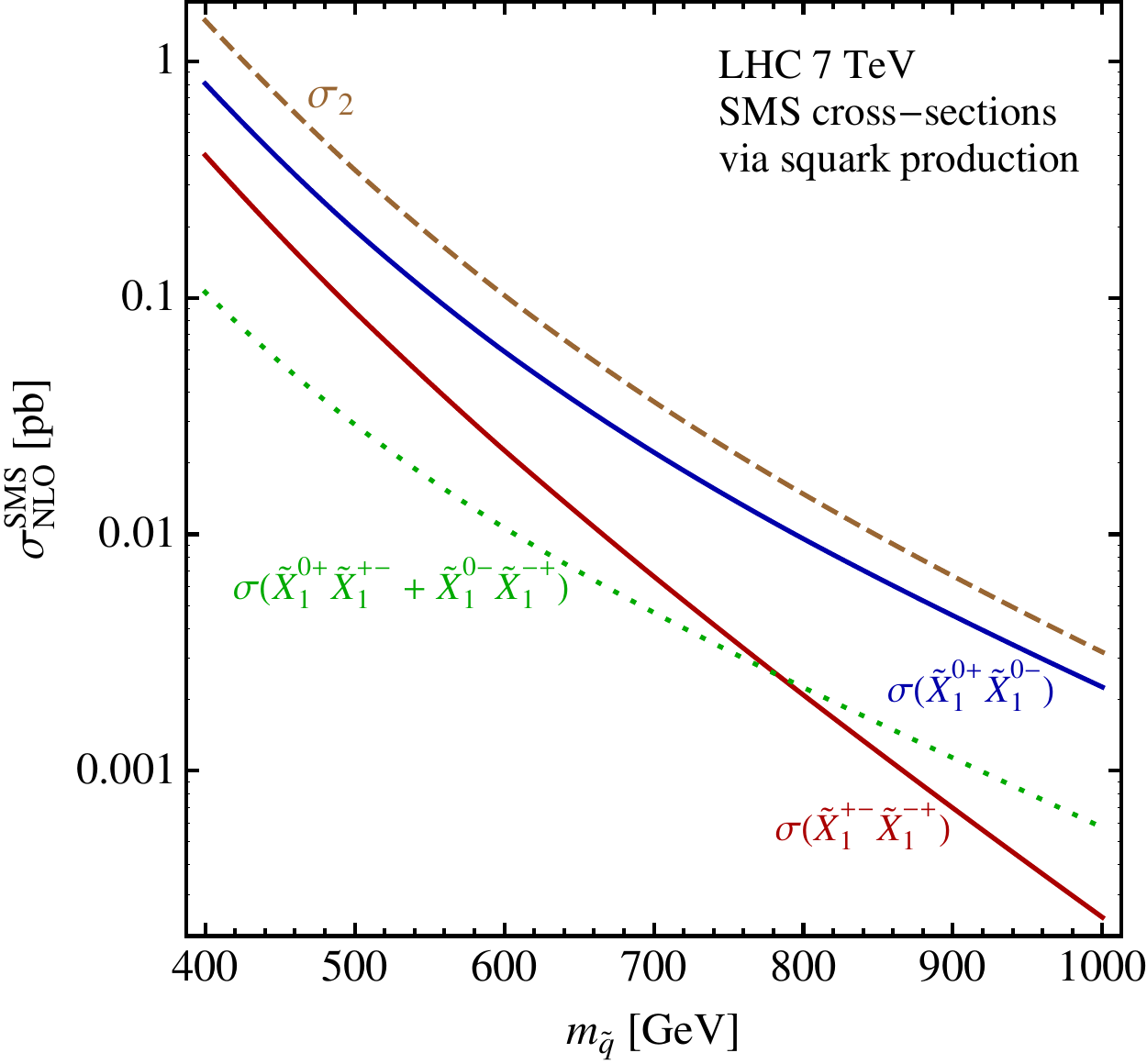}
\end{tabular}
\caption{\footnotesize{Cross-sections for the separate production of
$\tilde X^{0+}_1 \tilde X^{0-}_1$, $\tilde X^{+-}_1 \tilde X^{-+}_1$
and $\tilde X^{0+}_1 \tilde X^{+-}_1$ trough squark pair-production in
the SMS approach (see main text).  The solid and dotted lines
correspond to $\sigma_1$, according to the case.  The dashed line
marked as $\sigma_2$ corresponds to the full pair-production of
squarks, irrespective of how they decay.  The cross-sections are
computed for $M_{\tilde g}=2 $ TeV for a 7 TeV LHC run, with a
K-factor, $K=1.6$.}}
\label{crossdl}
\end{wrapfigure}
Here we emphasize that we regard the jets plus
$\tilde{X}^{0+}_1/\tilde{X}^{+-}_1$ stage as part of the
\textit{production}.  The point is that an important characteristic of
our scenario is that different types of squarks produce overwhelmingly
only one of these two states.  For instance, if we are interested in
two charginos in the squark cascade decays, this means that they must
have been produced through LH down-type squarks (with a smaller
contribution from $\tilde{c}_R \tilde{c}_R^*$ production), and the
production of any of the other squarks would not be relevant to this
topology.  Conversely, if we are interested in a topology with two
neutralinos, the LH down-type squarks do not contribute.  We denote by
$\sigma_1$ the corresponding cross sections, computed via
Eq.~(\ref{squarkproductionBR}) with $X = ``\tilde{X}^{+-}_1
\tilde{X}^{-+}_1 jj"$ or $X = ``\tilde{X}^{0+}_1 \tilde{X}^{0-}_1
jj"$, taking the ${\rm BR}$'s as exactly zero or one, according to the
type of squark pair $i$.\footnote{The only exception is the RH charm
squark, $\tilde{c}_R$, for which we take ${\rm BR}(\tilde{c}_R^* \to
\tilde{X}_1^{0+} j) = {\rm BR}(\tilde{c}_R \to \tilde{X}_1^{+-} j) =
0.5$, although the characteristics of the signal are not very
sensitive to this choice.  We also neglect the decays of $\tilde{s}_R$
into neutralino/neutrino plus jet.} At the same time, since in other
realizations of the $R$-symmetry these production patterns may not be
as clear-cut, we will also quote bounds based on a second production
cross section, denoted by $\sigma_2$, where it is assumed that
\textit{all} the squarks decay either into the lightest neutralino or
chargino channels with unit probability.  This second treatment is
closer to the pure SMS philosophy, but could be misleading in the case
that lepton number is an $R$-symmetry.  We show the corresponding
cross-sections in Fig.~\ref{crossdl}.

It should also be noted that the great majority of simplified models
studied by ATLAS and CMS consider either $m_{\tilde q} = M_{\tilde
g}$, or $m_{\tilde q} \gg M_{\tilde g}$.  Therefore, at the moment
there are only a handful of dedicated studies of our topologies,
although we will adapt studies performed for other scenarios to our
case.  In the most constraining cases, we will estimate the acceptance
by simulating the signal in our scenario\footnote{We have implemented
the full model in FeynRules~\cite{Christensen:2008py}, which was then
used to generate MadGraph 5 code~\cite{Alwall:2011uj}.  The parton
level processes are then passed through Pythia for hadronization and
showering, and through Delphes~\cite{Ovyn:2009tx} for fast detector
simulation.} and applying the experimental cuts, but for the most part
a proper mapping of the kinematic variables should suffice (provided
the topologies are sufficiently similar).  A typical SMS analysis
yields colored-coded plots for the upper bound on $\sigma \times {\rm
BR}$ (or $A \times \epsilon$) for the given process, in the plane of
the produced (strongly-interacting) particle mass (call it $m_{\tilde
q}$), and the LSP mass (call it $m_{\rm LSP}$).  In most cases, the
LSP is assumed to carry $\cancel{E}_T$.  Often, there is one
intermediate particle in the decay chain.  Its mass is parametrized in
terms of a variable $x$ defined by $ m_{\rm intermediate} = x
m_{\tilde q}+ (1-x) m_{\rm LSP}$.  In our ``neutralino LSP scenario",
the intermediate particle is either the lightest neutralino $\tilde
X^{0+}_1$ or the lightest chargino $ \tilde X_1^{+-}$, whose masses
are set by the $\mu$-term.  Since the particle carrying the
$\cancel{E}_T$ is the neutrino, i.e.~$m_{\rm LSP} = 0$, we have $x
\approx \mu/ m_{\tilde q}$.

\underline{\textit{We will set our bounds as follows}}: we compute our
theoretical cross section as described above (i.e.~based on the
$\sigma_1$ or $\sigma_2$ production cross-sections) as a function of
the squark mass, and considering the appropriate decay channel for the
$X^{0+}_1/X^{+-}_1$ (with ${\rm BR} = 1$ in the SMS approach).
Provided the topology is sufficiently similar, we identify the
$x$-axis on the color-coded plots in the experimental analyses
(usually $m_{\tilde{g}}$) with $m_{\tilde{q}}$, take $m_{LSP}=0$ (for
the neutrino), and identify ``$x$" as $\mu/m_{\tilde{q}}$ (from our
discussion above).  Then, we increase the squark mass until the
theoretical cross-section matches the experimental upper bound,
defining a lower bound on $m_{\tilde{q}}$.  In a few cases that have
the potential of setting strong bounds, but where the experimentally
analyzed topologies do not exactly match the one in our model, we
obtain the signal $\epsilon \times A$ from our own simulation and use
the 95\%~C.L.~upper bound on $\sigma \times \epsilon \times A$ to
obtain an upper bound on $\sigma$ that can be compared to our model
cross-section.  If there are several signal regions, we use the most
constraining one.

\subsection{Neutralino LSP Scenario}
\label{NeutralinoLSPScenario}

In the neutralino LSP scenario, and depending on the region of
parameter space (e.g.~the sneutrino vev or the values of the
$\lambda_S$ and $\lambda_T$ couplings), the lightest neutralino,
$\tilde X_1^{0+}$, can dominantly decay into $Z \bar{\nu}_e$, $h
\bar{\nu}_e$ or $W^- e^+_L$.  The ``lightest" chargino $\tilde
X_1^{+-}$ always decays into $W^+ \nu_e$.  Following the philosophy
explained in the previous subsection, we set separate bounds on four
simplified model scenarios:
\begin{multicols}{2}
\begin{itemize}
\item[$\bf (1)$] $\tilde{q} \to \tilde X_1^{0+} j \to (Z \bar{\nu}_e) \, j$~,

\item[$\bf (2)$] $\tilde{q} \to \tilde X_1^{0+} j \to (h \bar{\nu}_e) \, j$~,

\item[$\bf (3)$] $\tilde{q} \to \tilde X_1^{0+} j \to (W^- e^-_L) \, j$~,

\item[$\bf (4)$] $\tilde{q} \to \tilde X_1^{+-} j \to (W^+ \nu_e) \, j$~,

\end{itemize}
\end{multicols}
\noindent as well as on two benchmark scenarios (to be discussed in
Subsection~\ref{BenchmarkScenarios}) that illustrate the bounds on the
full model.

\begin{center}
\begin{table}
\center
\begin{tabular}{ |l  |c  |c  |c  |l |}
   \hline \rule{0mm}{5mm}
 \multirow{2}{*}{\bf Topology} & {\bf $\sigma_1$-bound} & {\bf $\sigma_2$-bound} &  \multirow{2}{*}{\bf Search} & \bf \multirow{2}{*}{Reference}
  \\
  & {\bf $m_{\tilde{q}}$~[GeV]} & {\bf $m_{\tilde{q}}$~[GeV]} & &
  \\ [0.2em]
  \hline \rule{0mm}{5mm}
 \multirow{2}{*}{$\tilde{q} \to \tilde X_1^{0+} j \to (Z \bar{\nu}_e) \, j$}  & 640 & 690 &  $Z(ll) $ + jets + $\cancel{E}_T$ & CMS~\cite{Chatrchyan:2012qka}  
 \\ [0.2em] 
 & 635 & 685 & jets + $\cancel{E}_T$ & ATLAS~\cite{Aad:2012hm}
 \\ [0.2em] 
 \hline \rule{0mm}{5mm}
$\tilde{q} \to \tilde X_1^{0+} j \to (h \bar{\nu}_e) \, j$ & 605 & 655 & jets + $\cancel{E}_T$ & ATLAS~\cite{Aad:2012hm}
\\ [0.2em] 
 \hline \rule{0mm}{5mm}
$\tilde{q} \to \tilde X_1^{0+} j \to (W^- e^-_L) \, j$  & 580 & 630 &  Multilepton &  ATLAS~\cite{ATLAS-CONF-2012-001}
\\ [0.2em] 
 \hline \rule{0mm}{5mm}
\multirow{3}{*}{$\tilde{q} \to \tilde X_1^{+-} j \to (W^+ \nu_e) \, j$}  & 530 & 650 & jets + $\cancel{E}_T$ & ATLAS~\cite{Aad:2012hm}  
 \\ [0.2em] 
 & 410  & 500 & Multilepton & ATLAS~\cite{:2012ms}
  \\ [0.2em] 
 & 350 & 430 &  $l$ + jets + $\cancel{E}_T$ & ATLAS~\cite{ATLAS-CONF-2012-104}
 \\ [0.2em] 
 \hline
 \hline \rule{0mm}{5mm}
 Benchmark 1 & $590 - 650$ & --- & jets + $\cancel{E}_T$ & ATLAS~\cite{Aad:2012hm}
 \\ [0.2em] 
 \hline \rule{0mm}{5mm}
 Benchmark 2 & $520 - 560$ & --- & jets + $\cancel{E}_T$ & ATLAS~\cite{Aad:2012hm}
 \\ [0.2em] 
 \hline
\end{tabular}
 \caption{\footnotesize{Bounds on $1^{\rm st}$ and $2^{\rm nd}$
 generation squark masses from squark pair production in the
 ``neutralino LSP scenario" for the Simplified Models $(1)$--$(4)$,
 and two benchmark scenarios.  See text for further details.}}
\label{summary}
\end{table}
\end{center}
\vspace{-6mm} 
There are several existing searches that can potentially constrain the
model:
\setlength{\columnsep}{-4cm}
\begin{multicols}{2}
\begin{itemize}
\item jets + $\cancel{E}_T$ ,
\item 1 lepton + jets + $\cancel{E}_T$ ,
\item $Z(ll) $ + jets + $\cancel{E}_T$~,
\item OS dileptons + $\cancel{E}_T$ + jets ,
\item multilepton + jets + $\cancel{E}_T$  (with or without Z veto).
\item[]
\end{itemize}
\end{multicols}
\setlength{\columnsep}{0.35cm}
We postpone the detailed description of how we obtain the
corresponding bounds to the appendix, and comment here only on the
results and salient features.  We find that typically the most
constraining searches are the generic jets + $\cancel{E}_T$ searches,
in particular the most recent ATLAS search with $5.8~{\rm
fb}^{-1}$~\cite{Aad:2012hm}.  In addition, some of the simplified
topologies can also be constrained by searches involving leptons +
jets + $\cancel{E}_T$.  For example, those involving a leptonically
decaying $Z$ are important for the $\tilde X_1^{0+} \rightarrow Z
\bar{\nu}_e$ case, while a number of multi-lepton searches can be
relevant for the topologies that involve a $W$.  We summarize our
findings in Table~\ref{summary}, where we exhibit the searches that
have some sensitivity for the given SMS topology.  We show the lower
bounds on the squark masses based on both the $\sigma_1$ and
$\sigma_2$ production cross-sections, as described in
Subsection~\ref{sec:Simplified}.  We see that these are below
$650$~GeV (based on $\sigma_1$; the bound from $\sigma_2$ is provided
only for possible application to other models).  We also show the
bounds for two benchmark scenarios (which depend on the sneutrino
vev), as will be discussed in the next subsection.  These are shown
under the $\sigma_1$ column, but should be understood to include the
details of the branching fractions and various contributing processes.
We have obtained the above results by implementing the experimental
analysis and computing the relevant $\epsilon \times A$ from our own
simulation of the signal, and using the model-independent 95\% CL
upper bounds on $\sigma \times \epsilon \times A$ provided by the
experimental analysis.  Whenever possible, we have also checked
against similar simplified model interpretations provided by the
experimental collaborations.  Such details are described in the
appendix, where we also discuss other searches that turn out to not be
sensitive enough, and the reasons for such an outcome.  In many cases,
it should be possible to optimize the set of cuts (within the existing
strategies) to attain some sensitivity.  This might be interesting,
for example, in the cases involving a Higgs, given that one might
attempt to reconstruct the Higgs mass.

We turn next to the analysis of the full model in
the context of two benchmark scenarios.

\subsubsection{Realistic Benchmark Points }
\label{BenchmarkScenarios}

%
\begin{figure}[t]
\begin{center}
\begin{tabular}{ccc}
\includegraphics[scale=0.6]{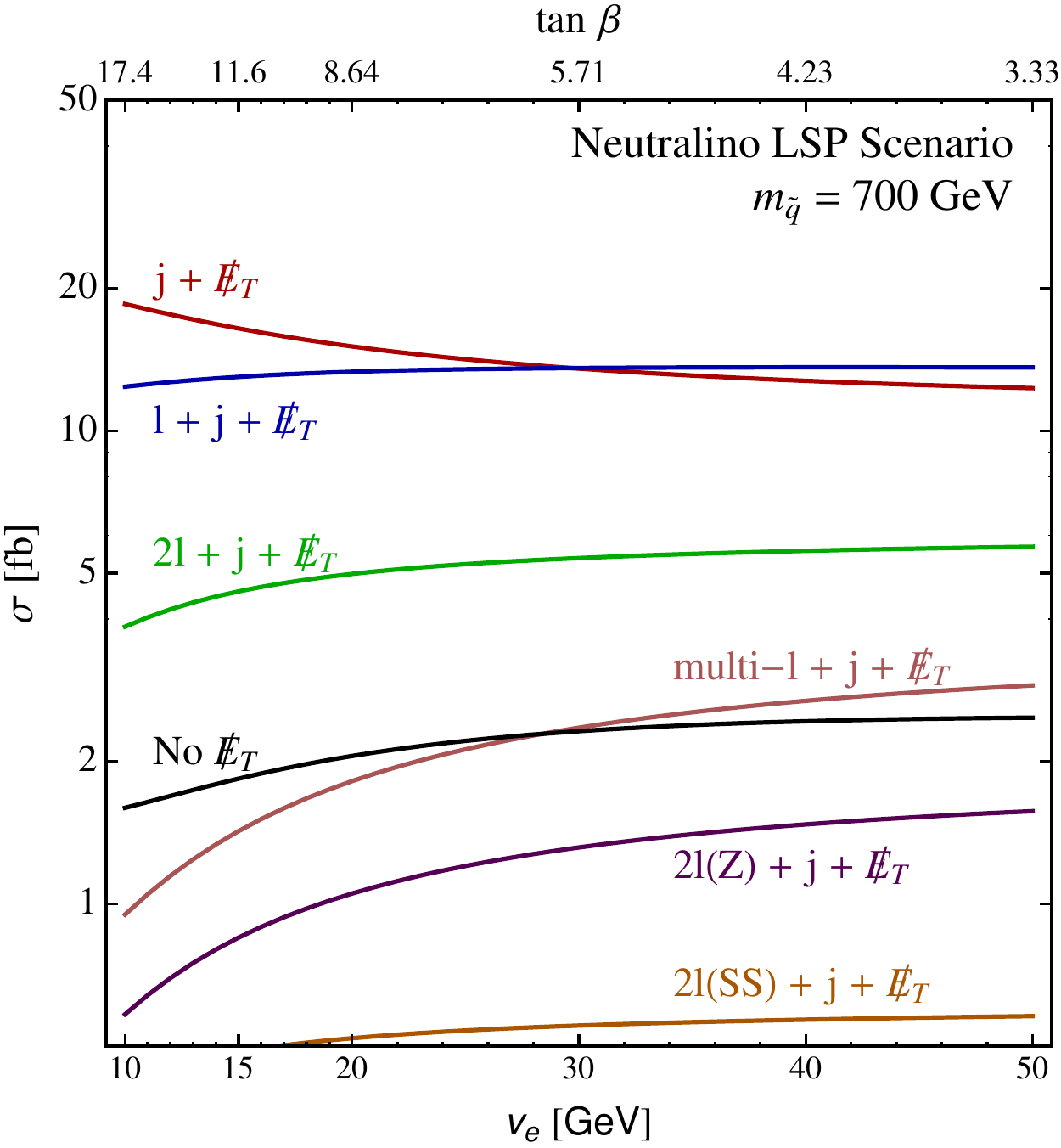} & &
\includegraphics[scale=0.6]{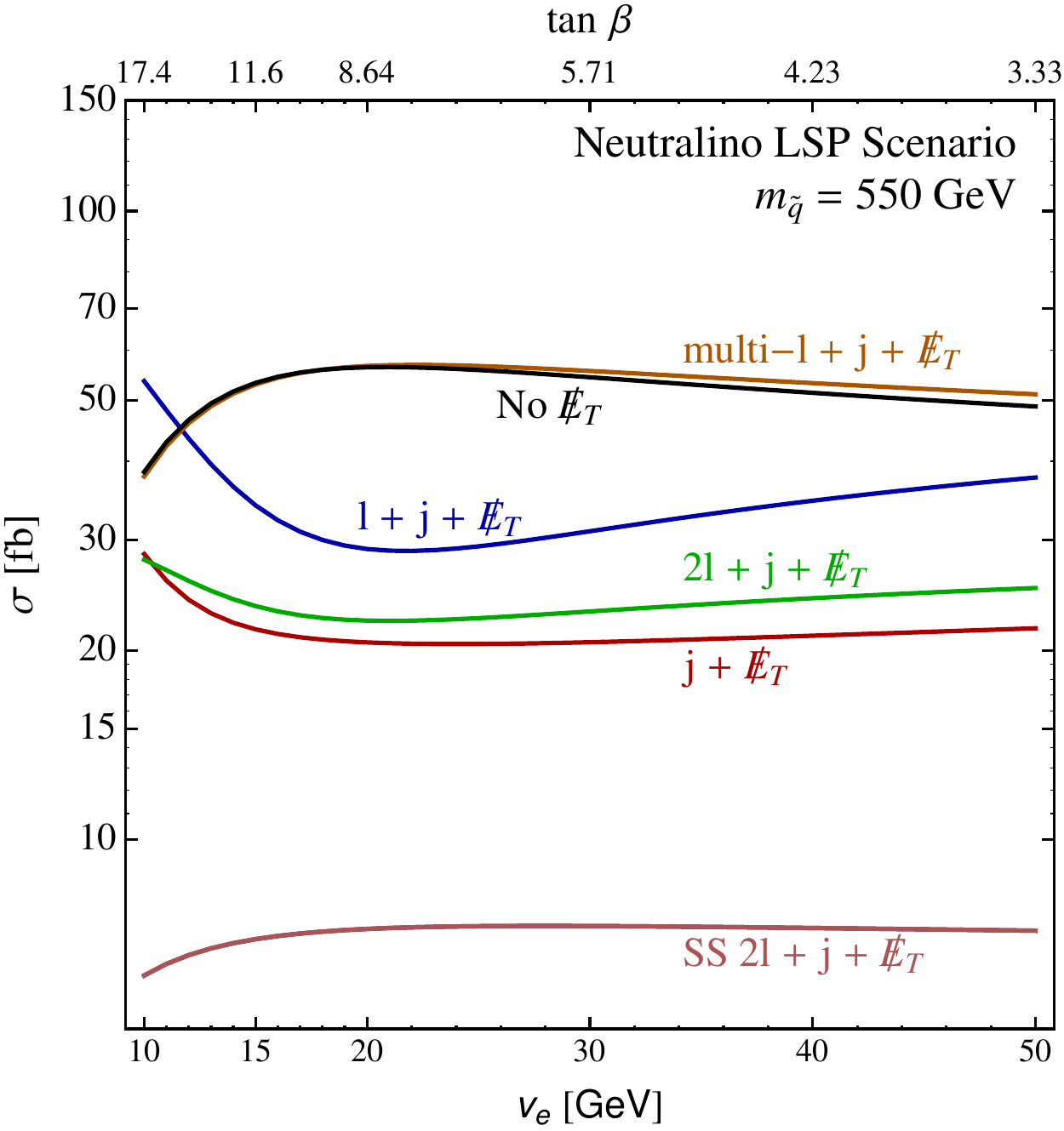}
\end{tabular}
\end{center}
\caption{\footnotesize{Cross-sections for a variety of signatures in
the ``neutralino LSP scenario".  All are computed for $M^D_{1} = 1$~TeV, 
$M^D_{2} = 1.5$~TeV, $\mu=200$~GeV, and assuming that
$\lambda'_{333}$ saturates Eq.~(\ref{lambdap333}).  In the left panel
we take $m_{\tilde{q}} = 700~{\rm GeV}$ and $\lambda^S_u = 0$,
$\lambda^T_u = 1$ (benchmark 1), while in the right panel we use
$m_{\tilde{q}} = 550~{\rm GeV}$ and $\lambda^S_u = \lambda^T_u = 0.4$
(benchmark 2).}}
\label{fig:LHCsign}
\end{figure}  
Besides the ``simplified model" type of bounds discussed above, it is
also interesting to present the bounds within benchmark scenarios that
reflect the expected branching fractions for the neutralinos/charginos
discussed in Subsections~\ref{sec:decayneutralinos} and
\ref{sec:decaycharginos}.  One difference with the analysis of the
previous subsections is that we can have all the combinations of
$\tilde X^{0+}_1 \tilde X_1^{0-} jj$, $\tilde X^{+-}_1 \tilde X_1^{-+}
jj$ and $\tilde X_1^{0+} \tilde X^{+-}_1 jj$ in squark decays, with
the corresponding BR's.  In Fig.~\ref{crossdl} we have shown the
individual cross-sections in the SMS approach.  These give a sense of
the relative contributions of the various channels.  In particular, we
see that the $\tilde X^{0+}_1 \tilde X_1^{0-}$ channel dominates.

\medskip
\noindent
\underline{\textit{Benchmark 1}}: ($M^D_{1} = 1$~TeV
$M^D_{2} = 1.5$~TeV, $\mu=200$~GeV, $\lambda^S_u = 0$,
$\lambda^T_u = 1$) corresponds to the case that the
$\tilde{X}^{0+}_1 \to h \bar{\nu}_e$ decay channel is important (in
fact, dominant at small sneutrino vev), while the gauge decay channels
of the $\tilde{X}^{0+}_1$ can be sizable (see left panel of
Fig.~\ref{fig:N1decayNeutrLSP}).  The LHC searches relevant to this
scenario are:
\setlength{\columnsep}{-2cm}
\begin{multicols}{2}
\begin{itemize}
\item jets + $\cancel{E}_T$ ,
\item 1 lepton + jets + $\cancel{E}_T$ ,
\item OS dileptons + $\cancel{E}_T$ + jets ,
\item dileptons (from $Z$ decay) + jets + $\cancel{E}_T$ ,
\item multilepton + jets + $\cancel{E}_T$  (without Z cut).
\item[]
\end{itemize}
\end{multicols}
\setlength{\columnsep}{0.35cm}
We apply the model-independent bounds discussed in the previous
sections, and find that the jets + $\cancel{E}_T$ search is the most
constraining one.  Using $\sigma_{j+\cancel{E}_T} \lesssim 20-40$~fb,
we find $m_{\tilde{q}} \gtrsim 620-690~{\rm GeV}$ ($m_{\tilde{q}}
\gtrsim 590-650~{\rm GeV}$) for $\ve = 10~{\rm GeV}$ ($\ve = 50~{\rm
GeV}$).  We show in the left panel of Fig.~\ref{fig:LHCsign} the
cross-sections for several processes, for $m_{\tilde{q}} = 700~{\rm
GeV}$.  These are computed from Eq.~(\ref{squarkproductionBR}) using
the actual BR's for the chosen benchmark.  Although there is some
dependence on the sneutrino vev, the global picture is robust against
$\ve$.

\medskip \noindent \underline{\textit{Benchmark 2}}: ($M^D_{1} =
1$~TeV, $M^D_{2} = 1.5$~TeV, $\mu=200$~GeV, $\lambda^S_u = \lambda^T_u
= 0.4$) corresponds to the case that the $\tilde{X}_1^{0+} \to W^-
e^+$ decay channel dominates (see right panel of
Fig.~\ref{fig:N1decayNeutrLSP}).  In the right panel of
Fig.~\ref{fig:LHCsign}, we show the cross-sections for the main
processes.  We see that, for this benchmark, the ``leptonic channels"
have the largest cross sections (especially the multilepton + jets +
$\cancel{E}_T$ one).  However, taking into account efficiencies of at
most a few percent for the leptonic searches (as we have illustrated
in the previous section), we conclude that the strongest bound on the
squark masses arises instead from the jets + $\cancel{E}_T$ searches
(as for benchmark 1).  Using $\sigma_{j+\cancel{E}_T} \lesssim
20-40$~fb, we find $m_{\tilde{q}} \gtrsim 520-580~{\rm GeV}$
($m_{\tilde{q}} \gtrsim 500-560~{\rm GeV}$) for $\ve = 10~{\rm GeV}$
($\ve = 50~{\rm GeV}$).  Note that there is a sizable ``no missing
energy" cross section.  However, this could be significantly lower
once appropriate trigger requirements are imposed.

 \subsection{Stau LSP Scenario}
 
In this scenario the dominant decay modes of $\tilde X_1^{0+} $ are
into $\tilde \tau^-_L \tau^+_L$ or $\tilde{\nu}_\tau \bar{\nu}_\tau$
(about 50-50), while the chargino $ \tilde X_1^{+-}$ decays into $
\tilde \tau^+_L \nu_{\tau}.$ The decay modes of $ \tilde \tau^-_L$
depend on the sneutrino vev: for large $\ve$ it decays dominantly into
$\bar{t}_L b_R$ (assuming $\lambda'_{333}$ is sizable), while for
smaller values of $\ve$ it decays dominantly into $\tau^-_R
\bar{\nu}_e$ trough the $\tau$ Yukawa coupling.  Similarly,
$\tilde{\nu}_\tau$ decays into $\bar{b}_L b_R$ for large sneutrino
vev, and into $\tau^-_R e^+_L$ for small sneutrino vev.  In the ``stau
LSP scenario" we prefer to discuss the two limiting cases of small and
large sneutrino vev, rather than present SMS bounds (recall from
Fig.~\ref{crossdl} that the squarks produce dominantly
$\tilde{X}^{0+}_1 \tilde{X}^{0-}_1$ pairs).  This scenario is,
therefore, characterized by \textit{third generation signals}.

\subsubsection{$\tilde{\tau}^-_L \rightarrow \tau^-_R
\bar{\nu}_e$ and $\tilde{\nu}_\tau \to \tau^-_R e^+_L$ decay modes}

These decays are characteristic of the small sneutrino vev limit.  In
this case all the final states would contain at least two taus: $i)$
for the $\tilde X_1^{0+} \tilde X_1^{0-}$ topology the final state
contains 2 jets, missing energy and $2\tau + 2e$, $3\tau + 1e$ or
$4\tau$'s; $ii)$ for the $\tilde X_1^{0+} \tilde X_1^{+-}$ topology
the final state contains 2 jets, missing energy and $2\tau + 1e$ or
$3\tau$'s; $iii)$ for the $\tilde X_1^{-+} \tilde X_1^{+-}$ topology
the final state contains 2 jets, missing energy and $2\tau$'s.  It is
important that cases $i)$ and $ii)$ can be accompanied by one or two
electrons, given that many searches for topologies involving
$\tau$'s~\footnote{Understood as hadronic $\tau$'s.} impose a lepton
($e$ or $\mu$) veto. 

Thus, for instance, a recent ATLAS study~\cite{ATLAS-CONF-2012-112}
with 4.7~fb$^{-1}$ searches for jets + $\cancel{E}_T$ accompanied by
exactly one (hadronically decaying) $\tau$ + one lepton ($e$ or
$\mu$), or by two $\tau$'s with a lepton veto.  Only the former would
apply to our scenario, setting a bound of $\sigma \times \epsilon
\times A = 0.68~{\rm fb}^{-1}$.  A previous ATLAS
search~\cite{ATLAS:2012ag} with 2.05~fb$^{-1}$ searches for \textit{at
least} $2\tau$'s (with a lepton veto), setting a bound of $\sigma
\times \epsilon \times A = 2.9~{\rm fb}^{-1}$.  However, the
efficiency of such searches is lower than the one for jets plus
missing energy (also with lepton veto).  Since in our scenario the
cross sections for these two signatures is the same, the latter will
set the relevant current bound.

There is also a CMS study \cite{CMS-PAS-SUS-12-004} sensitive to
$4\tau$ signals in the context of GMSB scenarios, which has a similar
topology to our case (SMS: $\tilde g \tilde g$ production with $\tilde
g \rightarrow q q \chi^0_1$ and $\chi^0_1 \to \tau^+\tau^-
\tilde{G}_\mu$).  From their Fig.~9b, we can see that the 95\% CL
upper limit on the model cross section varies between $0.3-0.03~{\rm
pb}$ for $400~{\rm GeV} < m_{\tilde{g}} < 700~{\rm GeV}$.  Including
the branching fractions, and reinterpreting the bound in the squark
mass plane,\footnote{As usual, the topology of this study contains two
additional hard jets at the parton level compared to our case.} we
find a bound of $m_{\tilde{q}} \gtrsim 600~{\rm GeV}$ at $\ve = 10$,
where the cross section is about 45~fb.  When the sneutrino vev
increases the bound gets relaxed so that for $\ve \gtrsim 20~{\rm
GeV}$ there is no bound from this study.

The generic searches
discussed in previous sections (not necessarily designed for
sensitivity to the third generation) may also be relevant:
\begin{multicols}{2}
\begin{itemize}
\item jets + $\cancel{E}_T$~,
\item jets + $\cancel{E}_T$ + 1 lepton~,
\item jets + $\cancel{E}_T$ + SS dileptons~,
\item jets + $\cancel{E}_T$ + OS dileptons~,
\item jets + $\cancel{E}_T$ + multi leptons~,
\item[]
\end{itemize}
\end{multicols}
\noindent where the leptons may arise from the $\tilde{\nu}_\tau$
decay as in cases $i)$ and $ii)$ above, or from leptonically decaying
$\tau$'s.\footnote{Note that when there are two taus and no additional
electrons, the SS dilepton searches do not apply.  This is a
consequence of the conserved $R$-symmetry.} It turns out that, as in
the ``neutralino LSP scenario", the strongest bound arises from the
jets + $\cancel{E}_T$ search.  We find from simulation of the signal
efficiency times acceptance for the ATLAS
analysis~\cite{Aad:2012hm} in our model that the most
stringent bound arises from signal region C (tight), and gives an
upper bound on the model cross section of about 120~fb.  Thus, we find
that $m_{\tilde{q}} \gtrsim 500~{\rm GeV}$ for $\ve = 10~{\rm GeV}$.

\subsubsection{$\tilde{\tau}^-_L \rightarrow
\bar{t}_L b_R$ and $\tilde{\nu}_\tau \to \bar{b}_L b_R$ decay modes}

When the third generation sleptons decay through these channels, as is
typical of the large sneutrino vev limit, the signals contain a $b
\bar b$ and/or a $t \bar t$ pair, as well as $\tau$'s.  Note that when
the $\tau$'s and tops decay hadronically one has a signal without
missing energy.  However, the branching fraction for such a process is
of order ${\rm BR}(\tilde{q} \to \tilde{X}^{0+}_1 j)^2 \times {\rm
BR}(\tilde{X}^{0+}_1 \to \tilde{\tau}_L^- \tau^+_L)^2 \times {\rm
BR}(\tilde{\tau}_L^- \to \bar{t}_L b_R)^2 \times {\rm BR}(t \to b
W^+)^2 \times {\rm BR}(W \to jj)^2 \times {\rm BR}(\tau \to jj)^2
\sim$ few per cent (in the large sneutrino vev limit, where all of
these branching fractions are sizable).  Indeed, we find that the ``no
$\cancel{E}_T$" cross section for $700~{\rm GeV}$ squarks in the
``stau LSP scenario" is of order $1~{\rm fb}$, which is relatively
small.  Rather, the bulk of the cross section shows in the jets +
$\cancel{E}_T$ and 1~lepton + jets + $\cancel{E}_T$ channels (with a
smaller 2~lepton + jets + $\cancel{E}_T$ contribution).  Simulation of
the ATLAS j+$\cancel{E}_T$ search~\cite{Aad:2012hm} in this
region of our model indicates that again the most stringent bound
arises from signal region C (tight) of this study, and gives an upper
bound on the model cross section of about 70~fb.  This translates into
a bound of $m_{\tilde{q}} \gtrsim 550~{\rm GeV}$ for $\ve = 50~{\rm
GeV}$.

\section{Third Generation Squark Phenomenology}
\label{3Pheno}

We turn now to the LHC phenomenology of the third generation squarks.
We start by studying the current constraints and then we will explain
how the third generation provides a possible smoking gun for our
model.  We separate our discussion into the signals arising from the
lepto-quark decay channels, and those that arise from the decays of the
third generation squarks into states containing $\tilde{X}^{0+}_1$ or
$\tilde{X}^{+-}_1$ (or their antiparticles).

\subsection{Lepto-quark Signatures}
\label{sec:leptoquarks}

\begin{figure}[t]
\begin{center}
\begin{tabular}{cc}
\includegraphics[scale=0.605]{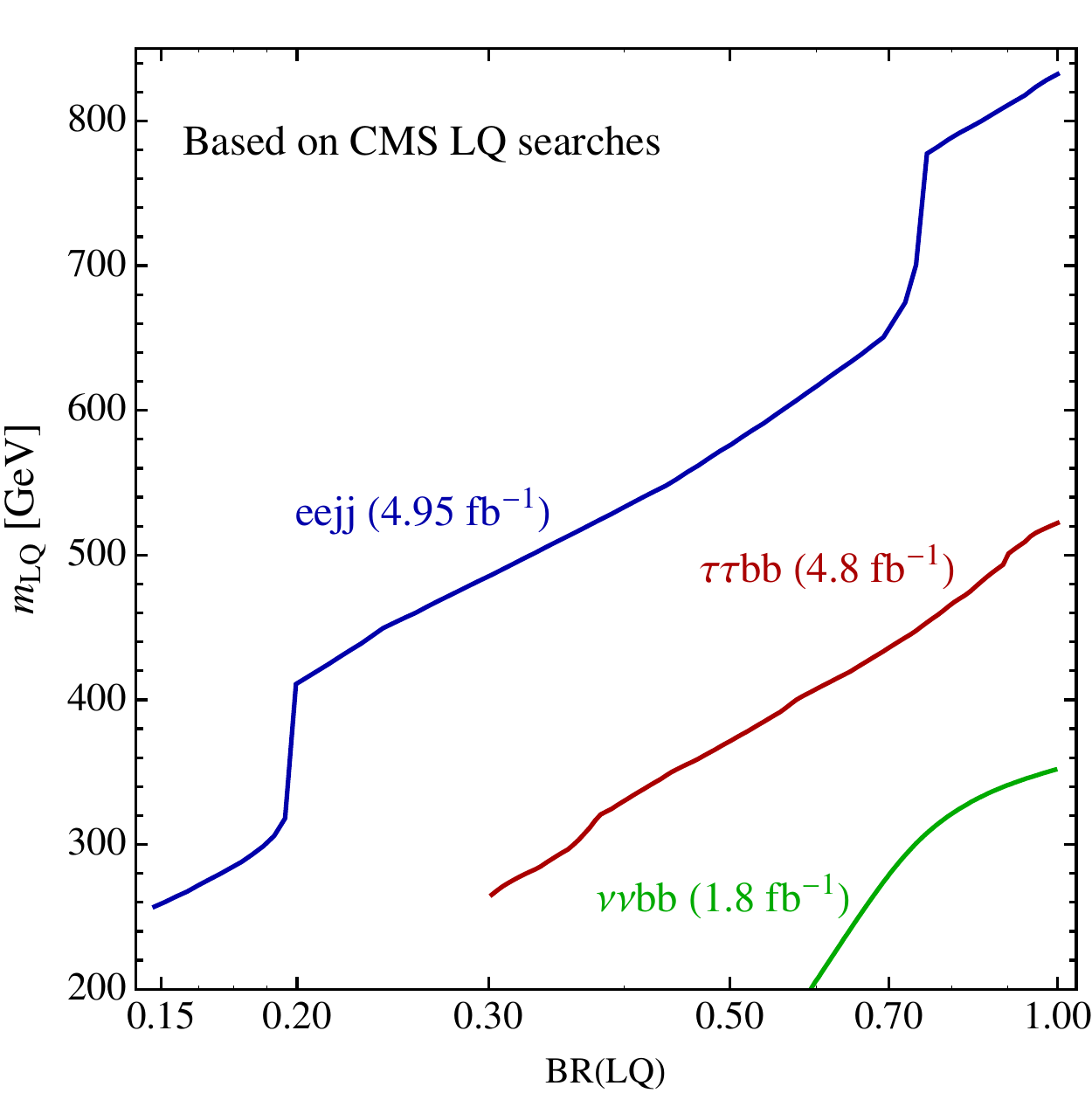} &
\includegraphics[scale=0.6]{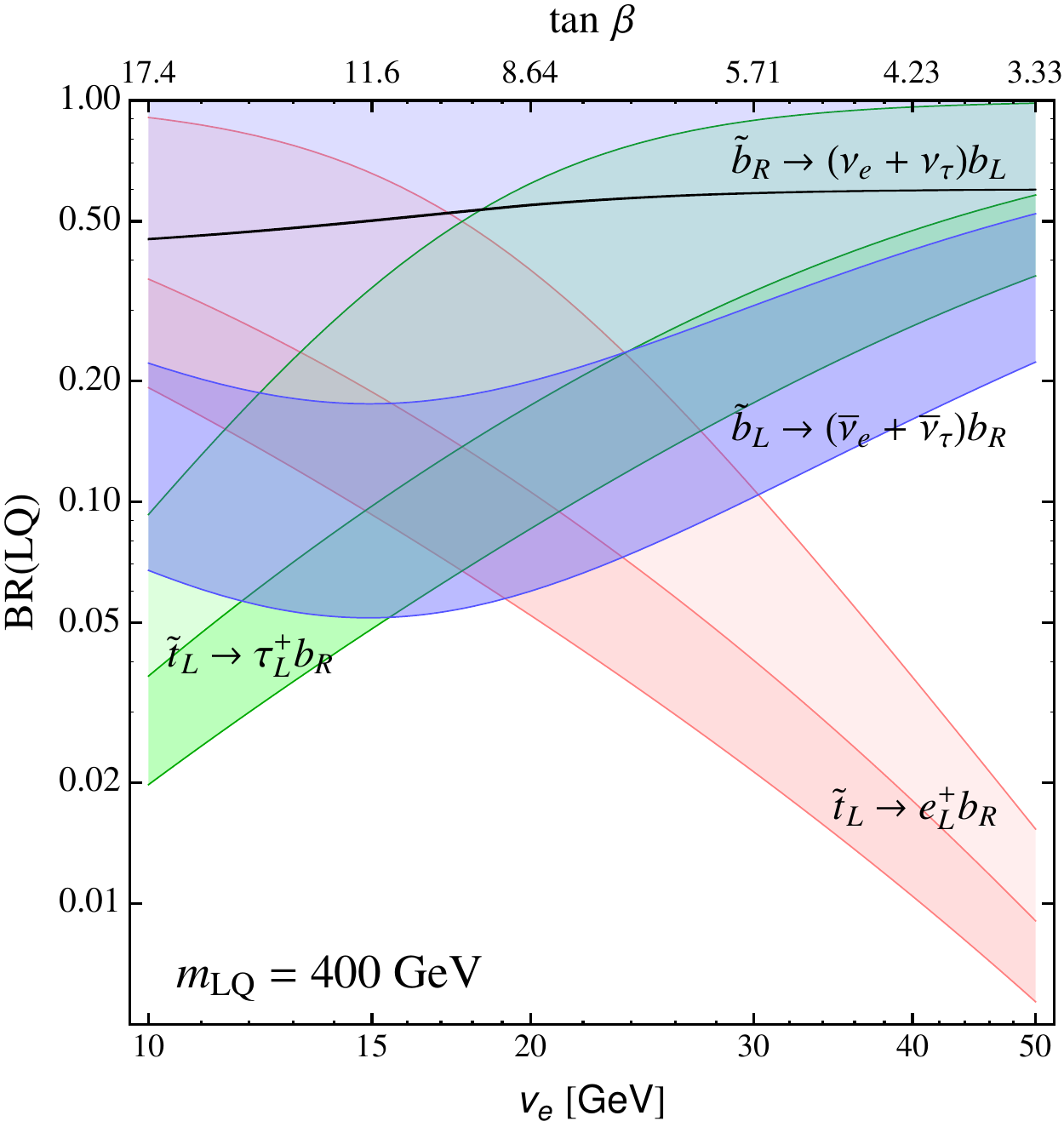}
\end{tabular}
\end{center}
\caption{\footnotesize{Left panel: current bounds on lepto-quark
masses from three channels: $eejj$ (blue), $\tau\tau bb$ (red) and
$\nu\nu bb$ (green), as a function of the lepto-quark branching
fraction into the corresponding channel (based on the CMS analyses
\protect\cite{:2012dn,CMS-PAS-EXO-11-003,CMS-PAS-EXO-12-002}).  Right
panel: Branching fractions into lepto-quark channels for $m_{\rm LQ} =
400~{\rm GeV}$, as a function of the sneutrino vev, for $M^D_{1} =
1$~TeV, $M^D_{2} = 1.5$~TeV, and scanning over $\lambda^S_u,
\lambda^T_u \in [0,1]$ and $|\mu| \in [0,200]$~GeV (darker areas) or
$|\mu| \in [200,300]$~GeV (lighter areas).  We do not show the
channels involving a top quark.}}
\label{fig:LQBounds}
\end{figure}

Due to the identification of lepton number as an $R$-symmetry, there
exist lepto-quark (LQ) decays proceeding through the $LQD^c$ couplings.
These can be especially significant for the third generation squarks.
As discussed in Subsection~\ref{sec:squark3decays}, in our scenario we
expect: $\tilde{t}_L \rightarrow e^+_L b_R$, $\tilde{t}_L \rightarrow
\tau^+_L b_R$, $\tilde{b}_L \rightarrow (\bar{\nu}_e + \bar{\nu}_\tau)
b_R$, $\tilde{b}_R \rightarrow (\nu_e + \nu_{\tau}) b_L$, $\tilde{b}_R
\rightarrow e^-_L t_L$ and $\tilde{b}_R \rightarrow \tau^-_L t_L$.  It
may be feasible to use the channels involving a top quark in the final
state~\cite{Gripaios:2010hv}, but such searches have not yet been
performed by the LHC collaborations.  Thus, we focus on the existing
$eejj$~\cite{Aad:2011ch,:2012dn}, $\nu\nu
bb$~\cite{CMS-PAS-EXO-11-003} and $\tau\tau
bb$~\cite{CMS-PAS-EXO-12-002} searches, where in our case the jets
are really $b$-jets.\footnote{It would be interesting to perform the
$eejj$ search imposing a $b$-tag requirement that would be sensitive
to our specific signature.} The first and third searches have been
performed with close to $5~{\rm fb}^{-1}$ by CMS, while the second has
been done with $1.8~{\rm fb}^{-1}$.  In the left panel of
Fig.~\ref{fig:LQBounds}, we show the bounds from these searches on the
LQ mass as a function of the branching fraction of the LQ into the
given channel.  The bounds are based on the NLO strong pair-production
cross-section.  We see that the most sensitive is the one involving
electrons, while the one involving missing energy is the least
sensitive.  This is in part due to the lower luminosity, but also
because in the latter case the search strategy is different since one
cannot reconstruct the LQ mass.

In the right panel of Fig.~\ref{fig:LQBounds} we show the
corresponding branching fractions in our scenario as a function of the
sneutrino vev, assuming $m_{\rm LQ} = 400~{\rm GeV}$ (which, as we
will see, turns out to be the mass scale of interest).  We have fixed
$M^D_{2} = 1.5$~TeV and $M^D_{1} = 1$~TeV, and scanned over $\mu \in
[-300,300]$~GeV and $\lambda^S_u, \lambda^T_u \in [0,1]$, which is
reflected in the width of the bands of different colors.  We assume
that $\lambda'_{333}$ saturates Eq.~(\ref{lambdap333}).  The BR's are
rather insensitive to $\lambda^S_u$ and $\lambda^T_u$, but depend
strongly on $\mu$, especially when $|\mu| \gtrsim 200~{\rm GeV}$.  The
reason is that for larger $\mu$ the neutralinos and charginos become
too heavy, the corresponding channels close, and the LQ channels can
dominate.  This affects the decays of $\tilde{t}_L$ and $\tilde{b}_L$,
but not those of $\tilde{b}_R$ as can be understood by inspecting
Figs.~\ref{fig:tL}, \ref{fig:bL} and the right panel of
Fig.~\ref{fig:tRbR}.\footnote{Note, in particular, that the neutralino
decay channel of $\tilde{b}_R$ is always suppressed, so that its
branching fractions are insensitive to $\mu$, unlike the cases of
$\tilde{t}_L$ and $\tilde{b}_L$.  This is why the ``$\tilde{b}_R \to
(\nu_e + \nu_\tau) b_L$ band" in Fig.~\ref{fig:LQBounds} appears
essentially as a line, the corresponding BR being almost independent
of $\mu$.} The darker areas correspond to the region $|\mu| \in
[0,200]$, while the lighter ones correspond to $|\mu| \in [200,300]$.
We can draw a couple of general conclusions:
\begin{enumerate}

\item The $\nu\nu bb$ branching fractions are below the sensitivity of
the present search, except when the neutralino/chargino channels are
suppressed or closed for kinematic reasons.  Even in such cases, the
lower bound on $m_{\tilde{b}_L}$ is at most $350~{\rm GeV}$.  Note
that $\tilde{b}_R$ is unconstrained.

\item The $\tau\tau bb$ search, which is sensitive to BR's above
$0.3$, could set some bounds at \textit{large} $\ve$ in some
regions of parameter space.  Such bounds could be as large as
$520~{\rm GeV}$, but there is a large region of parameter space that
remains unconstrained.

\item The $eejj$ search , which is sensitive to BR's above $0.15$,
could set some bounds at \textit{small} $\ve$ in some regions of
parameter space.  Such bounds could be as large as $815~{\rm GeV}$ if
$\ve \sim 10~{\rm GeV}$ and the neutralino/chargino channels are
kinematically closed.  However, in the more typical region with $\mu
\lesssim 200~{\rm GeV}$ the bounds reach only up to $550~{\rm GeV}$ in
the small $\ve$ region.  Nevertheless, there is a large region of
parameter space that remains completely unconstrained.

\end{enumerate}

The latter two cases are particularly interesting since the signals
arise from the (LH) stop, which can be expected to be light based on
naturalness considerations.  In addition to the lessons from the above
plots, we also give the bounds for our benchmark scenario with
$M^D_{1} = 1$~TeV, $M^D_{2} = 1.5$~TeV, $\mu=200$~GeV, $\lambda^S_u =
0$ and $\lambda^T_u = 1$, assuming again that Eq.~(\ref{lambdap333})
is saturated.  We find that the $\nu\nu bb$ search requires
$m_{\tilde{b}_L} \gtrsim 350~{\rm GeV}$, and gives no bound on
$m_{\tilde{b}_R}$.  The $\tau\tau bb$ search gives a bound on
$m_{\tilde{t}_L}$ that varies from $380$ to $400~{\rm GeV}$ as $\ve$
varies from $20-50~{\rm GeV}$.  The $eejj$ search gives a bound on
$m_{\tilde{t}_L}$ that varies from $470$ down to $300~{\rm GeV}$ as
$\ve$ varies from $10-30~{\rm GeV}$.  The other regions in $\ve$
remain unconstrained at present.  In our benchmark, when
$m_{\tilde{b}_L} \sim 350~{\rm GeV}$, we expect $m_{\tilde{t}_L} \sim
380-390~{\rm GeV}$, depending on the scalar singlet and (small)
triplet Higgs vevs (and with only a mild dependence on $\ve$).
We conclude that in the benchmark scenario a $400~{\rm GeV}$
LH stop is consistent with LQ searches, while offering the prospect of
a LQ signal in the near future, possibly in more than one channel.

\medskip
\noindent
\textbf{Comment on LQ signals from $2^{\rm nd}$ generation squarks}
\medskip

We have seen that the RH strange squark has a sizable branching
fraction into the LQ channel, $\tilde{s}_R \to e^-_L j$, of order
$0.4-0.65$.  From the left panel of Fig.~\ref{fig:LQBounds}, we see
that the $eejj$ CMS lepto-quark search gives a bound of $m_{\tilde{s}}
\approx 530-630~{\rm GeV}$, which is quite comparable to (but somewhat
weaker than) the bounds obtained in Section~\ref{12Pheno}.  Thus, a LQ
signal associated to the RH strange squark is also an exciting
prospect within our scenario.

\subsection{Other Searches}

There are a number of searches specifically optimized for third
generation squarks.  In addition, there are somewhat more generic
studies with b-tagged jets (with or without leptons) that can have
sensitivity to our signals.  We discuss these in turn.

\medskip \underline{\textit{Direct stop searches}}: 
In the case of the top squark, different strategies are used to
suppress the $t\bar{t}$ background depending on the stop mass.
However, the searches are tailored to specific assumptions that are
not necessarily satisfied in our framework:

\vspace{-2mm}
\begin{itemize}

\item Perhaps the most directly applicable search to our scenario is
an ATLAS GMSB search~\cite{Aad:2012cz} ($\tilde{t}_1 \tilde{t}^*_1$
pair production with $\tilde{t}_1 \to t \tilde{\chi}^0_1$ or
$\tilde{t}_1 \to b \tilde{\chi}^+_1$ and finally $\tilde{\chi}^0_1 \to
Z \tilde{G}$ or $\tilde{\chi}^+_1 \to W^+ \tilde{G}$), so that the
topologies are identical to those for LH and RH stop pair-production
in our model, respectively, with the replacement of the light
gravitino by $\nu_e$ (although the various branching fractions are
different; see Figs.~\ref{fig:tL} and \ref{fig:tRbR} for our benchmark
scenario).  Ref.~\cite{Aad:2012cz} focuses on the decays involving a
$Z$, setting bounds of $\sigma \times \epsilon \times A =
18.2~(9.7)~{\rm fb}$ for their signal region SR1 (SR2).  Simulation of
our signal for our benchmark parameters and taking 400~GeV LH stops
gives $\epsilon \times A \approx 1.9\%~(1.7\%)$ for SR1 (SR2), which
include all the relevant branching fractions.  The corresponding bound
on the model cross section would then be $\sigma_{\tilde{t}_L
\tilde{t}_L} \approx 1~(0.6)$~pb.  However, a cross-section of 0.6~pb
is only attained for stops as light as 300~GeV, and in this case the
efficiency of the search is significantly smaller, as the phase space
for the $\tilde{t}_L \to \tilde{X}^{0+}_1 t$ decay closes (recall that
due to the LEP bound on the chargino, and the Higgsino-like nature of
our neutralino, the mass of $\tilde{X}^{0+}_1$ must be larger than
about 100~GeV).  We conclude that this search is not sufficiently
sensitive to constrain the LH stop mass.  Also, the requirement that
the topology contain a $Z$ gauge boson makes this search very
inefficient for the RH stop topology: $\tilde{t}_R \to
\tilde{X}^{+-}_1 b_R$, $\tilde{X}^{+-}_1 \to W^+ \nu_e$, so that no
useful bound can be derived on $m_{\tilde{t}_R}$.

\item There is a search targeted for stops lighter than the top
($\tilde{t} \to b \tilde{\chi}^+_1$, followed by $\tilde{\chi}^+_1 \to
W^+ \tilde{\chi}^0_1$).  This is exactly the topology for
$\tilde{t}_R$ production in our scenario (with $m_{\rm LSP} = 0$ for
the neutrino), but does not apply to $\tilde{t}_L$ since its decays
are dominated by lepto-quark modes in this mass region.  Fig.~4c
in~\cite{ATLAS-CONF-2012-070} shows that for a chargino mass of
$106~{\rm GeV}$, stop masses between 120 and 164~GeV are excluded.  As
the chargino mass is increased, the search sensitivity decreases, but
obtaining the stop mass limits would require detailed simulation in
order to compare to their upper bound $\sigma \times \epsilon \times A
= 5.2-11~{\rm fb}$.

\item There are also searches for stop pair production with $\tilde{t}
\to t \tilde{\chi}^0_1$.  A search where both tops decay
leptonically~\cite{ATLAS-CONF-2012-071} would yield the same final
state as for $\tilde{t}_R \tilde{t}^*_R$ production in our case
($b\bar{b}W^+W^-+\cancel{E}_T$ with the $W$'s decaying leptonically).
However, the kinematics is somewhat different than the one assumed
in~\cite{ATLAS-CONF-2012-071} which can impact the details of the
discrimination against the $t \bar{t}$ background, which is based on a
$M_{T2}$ analysis.  Indeed, we find from simulation that the $M_{T2}$
variable in our case tends to be rather small, and $\epsilon \times A$
for this analysis is below 0.1\% (including the branching ratios).
Therefore, this search does not set a bound on the RH stop in our
scenario.  

There is a second search focusing on fully hadronic top
decays~\cite{:2012si}, that can be seen to apply for $\tilde{t}_L
\tilde{t}_L^*$ production with $\tilde{t}_L \to t \tilde{X}_1^{0+}$
followed by $\tilde{X}_1^{0+} \to \bar{\nu}_e Z/h$.  For instance,
when both Z gauge bosons decay invisibly the topology becomes
identical to the one considered in the above analysis (where
$\tilde{t}_L \to t + \cancel{E}_T$).  Also when both $Z$'s decay
hadronically one has a jet + $\cancel{E}_T$ final state.  In fact,
although the analysis attempts to reconstruct both tops, the required
3-jet invariant mass window is fairly broad.  We find from simulation
that when ${\rm BR}(\tilde{X}_1^{0+} \to Z \bar{\nu}_e) = 1$, the
$\epsilon \times A$ of our signal is very similar to that in the
simplified model considered in~\cite{:2012si}.  However, when ${\rm
BR}(\tilde{X}_1^{0+} \to h \bar{\nu}_e) = 1$ we find that $\epsilon
\times A$ is significantly smaller.  Due to the sneutrino vev
dependence of these branching fractions in our model, we find (for
\textit{benchmark 1}) that this search can exclude $m_{\tilde{t}_L}$
in a narrow window around $400~{\rm GeV}$ for a large sneutrino vev
($\ve \sim 50~{\rm GeV}$).  For lower stop masses the search is
limited by phase space in the decay $\tilde{t}_L \to t
\tilde{X}_1^{0+}$, while at larger masses the sensitivity is limited
by the available ${\rm BR}(\tilde{X}_1^{0+} \to Z \bar{\nu}_e)$ (see
Fig.~\ref{fig:N1decayNeutrLSP}).  At small sneutrino vev no bound on
$m_{\tilde{t}_L}$ can be derived from this search due to the
suppressed branching fraction of the $Z$-channel.  We also find that
the RH stop mass can be excluded in a narrow window around $380~{\rm
GeV}$ from the decay chain $\tilde{t}_R \to \tilde{X}^{+-}_1 b_R$
followed by $\tilde{X}^{+-}_1 \to W^+ \nu_e$.  Although there are no
tops in this topology, it is possible for the 3-jet invariant mass
requirement to be satisfied, and therefore a bound can be set in
certain regions of parameter space.

There is a third search that allows for one hadronic and one leptonic
top decay~\cite{:2012ar}.  We find that it is sensitive to
the LH stop in a narrow window around $m_{\tilde{t}_L} \sim 380~{\rm
GeV}$ (for \textit{benchmark 1}).  However, we are not able to set a
bound on $m_{\tilde{t}_R}$ from this search.

\end{itemize}

We conclude that the present dedicated searches for top squarks are
somewhat inefficient in the context of our model, but could be
sensitive to certain regions of parameter space.  The most robust
bounds on LH stops arise rather from the lepto-quark searches
discussed in the previous section.  However, since the latter do not
constrain the RH stop, it is interesting to notice that there exist
relatively mild bounds (below the top mass) for $\tilde{t}_R$, as
discussed above, and perhaps sensitivity to masses around $400~{\rm
GeV}$.

\medskip 
\underline{\textit{Direct sbottom searches}}:
Ref.~\cite{CMS-PAS-SUS-12-017} sets a limit on the sbottom
mass of about $420~{\rm GeV}$, based on $\tilde{b} \tilde{b}^*$ pair
production followed by $\tilde{b} \to t \tilde{\chi}_1^-$ and
$\tilde{\chi}_1^- \to W^- \tilde{\chi}^0_1$ (for $m_{\tilde{\chi}^0_1}
= 50~{\rm GeV}$, and assuming BR's~$=1$).  This is essentially our
topology when $\tilde{b}_L \to t \tilde{X}^{-+}_1$ and
$\tilde{X}^{-+}_1 \to W^- \bar{\nu}_e$.  When kinematically open,
these channels indeed have BR close to one, so that the previous mass
bound would approximately apply (the masslessness of the neutrino
should not make an important difference).  However, ${\rm
BR}(\tilde{b}_L \to t \tilde{X}^{-+}_1)$ can be suppressed near
threshold, as seen in Fig.~\ref{fig:bL}.  For instance, if ${\rm
BR}(\tilde{b}_L \to t \tilde{X}^{-+}_1) = 0.5$, the mass bound becomes
$m_{\tilde{b}_L} \gtrsim 340~{\rm GeV}$.  The RH sbottom in our model
does not have a normal chargino channel (but rather a decay involving
an electron or tau, which falls in the lepto-quark category), so this
study does not directly constrain $m_{\tilde{b}_R}$.

\begin{wrapfigure}[29]{r}{0.55\textwidth}
\vspace{-5pt}
\centering
\begin{tabular}{cc}
\includegraphics[scale=0.7]{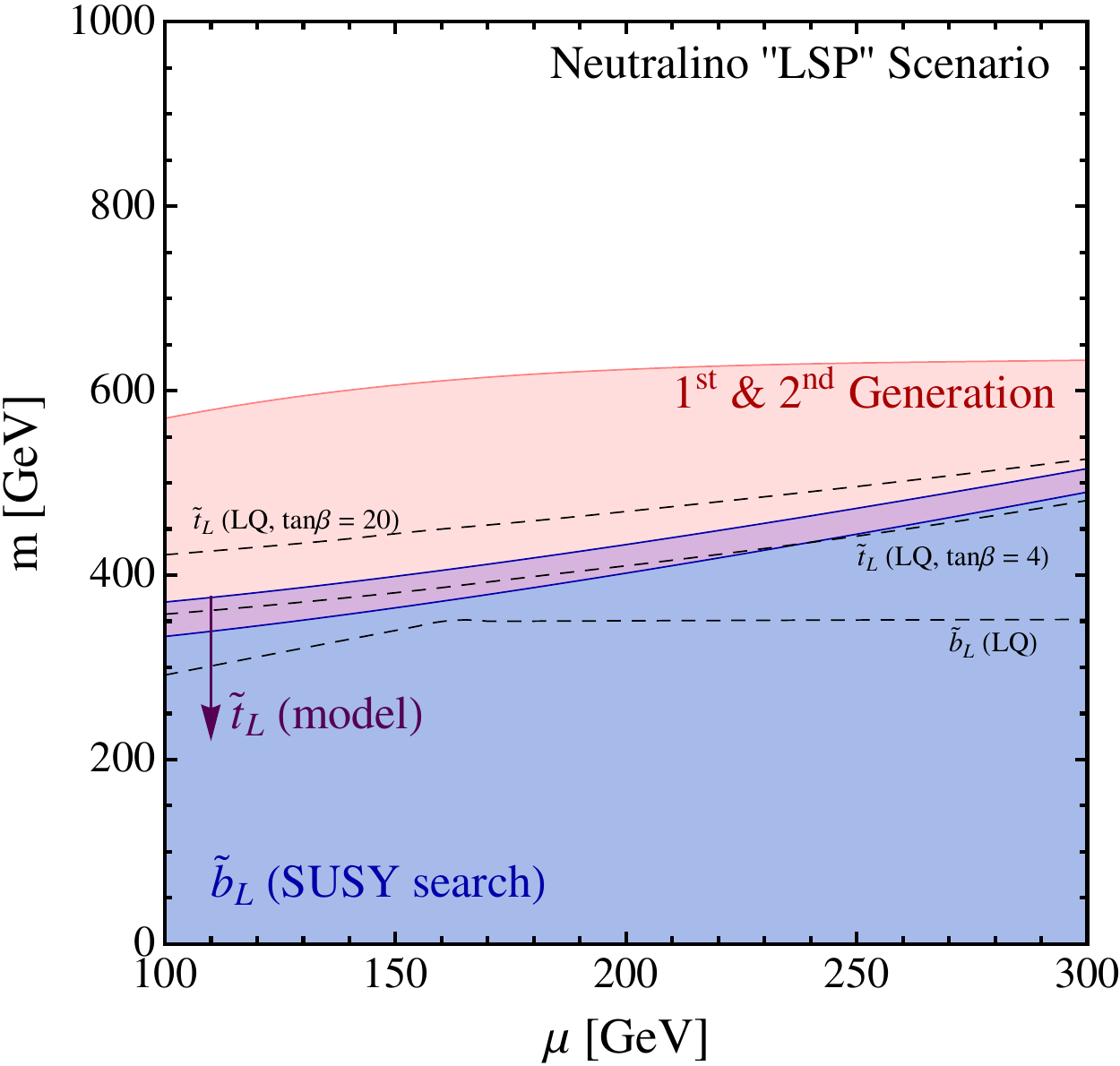}
\end{tabular}
\caption{\footnotesize{Summary of exclusions for the ``neutralino LSP
scenario" with $M^D_{1} = 1$~TeV, $M^D_{2} = 1.5$~TeV, $\lambda^S_u =
0$ and $\lambda^T_u = 1$, as a function of $\mu$ (approximately the
$\tilde{X}^{0+}_1$/$\tilde{X}^{+-}_1$ mass).  The exclusion on first
and second generation squarks comes from jets + $\cancel{E}_T$
searches.  The bound on $\tilde{b}_L$ come from direct
$b\,b\chi^0\chi^0$ SUSY searches, which are somewhat stronger than the
corresponding lepto-quark searches [dashed line marked $\tilde{b}_L
(LQ)$].  This implies, indirectly, a bound on $\tilde{t}_L$ about
30-50~GeV larger.  (We do not show the bound of $m_{\tilde{b}_R}
\approx 470~{\rm GeV}$, which is independent of $\mu$.)  We also
indicate by dashed lines the $\tilde{t}_L$ lepto-quark searches in the
most constraining cases: small $\tan\beta$ ($\tau\tau bb$ search) and
large $\tan\beta$ ($eejj$ search).  However, these can be completely
evaded for other values of $\tan\beta$.}}
\label{fig:AllBounds}
\end{wrapfigure}
CMS has recently updated their $\alpha_T$-based search for sbottom
pair production decaying via $\tilde{b} \to b + \cancel{E}_T$~
\cite{CMS-summary}.  For $m_{\rm LSP} = 0$ and ${\rm BR}(\tilde{b} \to
b + \cancel{E}_T) = 1$, they set an impressive bound of $m_{\tilde{b}}
\gtrsim 550~{\rm GeV}$.  Taking into account the branching fraction
for the $\tilde{b}_L \to (\bar{\nu}_e + \bar{\nu}_\tau) b_R$ decay
mode in our model, we find a lower bound that ranges from $m^{\rm
min}_{\tilde{b}_L} \approx 330~{\rm GeV}$ to $m^{\rm
min}_{\tilde{b}_L} \approx 490~{\rm GeV}$ as $\mu$ ($\approx
m_{\tilde{X}^0_1}$) ranges from $100~{\rm GeV}$ to $300~{\rm GeV}$
(for our benchmark values of the other model parameters).  The
corresponding lower bound on the RH sbottom mass is $m^{\rm
min}_{\tilde{b}_R} \approx 470~{\rm GeV}$, independent of $\mu$.  Here
we have assumed that $\lambda'_{333}$ saturates the bound in
Eq.~(\ref{lambdap333}), as we have been doing throughout.  If this
coupling is instead negligible, thus closing the $\nu_\tau$ channel,
we find that $m^{\rm min}_{\tilde{b}_L} \approx 290~{\rm GeV}$ to
$m^{\rm min}_{\tilde{b}_L} \approx 490~{\rm GeV}$ as $\mu$ ranges from
$100~{\rm GeV}$ to $300~{\rm GeV}$, while $m^{\rm min}_{\tilde{b}_R}
\approx 430~{\rm GeV}$, again independent of $\mu$.  We note that the
bound on $m_{\tilde{b}_L}$ sets indirectly, within our model, a bound
on the LH stop, since the latter is always heavier than $\tilde{b}_L$
(recall that the LR mixing is negligible due to the approximate
$R$-symmetry).  Typically, $m_{\tilde{t}_L} - m_{\tilde{b}_L} \sim
30-50~{\rm GeV}$.

\begin{center}
\begin{table}[t]
\center
    \begin{tabular}{ |l  |c  |c  |c |}
    \hline \rule{0mm}{5mm}
 \bf Signature & $\sigma \times \epsilon \times A$  [fb] & ${\cal L}~[{\rm fb}^{-1}$]  & \bf Reference 
  \\ [0.2em] 
  \hline \rule{0mm}{5mm}
  $\geq 1b$ + $\geq$ 4 jets + $1$ lepton + $\cancel{E}_T $ & $8.5-22.2$ & 2.05 & ATLAS \cite{ATLAS:2012ah}   
  \\ [0.2em]
   \hline \rule{0mm}{5mm}
  $\geq 2b$ + jets + $ \cancel{E}_T $ & $4.3-61$ & 2.05 & ATLAS  \cite{ATLAS:2012ah}
  \\ [0.2em] 
  \hline \rule{0mm}{5mm}
  $\geq 3b$ + jets + $ \cancel{E}_T $ & $1.5-5.1$ & 4.7 & ATLAS \cite{:2012pq}
  \\ [0.2em] 
  \hline
  \end{tabular}
\caption{\footnotesize{Generic searches for events with $ b$ tagged jets.}}
\label{btag}  
\end{table}  
\end{center}

\vspace{-7mm}

\medskip \underline{\textit{Generic searches sensitive to third
generation squarks}}: In Table~\ref{btag}, we summarize a number of
generic searches with b-tagging and with or without leptons.  We see
that the bounds on $\sigma \times \epsilon \times A$ range from a few
fb to several tens of fb.  We find that our model cross section for
these signatures (from pair production of $400~{\rm GeV}$
$\tilde{t}_L$, $\tilde{t}_R$, $\tilde{b}_L$ or $\tilde{b}_R$) are in
the same ballpark, although without taking into account efficiencies
and acceptance.  Thus, we regard these searches as potentially very
interesting, but we defer a more detailed study of their reach in our
framework to the future.

\medskip 
We summarize the above results in Fig.~\ref{fig:AllBounds}, where we
also show the bounds on the first two generation squarks
(Section~\ref{12Pheno}), as well as the lepto-quark bounds discussed
in Section~\ref{sec:leptoquarks} (shown as dashed lines).  The blue
region, labeled ``$\tilde{b}_L (\textrm{SUSY search})$", refers to the
search via two $b$-tagged jets plus $\cancel{E}_T$, which has more
power than the LQ search that focuses on the same final state.  The
region labeled ``$\tilde{t}_L (\textrm{model})$" refers to the bound
on $\tilde{t}_L$ inferred from the SUSY search on $\tilde{b}_L$.  We
do not show the less sensitive searches, nor the bound on
$\tilde{b}_R$, which is independent of $\mu$, and about $470~{\rm
GeV}$ in our benchmark scenario.

\section{Summary and Conclusions}
\label{predictions}

We end by summarizing our results, and emphasizing the most important
features of the framework.  We also discuss the variety of signals
that can be present in our model.  Although some of the individual
signatures may arise in other scenarios, taken as a whole, one may
regard these as a test of the leptonic $R$-symmetry.  The model we
have studied departs from ``bread and butter" SUSY scenarios (based on
the MSSM) in several respects, thereby illustrating that most of the
superpartners could very well lie below the 1~TeV threshold in spite
of the current ``common lore" that the squark masses have been pushed
above it.

There are two main theoretical aspects to the scenario: $a)$ the
presence of an approximate $U(1)_R$ symmetry at the TeV scale, and
$b)$ the identification of \textit{lepton number} as the $R$-symmetry
(which implies a ``non-standard" extension of lepton number to the new
physics sector).  The first item implies, in particular, that all BSM
fermions are Dirac particles.  A remarkable phenomenological
consequence is manifested, via the Dirac nature of gluinos, as an
important suppression of the total production cross section of the
strongly interacting BSM particles (when the gluino is somewhat
heavy).  This was already pointed out in Ref.~\cite{Kribs:2012gx} in
the context of a simplified model analysis.  We have seen here that
the main conclusion remains valid when specific model branching
fractions are included, and even when the gluino is not super-heavy
(we have taken as benchmark a gluino mass of 2~TeV).  We find that:
\begin{itemize}

\item The bounds on the first two generation squarks (assumed
degenerate) can be as low as $500-700~{\rm GeV}$, depending on whether
a slepton (e.g.~$\tilde{\tau}_L$) is lighter than the lightest
neutralino [$\tilde{X}^{0+}_1$ in our notation; see comments after
Eq.~(\ref{eq:neucomp})].  There are two important ingredients to this
conclusion.  The first one is the above-mentioned suppression of the
strong production cross section.  Equally important, however, is the
fact that the efficiencies of the current analyses deteriorate
significantly for lower squark masses.  For example, the requirements
on missing energy and $m_{\rm eff}$ (a measure of the overall energy
involved in the event) were tightened in the most recent jets +
$\cancel{E}_T$ analyses ($\sim 5~{\rm fb}^{-1}$) compared to those of
earlier analyses with $\lesssim 1~{\rm fb}^{-1}$.  As a result, signal
efficiencies of order one (for $1.4~{\rm TeV}$ squarks and $2~{\rm
TeV}$ gluinos in the MSSM) can easily get diluted to a few percent (as
we have found in the analysis of our model with $700~{\rm GeV}$
squarks and $2~{\rm TeV}$ gluinos).  This illustrates that the desire
to probe the largest squark mass scales can be unduly influenced by
our prejudices regarding the expected production cross sections.
\textit{We would encourage the experimental collaborations to not
overlook the possibility that lighter new physics in experimentally
accessible channels might be present with reduced production cross
sections.} Models with Dirac gluinos could offer a convenient SUSY
benchmark for optimization of the experimental analyses.  It may be
that a dedicated analysis would strengthen the bounds we have found,
or perhaps result in interesting surprises.

\end{itemize}

It is important to keep in mind that the previous phenomenological
conclusions rely mainly on the Dirac nature of gluinos, which may be
present to sufficient approximation even if the other gauginos are not
Dirac, or if the model does not enjoy a full $U(1)_R$ symmetry.
Nevertheless, the presence of the $U(1)_R$ symmetry has further
consequences of phenomenological interest, e.g.~a significant
softening of the bounds from flavor physics or
EDM's~\cite{Kribs:2007ac} (the latter of which could have important
consequences for electroweak
baryogenesis~\cite{Kumar:2011np,Fok:2012fb}).  In addition, the
specific realization emphasized here, where the $R$-symmetry coincides
with lepton number in the SM sector, has the very interesting
consequence that:
\begin{itemize}

\item A sizable sneutrino vev, of order tens of GeV, is easily
consistent with neutrino mass constraints (as argued
in~\cite{Frugiuele:2011mh}, \cite{FGKP}; see also
Ref.~\cite{Bertuzzo:2012su} for a detailed study of the neutrino
sector).  The point is simply that lepton number violation is tied to
$U(1)_R$ violation, whose order parameter can be identified with the
gravitino mass.  When the gravitino is light, neutrino Majorana masses
can be naturally suppressed (if there are RH neutrinos, the associated
Dirac neutrino masses can be naturally suppressed by small Yukawa
couplings).  We have also seen that there are interesting consequences
for the collider phenomenology.  Indeed, the specifics of our LHC
signatures are closely tied to the non-vanishing sneutrino vev (in
particular the neutralino decays: $\tilde{X}^{0+}_1 \to Z \bar{\nu}_e
/ h \bar{\nu}_e / W^- e^+_L$, or the chargino decay: $\tilde{X}^{+-}_1
\to W^+ \nu_e$).

\end{itemize}

This should be contrasted against possible sneutrino vevs in other
scenarios, such as those involving bilinear $R$-parity violation,
which are subject to stringent constraints from the neutrino sector.
Note also that the prompt nature of the above-mentioned decays may
discriminate against scenarios with similar decay modes arising from a
very small sneutrino vev (thus being consistent with neutrino mass
bounds in the absence of a leptonic $U(1)_R$ symmetry).  In addition,
the decays involving a $W$ gauge boson would indicate that the
sneutrino acquiring the vev is LH, as opposed to a possible vev of a
RH sneutrino (see e.g.~\cite{FileviezPerez:2012mj} for such a
possibility).

A further remarkable feature --explained in more detail in the
companion paper~\cite{FGKP}-- is that in the presence of
\textit{lepto-quark} signals, the connection to neutrino physics can
be an important ingredient in making the argument that an approximate
$U(1)_R$ symmetry is indeed present at the TeV scale.  In short:
\begin{itemize}

\item If lepto-quark signals were to be seen at the LHC (these arise
from the $LQD^c$ ``RPV" operator), it would be natural to associate
them to third generation squarks (within a SUSY interpretation, and
given the expected masses from naturalness considerations).  In such a
case, one may use this as an indication that some of the
$\lambda'_{i33}$ couplings are not extremely suppressed.  The neutrino
mass scale then implies a suppression of LR mixing in the LQ sector.
From here, RG arguments allow us to conclude that the \textit{three}
Majorana masses, several $A$-terms and the $\mu$-term linking the
Higgs doublets that give mass to the up- and down-type fermions (see
footnote~\ref{muterm}) are similarly suppressed relative to the
overall scale of superpartners given by $M_{\rm SUSY}$, which is the
hallmark of a $U(1)_R$ symmetry.  Therefore, the connection to
neutrino masses via a LQ signal provides strong support for an
approximate $U(1)_R$ symmetry in the {\it full} TeV scale Lagrangian,
and that this symmetry is tied to lepton number, which goes far beyond
the Dirac nature of gluinos.  In particular, it also implies a Dirac
structure in the fermionic electroweak sector, which would be hard to
test directly in many cases.  Indeed, in the benchmark we consider,
the lightest electroweak fermion states are Higgsino-like and hence
have a Dirac nature anyway, while the gaugino like states are rather
heavy and hence difficult to access.  What we have shown is that the
connection to neutrino masses can provide a powerful probe of the
Dirac structure even in such a case.

\end{itemize}

The (approximately) conserved $R$-charge, together with electric
charge conservation can impose interesting selection rules
(e.g.~allowing 2-body decays of the LH squarks, including
$\tilde{t}_L$, into a state involving an electron but not involving
the next lightest chargino, $\tilde{X}_1^{+-}$).  Of course,
eventually the approximate $R$-symmetry should become evident in the
decay patterns of the BSM physics.  The above lepto-quark signals, and
perhaps signals from resonant single slepton production~\cite{FGKP}
that may be present in more general RPV scenarios, can be amongst the
first new physics signals discovered at the LHC. Although by
themselves, these may admit interpretations outside the present
framework, the ``$L=R$" model has a variety of signals that provide
additional handles.  Some of them are summarized below.

\begin{figure}[t]
\begin{center}
\begin{tabular}{ccc}
\includegraphics[scale=0.8]{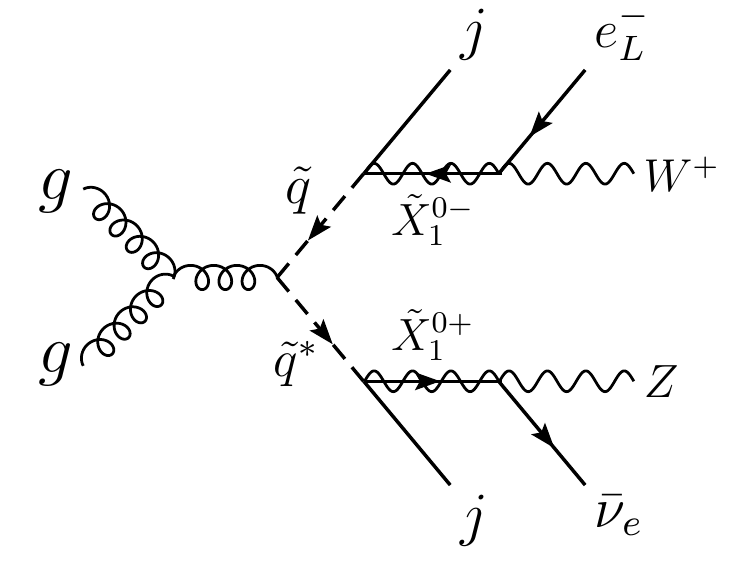} 
& \hspace{5mm} &
\includegraphics[scale=.8]{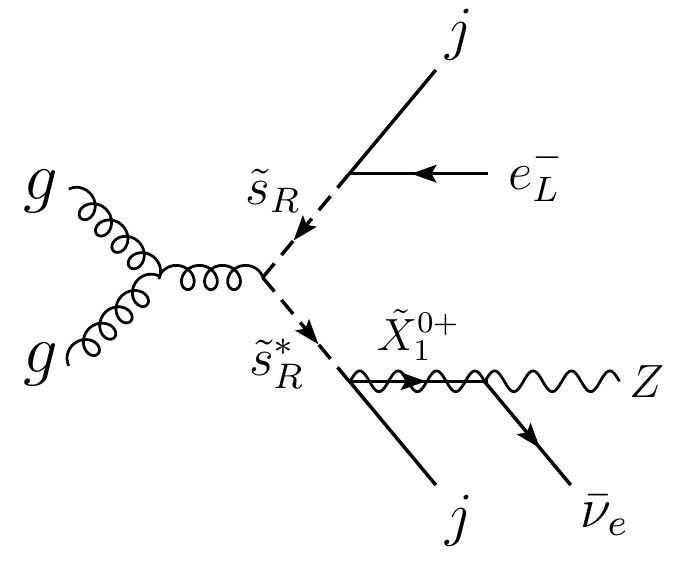} 
\end{tabular}
\end{center}
\caption{\footnotesize{Examples of processes with one fully visible
decay chain (thus allowing for mass reconstructions), while containing
a significant amount of missing energy from the second decay chain
(that can help for triggering and discrimination against
backgrounds). The arrows indicate the flow of $L=R$ number.}}
\label{CascadeDecays}
\end{figure}

The presence of fully visible decay modes, in addition to those
involving neutrinos, may give an important handle in the
reconstruction of SUSY events.  An example is displayed in the left
diagram of Fig.~\ref{CascadeDecays}, where one of the squarks decays
via $\tilde{q} \to j \tilde{X}^{0-}$ followed by $\tilde{X}^{0-} \to
e^-_L W^+$ (with a hadronic $W$), while the other squark gives off
missing energy in the form of neutrino(s), which can help in
increasing the signal to background ratio.  Although the combinatorics
might be challenging, there are in principle sufficient kinematic
constraints to fully reconstruct the event.

Perhaps more striking would be the observation of the lepto-quark
decay mode of the RH strange squark, as discussed at the end of
Section~\ref{sec:leptoquarks}.  The pure LQ event ($eejj$) would allow
a clean measurement of $m_{\tilde{s}_R}$, which could then be used in
the full reconstruction of ``mixed" events involving missing energy,
such as displayed in the right diagram of Fig.~\ref{CascadeDecays}.
Furthermore, if the gluinos are not too heavy, associated production
of different flavor squarks (one being $\tilde{s}_R$) through gluino
$t$-channel exchange, may allow an interesting measurement of the
second squark mass.  Both of these would offer discriminatory power
between scenarios with relatively light squarks ($\sim 700~{\rm GeV}$,
as allowed by the $R$-symmetry) versus scenarios with heavier squarks
(e.g.~$\gtrsim 1~{\rm TeV}$ with ultra-heavy gluinos, as might happen
within the MSSM), by providing information on the scale associated
with a putative excess in, say, the jets + $\cancel{E}_T$ channel.

An important possible feature of the present scenario is the presence
of final states with large third-generation multiplicities.  We have
seen how in the ``stau LSP scenario" squark pair production can result
in final states with multiple $\tau$'s, often accompanied by one or
more leptons ($e$ or $\mu$).  Although these may not be the discovery
modes due to a reduced efficiency compared to more standard squark
searches, they remain as an extremely interesting channel to test the
present scenario.  Similarly, processes such as $\tilde{q} \tilde{q}^*
\to j j \tilde{X}^{0+}_1 \tilde{X}^{0-}_1 \to j j \tau_L^+ \tau_L^-
\tilde{\tau}^+_L \tilde{\tau}^-_L \to j j \tau_L^+ \tau_L^- b_R
\bar{b}_R t_L \bar{t}_L$, display all the heavy third generation
fermions in the final state, and it would be extremely interesting to
conduct dedicated searches for this kind of topology.  Another
extremely interesting `no missing energy" topology arises in the
``neutralino LSP scenario": $\tilde{q} \tilde{q}^* \to j j
\tilde{X}^{0+}_1 \tilde{X}^{0-}_1 \to j j e^+_L e^-_L W^+ W^-$.  In
the lepto-quark sector, signals such as $\tilde{b}_R \tilde{b}_R^* \to
e^+_L e^-_L t_L \bar{t}_L / \tau^+_L \tau^-_L t_L \bar{t}_L$ have not
been looked for experimentally, but have been claimed to be feasible
in Ref.~\cite{Gripaios:2010hv}.  Needless to say, experimentalists are
strongly encouraged to test such topologies given the expected
importance of the third generation in connection to the physics of
electroweak symmetry breaking.

Finally, it is important to note that there may be alternate
realizations of an approximate $U(1)_R$ symmetry at the TeV scale.
For example, in the realization in which the $R$-symmetry is
identified with baryon number~\cite{Brust:2011tb,Brust:2012uf}, the
``LSP" decays predominantly to jets, giving rise to events with very
little missing energy and hence evading most of the current LHC
bounds.  So, these models may hide the SUSY signals under SM
backgrounds.  A remarkable feature of the realization studied in this
paper is that fairly ``visible" new physics could still be present
just were naturalness arguments could have indicated.  It is certainly
essential to test such (and possibly other) realizations if we are to
address one of the most important questions associated to the weak
scale: whether, and to what extent, EWSB is consistent with
naturalness concepts as understood within the well-tested effective
field theory framework.

\section*{Acknowledgments}

C.F. and T.G. are supported in part by the Natural Sciences and
Engineering Research Council of Canada (NSERC).  E.P. is supported by
the DOE grant DE-FG02-92ER40699.  P.K. has been supported by the DOE
grant DE-FG02-92ER40699 and the DOE grant DE-FG02-92ER40704 during the
course of this work.

\appendix

\section{Simplified Model Analysis}

In this appendix we provide details of the interpretation of a number
of ATLAS and CMS analysis within the simplified models defined for the
``neutralino LSP scenario" in Subsection~\ref{NeutralinoLSPScenario}.

\subsection{Topology (1) : $\tilde X_1^{0+} \rightarrow Z \bar{\nu}_e$}

The LHC searches  relevant  for this topology are: 
\begin{itemize}
\item jets + $\cancel{E}_T$~,
\item $Z(ll) $ + jets + $\cancel{E}_T$~,
\item multilepton ($\geq 3l$) + jets + $\cancel{E}_T$ (without Z veto).
\end{itemize}

\begin{wrapfigure}[28]{r}{0.55\textwidth}
\vspace{-5pt}
\centering
\includegraphics[scale=0.65]{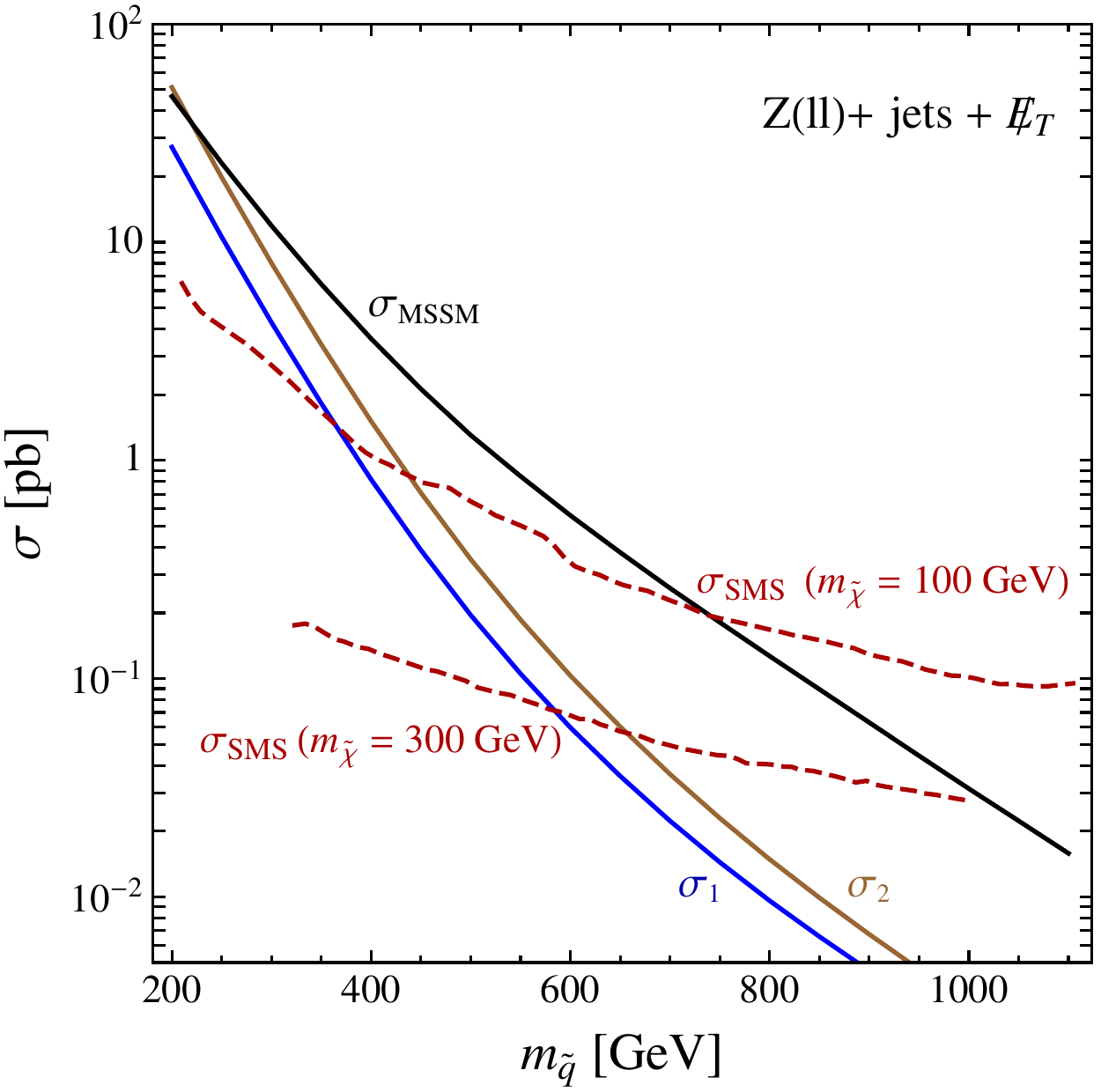}
\caption{Production cross-sections of $\tilde X^{0+}_1 \tilde
X^{0-}_1$ via squark decays, for 2 TeV gluinos (see
Subsection~\ref{sec:Simplified} for the definition of $\sigma_1$ and
$\sigma_2$, where $\sigma_1$ is the relevant one in our scenario).
For reference, we show the MSSM \textit{total} strong production
cross-section (squarks and gluinos).  The dashed lines are the SMS
upper limit from the CMS searches for the channel $Z(ll) +
\textrm{jets} + \cancel{E}_T$, assuming $m_{\tilde{\chi}}=100~{\rm
GeV}$ and $m_{\tilde{\chi}}=300~{\rm
GeV}$\protect\cite{CMS-PAS-SUS-11-016}.}
\label{Zll}
\end{wrapfigure}
We start with the dilepton $Z( l l)$ + jets + $\cancel{E}_T$ channel,
basing our discussion on a CMS analysis with $4.98~{\rm fb}^{-1}$
($\tilde g \tilde g$ production with $\tilde g \rightarrow q q \chi$
and $\chi \rightarrow Z$ LSP~\footnote{Note that this topology is not
identical to ours, having two extra
jets.})~\cite{Chatrchyan:2012qka,ICHEP-CMS}.  Bounds are given for $x
= 1/4, 1/2$ and $3/4$.  Identifying $m_{\tilde{q}}$ with the ``gluino
mass", taking $m_{\rm LSP} = 0$, and adjusting the squark mass until
the experimental upper bound on $\sigma$ is matched by our theoretical
cross section, we find for $x=1/2$: $\sigma_1(m_{\tilde{q}} \approx
585~{\rm GeV}) \approx 0.07~{\rm pb}$ and $\sigma_2(m_{\tilde{q}}
\approx 650~{\rm GeV}) \approx 0.06~{\rm pb}$.  What this means is
that this topology/analysis gives a lower bound of $m_{\tilde{q}}
\approx 585~{\rm GeV}$ when $\tilde X_1^{0+}$ is produced as in our
scenario, and of $m_{\tilde{q}} \approx 650~{\rm GeV}$ in a scenario
where \textit{all} the squarks decay into neutralino plus jet,
followed by the decay $\tilde X_1^{0+} \rightarrow Z \bar{\nu}_e$ with
${\rm BR} = 1$ ($\sigma_1$ and $\sigma_2$ are computed as explained in
Subsection~\ref{sec:Simplified}).  For a lighter $\tilde X_1^{0+}$
($x=1/4$), the corresponding bounds are $m_{\tilde{q}} \approx
360~{\rm GeV}$ and $440~{\rm GeV}$, respectively.  All of these can be
read also from Fig.~\ref{Zll}.

As a check, and to evaluate the effect of the additional two jets in
the topology considered in~\cite{Chatrchyan:2012qka} compared to the
squark pair-production of our case, we have obtained the $\epsilon
\times A$ from simulation of our signal ($\tilde{q}\tilde{q}$
production with $\tilde{q}\to q X_1^{0+}$ and $\tilde X_1^{0+}
\rightarrow Z \bar{\nu}_e$, taking $m_{\tilde{X}^0} = 200~{\rm GeV}$)
in the various signal regions of the CMS analysis.\footnote{We have
also simulated the case of $\tilde g \tilde g$ production with $\tilde
g \rightarrow q q \chi$ and $\chi \rightarrow Z$ LSP, taking 900~GeV
gluinos, heavy (5~TeV) squarks, a massless LSP and $x=1/2$,
i.e.~$m_\chi = 450~{\rm GeV}$.  We reproduce the $\epsilon \times A$
in~\cite{Chatrchyan:2012qka} within 30\%.} We find that the strongest
bound arises from the ``MET Search" with $\cancel{E}_T > 300~{\rm
GeV}$ (with $\epsilon \times A \approx 1.5\%$, including the branching
fractions of the $Z$), and corresponds to a model cross-section of
about $40~{\rm fb}$.  This translates into the bounds $m_{\tilde{q}}
\gtrsim 640~{\rm GeV}$ (based on $\sigma_1$) and $m_{\tilde{q}}
\gtrsim690~{\rm GeV}$ (based on $\sigma_2$), which are somewhat
stronger than above.

For the jets + $\cancel{E}_T$ signal we use an ATLAS search with
$5.8~{\rm fb}^{-1}$~\cite{Aad:2012hm}, which includes five different
signals regions depending on the jet multiplicity.  In order to apply
this analysis, we estimate the efficiency times acceptance in our
model in the different signal regions by simulating our signal
($\tilde{X}_1^{0+} \tilde{X}_1^{0-} j
j$ production via the processes defining $\sigma_1$, followed by
$\tilde{X}_1^{0+} \to Z \bar{\nu}_e$ with $\textrm{BR} = 1$), and then
applying the cuts in~\cite{Aad:2012hm}.  Our topology, and our model
in general, is distinguished by long cascade decays, and we find that
the strongest bound arises from signal region D (tight), i.e.~a 5 jet
region, setting a bound on the signal cross-section of about $20~{\rm
fb}$.  We find a lower limit of $m_{\tilde q} \sim 635$~GeV (based on
$\sigma_1$), and $m_{\tilde q} \sim 685$~GeV (based on $\sigma_2$).
We conclude that the bounds from this analysis are very comparable to
those from the $Z( l l)$ + jets + $\cancel{E}_T$ channel.  We note
that CMS has a $M_{T2}$-based Simplified Model analysis of the jets +
$\cancel{E}_T$ signature with $4.73$~fb$^{-1}$~\cite{:2012jx} (SMS:
$\tilde g \tilde g$ production with $\tilde g \rightarrow q q +
\textrm{LSP}$).  Applying the procedure detailed at the end of
Subsection~\ref{sec:Simplified}, we find that $\sigma_1(m_{\tilde{q}}
\approx 350~{\rm GeV}) \times {\rm BR}(Z \to jj)^2\approx 0.9~{\rm
pb}$ and $\sigma_2(m_{\tilde{q}} \approx 440~{\rm GeV}) \times {\rm
BR}(Z \to jj)^2 \approx 0.4~{\rm pb}$.  The fact that these limits are
much weaker than those obtained from the ATLAS study may be related in
part to the additional hard jets that differentiate the gluino from
the squark pair production topology.

There are no SMS limits on multilepton searches applicable to our
topologies, but there are a number of model-independent upper bounds
on $\sigma \times \epsilon \times A$, as summarized in
Table~\ref{sigma}.  However, putting in the ${\rm BR}(Z \to l^+ l^-)$
and taking into account the general lessons from the computed
efficiencies for ``Topologies (3) and (4)" below, we conclude that
such searches are less sensitive than the previous two searches.

\begin{center}
\begin{table}
\center
\begin{tabular}{ |l  |c  |c  |l |}
   \hline \rule{0mm}{5mm}
 \bf Search &  $ \sigma \times \epsilon \times A $ [fb]  & ${\cal L}~[{\rm fb}^{-1}$]  & \bf Reference
  \\ [0.2em]
  \hline \rule{0mm}{5mm}
 \multirow{2}{*}{1 lepton}   &  $1.1-1.7$ & 5.8 & ATLAS~\cite{ATLAS-CONF-2012-104}
 \\ [0.2em] 
 &  $1-2$ & 4.7 & ATLAS~\cite{:2012ms}
 \\ [0.2em] 
 \hline \rule{0mm}{5mm}
2 OS leptons &  $1-5$ &   4.98 &  CMS~\cite{Chatrchyan:2012te} 
\\ [0.2em] 
 \hline \rule{0mm}{5mm}
2 SS leptons  &  1.6 &   2.05  & ATLAS~\cite{ATLAS:2012ai}
\\ [0.2em] 
 \hline \rule{0mm}{5mm}
$Z (l^+ l^-)$ &  $0.6-8$ &   4.98  &  CMS~\cite{Chatrchyan:2012qka}  
 \\ [0.2em] 
 \hline \rule{0mm}{5mm}
   \multirow{2}{*}{Multilepton}  &  1.5 (no $Z$), 3.5 ($Z$) &  2.06  & ATLAS~\cite{ATLAS-CONF-2012-001}
  \\ [0.2em] 
 &  $1-2$ & 4.7 & ATLAS~\cite{:2012ms}
  \\ [0.2em] 
 \hline
\end{tabular}
 \caption{\footnotesize{Upper limits on $\sigma \times \epsilon \times A$ for a
 number of leptonic channels, with the corresponding luminosity and
 the ATLAS or CMS reference.}}
\label{sigma}
\end{table}
\end{center}
%

\subsection{Topology (2) : $\tilde X_1^{0+} \rightarrow h  \bar{\nu}_e$}

For this topology we use the ATLAS jets + $\cancel{E}_T$
search~\cite{Aad:2012hm} since the Higgs decays mostly into hadrons.
This topology is characterized by a high jet multiplicity, as was the
case with the hadronic $Z$ of the previous topology.  The efficiency
times acceptance is the same as in the case studied above (with a $Z$
instead of $h$), so that the bound on the model cross section is about
$20$~fb.  We find a lower limit of $m_{\tilde q} \sim 605$~GeV (based on
$\sigma_1$), and $m_{\tilde q} \sim 655$~GeV (based on $\sigma_2$).

Note that, since the Higgs decays predominantly into $b \bar b$,
searches with $b$ tagged jets are interesting for this topology.  In a
search for final states with $\cancel{E}_T$ and at least three b-jets
(and no leptons), ATLAS sets a bound on the corresponding visible
cross section of about 2~fb~\cite{:2012pq}.  However, simulating our
signal (for $700$~GeV squarks) in MG5 $+$ Pythia $+$ Delphes, we find
an extremely small efficiency for the present topology: $\epsilon
\times A \approx 10^{-4}$ for their signal regions SR4-L and SR4-M
(and much smaller efficiencies for the other SR's).  This arises from
the aggressiveness of the $\cancel{E}_T$ requirement and the combined
efficiency of tagging three $b$-jets.  As a result we infer a very
mild bound on the model cross section of about $18$~pb, and no
meaningful bound on the squark masses, as such a cross section can be
reached only for squarks as light as a couple hundred GeV, where the
$\epsilon \times A$ would be even smaller.  Nevertheless, it would be
interesting to optimize such an analysis for the present model (with
suppressed production cross sections), and furthermore to try to
reconstruct $b\bar{b}$ resonances at about $125~{\rm GeV}$.

The leptonic searches are not constraining due to the significant
suppression from the Higgs branching fraction into final states that
might involve leptons.

\subsection{Topology (3) : $\tilde X_1^{0+} \rightarrow  W^- e^+_L$}

In this case the two relevant searches are: jets + two leptons
\textit{without} $\cancel{E}_T$, and multileptons + jets +
$\cancel{E}_T$.  The first signal has a branching fraction of ${\rm
BR}(W \to jj)^2 \approx 0.45$.  However, at the moment there are no
searches that constrain this topology, since these typically include
important cuts on the missing transverse energy.  It would be
interesting to perform a dedicated search for this signal.  Here we
focus on the existing multilepton searches.  CMS has a detailed
analysis including a large number of
channels~\cite{Chatrchyan:2012mea}.  Unfortunately, the results are
model-dependent and no information on $\sigma \times \epsilon \times
A$ for the different signal regions is provided.  ATLAS has a $\geq 4$
leptons ($\!  \mbox{} + \textrm{jets} + \cancel{E}_T$) search with and
without $Z$ veto~\cite{ATLAS-CONF-2012-001}.  Their upper limit (with
a $Z$ veto) is $\sigma \times \epsilon \times A \approx 1.5~{\rm
fb}$.~\footnote{This corresponds to combining a number of channels
with different flavor composition, not all of which are present in our
model.  Thus, this result provides only an estimate for the possible
bound in our model from such a multi-lepton search.} We find from
simulation of our signal that, for this analysis, $\epsilon \times A
\approx 0.02$ (which includes the branching fractions of the $W$
decays).  We can therefore set a limit of $m_{\tilde{q}} \gtrsim
580$~GeV (based on $\sigma_1$) and $m_{\tilde{q}} \gtrsim 630$~GeV
(based on $\sigma_2$), corresponding to a model cross section of about
$75$~fb.

\subsection{Topology (4) : $\tilde X_1^{+-} \rightarrow  W^{+} \nu_e$}

In this case the relevant LHC searches are:
\begin{itemize}
\item jets + $\cancel{E}_T$~,
\item 1 lepton + jets + $\cancel{E}_T$~,
\item OS dileptons + jets + $\cancel{E}_T$~.
\end{itemize}
We can use again the ATLAS bound on jets + $\cancel{E}_T$ discussed
above.  In this case, however, the efficiency times acceptance turns
out to be smaller.\footnote{From simulation via MG5 $+$ Pythia $+$
Delphes of $\tilde{X}^{+-}_1 \tilde{X}^{-+}_1 j j$ via the processes
in the definition of $\sigma_1$ (see
Subsection~\ref{sec:Simplified}).} The strongest constraint arises
again from signal region D (tight) in~\cite{Aad:2012hm}, and
gives an upper bound on our model cross section of about 60~fb.
This translates into $m_{\tilde{q}} \gtrsim 530$~GeV (based on $\sigma_1$)
and $m_{\tilde{q}} \gtrsim 650$~GeV (based on $\sigma_2$).

In a multi-lepton study, the ATLAS collaboration has considered our
simplified model (model C in~\cite{:2012ms}), except that \textit{all}
the squarks are assumed to decay with unit branching fraction through
the chargino channel (i.e.~the process characterized by $\sigma_2$).
If we assume the same efficiency times acceptance for the process in
our scenario, i.e.~based on $\sigma_1$, we read from their Fig.~10 the
bounds $m_{\tilde q} \gtrsim 410$~GeV (based on $\sigma_1$), and
$m_{\tilde q} \gtrsim 500$~GeV (based on $\sigma_2$), which correspond
to model cross sections of about $0.35~{\rm pb}$.  

ATLAS also has a search for exactly 1 lepton + $\geq 4$ jets +
$\cancel{E}_T$, setting a bound on $\sigma \times \epsilon \times A
\approx 1.1-1.7~{\rm fb}$, depending on whether the lepton is an
electron or a muon~\cite{ATLAS-CONF-2012-104}.  Simulation of the
above process ($\tilde{q}\tilde{q}$ production with $\tilde{q}\to q
X_1^{+-}$ followed by $\tilde X_1^{+-} \rightarrow W^{+} \nu_e$,
taking $m_{\tilde{X}^\pm} = 200~{\rm GeV}$) gives that the ATLAS
analysis has $\epsilon \times A \approx 10^{-3}$ (this includes the
branching fractions for the $W$ decays).  We see that the efficiency
is quite low.  This is due, in part, to the fact that the analysis
requires at least four jets with $p_T > 80~{\rm GeV}$.  While the two
jets from squark decays easily pass the $p_T$ cut, the other two jets
arise from a $W$ decay and are softer (the other $W$ decaying
leptonically).  But when the quarks are sufficiently boosted to pass
the $p_T$ cut, they also tend to be collimated, and are likely to be
merged into a single jet.  As a result, using an upper bound on the
model cross section of order $1~{\rm pb}$, we get a rather mild bound
of $m_{\tilde q} \gtrsim 350$~GeV (based on $\sigma_1$), and
$m_{\tilde q} \gtrsim 430$~GeV (based on $\sigma_2$).

For the OS dilepton signal CMS sets a bound of $\sigma \times \epsilon
\times A \lesssim 1-5~{\rm fb}$ with
4.98~fb$^{-1}$~\cite{Chatrchyan:2012te}.  Our simulation gives
$\epsilon \times A \sim 10^{-3}$ (including the $W$ BR's), resulting
again in an upper bound on the model cross section of about $1~{\rm
pb}$, and the same mild bounds as above.

CMS studies a simplified model ($\tilde g \tilde g$ production with
$\tilde g_1 \rightarrow q q \chi^0$ and $\tilde g_2 \rightarrow q q
\chi^\pm$) with $4.98~{\rm fb}^{-1}$~\cite{CMS-PAS-SUS-11-016}, where
the neutralino $\chi^0$ is the LSP, while $\chi^{\pm}$ decays into
$W^{\pm} \chi^0$.  Therefore, as in our scenario, a single lepton is
produced via $W$ decay, although there are two extra hard jets
compared to our case from the gluino versus squark production.  From
Fig.~8 of~\cite{CMS-PAS-SUS-11-016}, with $m_{\rm LSP} = 0$, we find
that our cross section, in the range $300~{\rm GeV} < m_{\tilde{q}} <
800~{\rm GeV}$, is more than an order of magnitude below the current
sensitivity.  Here we used our $\sigma_1$ including the branching for
exactly one of the W's to decay leptonically. 

Finally, our model has very suppressed SS dilepton signals due to the
Dirac nature of the gluino,\footnote{Two SS positrons can be obtained
through $u_L u_R$ production consistent with the Dirac nature of
gluinos, but the SS dilepton cross section is small, at the $0.2~{\rm
fb}$ level for $700~{\rm GeV}$ squarks in the ``neutralino LSP"
scenario.  In the ``stau LSP" scenario the SS dilepton + jets +
$\cancel{E}_T$ signal can reach $1-2$~fb.} so that no interesting
bounds arise from this search.  In conclusion, for this simplified
topology, the strongest bounds again arise from the generic jets +
$\cancel{E}_T$ searches, although it should be possible to optimize
the leptonic searches to our signal topologies to obtain additional
interesting bounds.

\bibliographystyle{utphys}
\bibliography{Paper2}

\end{document}